\documentclass[11pt]{article}
\pdfoutput=1

\usepackage[utf8]{inputenc}
\usepackage{multirow}
\usepackage{amsmath, amsfonts, amssymb}
\usepackage{comment}
\usepackage{graphicx}
\usepackage{float}
\usepackage{psfrag}
\usepackage{amsthm}
\usepackage{caption}
\usepackage{subcaption}
\usepackage{empheq}
\newcommand*\widefbox[1]{\fbox{\hspace{0.5em}#1\hspace{0.5em}}}
\usepackage{import}
\usepackage{mathrsfs}
\usepackage{bm}
\usepackage[usenames,dvipsnames,svgnames,table]{xcolor}
\usepackage{enumerate}
\usepackage{arydshln}
\usepackage{soul}
 \usepackage{slashed}
 \usepackage{adjustbox}
\usepackage{mathrsfs}
\usepackage{mathalfa}
\usepackage{longtable}
\allowdisplaybreaks
 \usepackage{a4wide}
  \usepackage{tikz}
  \usepackage{tikz-cd}
  \usetikzlibrary{shapes.geometric,arrows}
  \usepackage{tcolorbox}
\tikzstyle{startstop} = [rectangle, rounded corners, minimum width=3cm, minimum height=1cm,text centered, draw=black, fill=red!30]
\tikzstyle{io} = [trapezium, trapezium left angle=70, trapezium right angle=110, minimum width=3cm, minimum height=1cm, text centered, draw=black, fill=blue!30]
\tikzstyle{process} = [rectangle, minimum width=3cm, minimum height=1cm, text centered, draw=black, fill=orange!30]
\tikzstyle{decision} = [diamond, minimum width=3cm, minimum height=1cm, text centered, draw=black, fill=green!30]
\tikzstyle{arrow} = [thick,->,>=stealth]
\tikzcdset{row sep/normal=0.75cm}

\usepackage[labelformat=simple]{subcaption}    %%Adding option to remove parenthesis
  \usepackage{color}
  \definecolor{dark-gray}{gray}{0.20}
  \definecolor{gray}{gray}{0.30}
  \definecolor{light-gray}{gray}{0.80}
  \definecolor{dark-red}{rgb}{0.7,0,0}
  \definecolor{dark-green}{rgb}{0,0.5,0}
  \definecolor{dark-blue}{rgb}{0.3,0.3,0.7}
  \definecolor{light-blue}{rgb}{0.8,0.8,1}
      \definecolor{swamp}{RGB}{240, 199, 197}

  \usepackage{pifont}

\usepackage{newunicodechar} % To write the definition on the next line
\usepackage{setspace}

\newcommand{\be}{\begin{equation}}
\newcommand{\ee}{\end{equation}}

\def\be{\begin{equation}}
\def\ee{\end{equation}}
\def\bea{\begin{eqnarray}}
\def\eea{\end{eqnarray}}

\newcommand{\beq}{\begin{equation}}  \newcommand{\eeq}{\end{equation}}
\newcommand{\bal}{\begin{aligned}}   \newcommand{\eal}{\end{aligned}}
\def\beqa{\begin{eqnarray}}
\def\eeqa{\end{eqnarray}}

\newcommand{\dd}{\mathrm{d}}

\newcommand{\I}{\text{Im}\,}

\newcommand{\volume}{\text{{\small} vol}\, }

\usepackage{stackengine}
\usepackage{calc}
\newlength\shlength
\newcommand\vv[2][0]{\setlength\shlength{#1pt}%
  \stackengine{-5.6pt}{$#2$}{\smash{$\kern\shlength%
    \stackengine{7.55pt}{$\mathchar"017E$}%
      {\rule{\widthof{$#2$}}{.57pt}\kern.4pt}{O}{r}{F}{F}{L}\kern-\shlength$}}%
      {O}{c}{F}{T}{S}}

\def\simleq{\; \raise0.3ex\hbox{$<$\kern-0.75em
      \raise-1.1ex\hbox{$\sim$}}\; }
   \def\simgeq{\; \raise0.3ex\hbox{$>$\kern-0.75em
      \raise-1.1ex\hbox{$\sim$}}\; }

\numberwithin{equation}{section}

\usepackage{jheppub}
\usepackage{hyperref}
\usepackage{cleveref}

\hypersetup{
	colorlinks=true,
	linkcolor=dark-blue,
	citecolor=dark-red,
	urlcolor=dark-green,
	linktoc=page
}

\theoremstyle{remark}

\crefname{appendix}{Appendix}{Appendices}

\title{\centering Morse-Bott inequalities, Topology Change and Cobordisms to Nothing }

\author{Ignacio Ruiz} 

\affiliation{Instituto de F\'{i}sica Te\'{o}rica UAM-CSIC, Universidad Aut\'{o}noma de Madrid, Cantoblanco, 28049 Madrid, Spain}
\affiliation{Departamento de F\'{i}sica Te\'{o}rica, Universidad Aut\'{o}noma de Madrid, Cantoblanco, 28049 Madrid, Spain}

\preprint{IFT-UAM/CSIC-24-154}
\emailAdd{ignacio.ruiz@uam.es}

\abstract{The Cobordism Conjecture predicts spacetime-ending configurations, such as Bubbles of Nothing (BoN), being commonplace. These correspond to vacuum decays in which the compactification manifold $\mathcal{C}_n$ shrinks to a point, with the instability expanding at the speed of light and leaving nothing (not even spacetime) behind. Most constructions of BoN or cobordisms to nothing found in the literature feature simple instances of $\mathcal{C}_n$ or singular cobordisms, which cannot be approached from the effective field theory. Assuming the solution mediating such decay to nothing is homeomorphic to a smooth description, we are able to go a step further, and obtain topological bounds on its homology for generic $\mathcal{C}_n$. Through the use of Morse-Bott theory we then translate this into information on the number and types of topology changes the compact manifold experiences as we move towards the tip of the bordism, as well as the location of possible cobordism defects. We illustrate our results with different detailed examples coming from String Theory. Furthermore, with this approach, we are able to study more complicated arrangements such as BoN collisions or intersection of End of the World branes. 

}

\begin{document}
\hypersetup{pageanchor=false}
\makeatletter
\let\old@fpheader\@fpheader

\makeatother
\maketitle

\newcommand{\remove}[1]{\textcolor{red}{\sout{#1}}}

\pagenumbering{roman}
\newpage
\pagenumbering{arabic}
\setcounter{page}{1} 
\section{Introduction}
\label{sec:intro}

One of the most surprising aspects of superstring theories is the prediction of $D=10$ as the critical dimension ($D=26$ for the bosonic string) of spacetime. In order to match the observed $d=4$ of our universe, it is necessary to somehow get rid of the $n=D-d=6$ extra directions. A straightforward way to do this is by \emph{compactifying} them, in such a way that they take values on a compact $n$-dimensional manifold $\mathcal{C}_n$.

This procedure of \emph{dimensional reduction} results in a low-energy $d$-dimensional effective field theory (EFT), which is valid for energies lower than the Kaluza-Klein scale. The choice of internal manifold, as well as other ingredients on it, such as fluxes or branes, will determine the field content of the EFT and its dynamics. The enormous set of different possible compactifications results in a large \emph{landscape} of potential theories.

In the presence of such many vacua/low energy descriptions, the decay or transition into another is a general expectation, associated for example to different local minima in flux compactifications \cite{Kachru:2002gs,Blanco-Pillado:2009lan}, as well as topology changing transitions to a different compact manifold \cite{Atiyah:2000zz,Acharya:2004qe}, such as flops \cite{Aspinwall:1993yb,Witten:1993yc,Greene:2000yb} or conifolds \cite{Strominger:1995cz,Greene:1995hu}. Generally speaking, topology changes are a generic expectation of Quantum Gravity, \cite{Giddings:1987cg,Greene:1990ud,Giveon:1994fu,Witten:1998zw,Garcia-Etxebarria:2018ajm,Demulder:2023vlo}.

Another type of instability was shown by Witten \cite{Witten:1981gj}, where the decay does not occur to a different vacuum, but rather to \emph{nothing}, this is, to a configuration without spacetime. This construction, which will be reviewed in more detail in section \ref{ss: BoN}, consists in a decay channel for a Minkowski vacuum with a single extra dimension compactified on a circle of radius $\mathsf{R}$. A sphere of size  $\mathsf{R}$ pops up, with spacetime ending smoothly on its surface, in such a way that its interior is non-accessible, with ``nothing'' (not even spacetime) being inside. The bubble expands with a velocity asymptotic to that of light, and as we approach its surface both the lower-dimensional curvature and the (normalized) scalar controlling the size of the circle blow up. This type of vacuum decay, dubbed \emph{Bubble of Nothing} (BoN) has been generalized to other compactification manifolds and internal ingredients \cite{Fabinger:2000jd,Dine:2004uw,Horowitz:2007pr,Yang:2009wz,Blanco-Pillado:2010xww,Blanco-Pillado:2010vdp,Brown:2010mf,Blanco-Pillado:2011fcm,Brown:2011gt,deAlwis:2013gka,Brown:2014rka,Blanco-Pillado:2016xvf,Ooguri:2017njy,Dibitetto:2020csn,GarciaEtxebarria:2020xsr,Bomans:2021ara,Draper:2021ujg,Draper:2021qtc,Petri:2022yhy,Bandos:2023yyo,Ookouchi:2024tfz,Heckman:2024zdo}.

\begin{figure}[h]
\begin{center}
\begin{subfigure}[b]{0.55\textwidth}
\center
\includegraphics[width=0.90\textwidth]{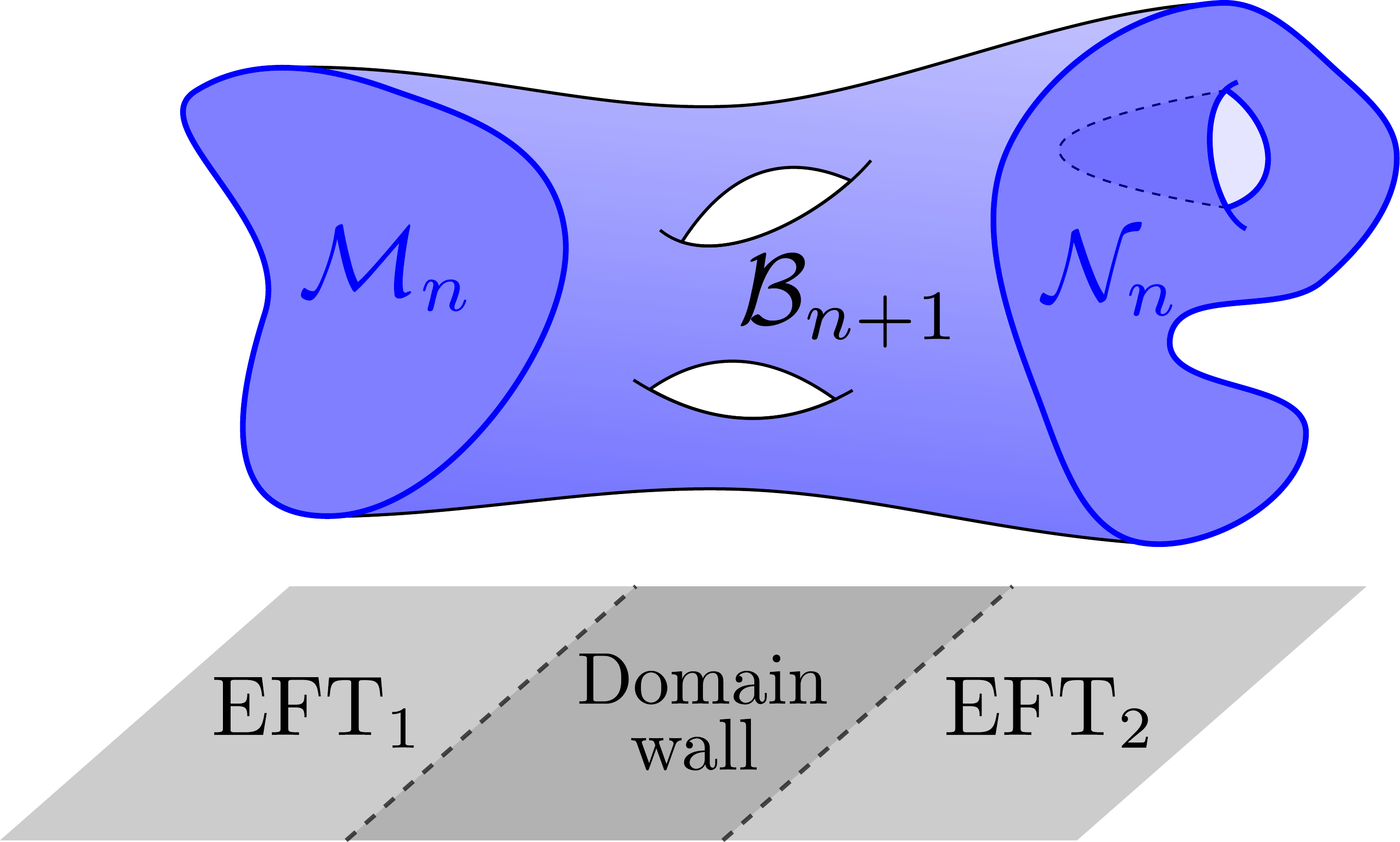}
\caption{\hspace{-0.3em}) Cobordism between two manifolds.} \label{ff.cob1}
\end{subfigure}
\begin{subfigure}[b]{0.44\textwidth}
\center
\includegraphics[width=0.87\textwidth]{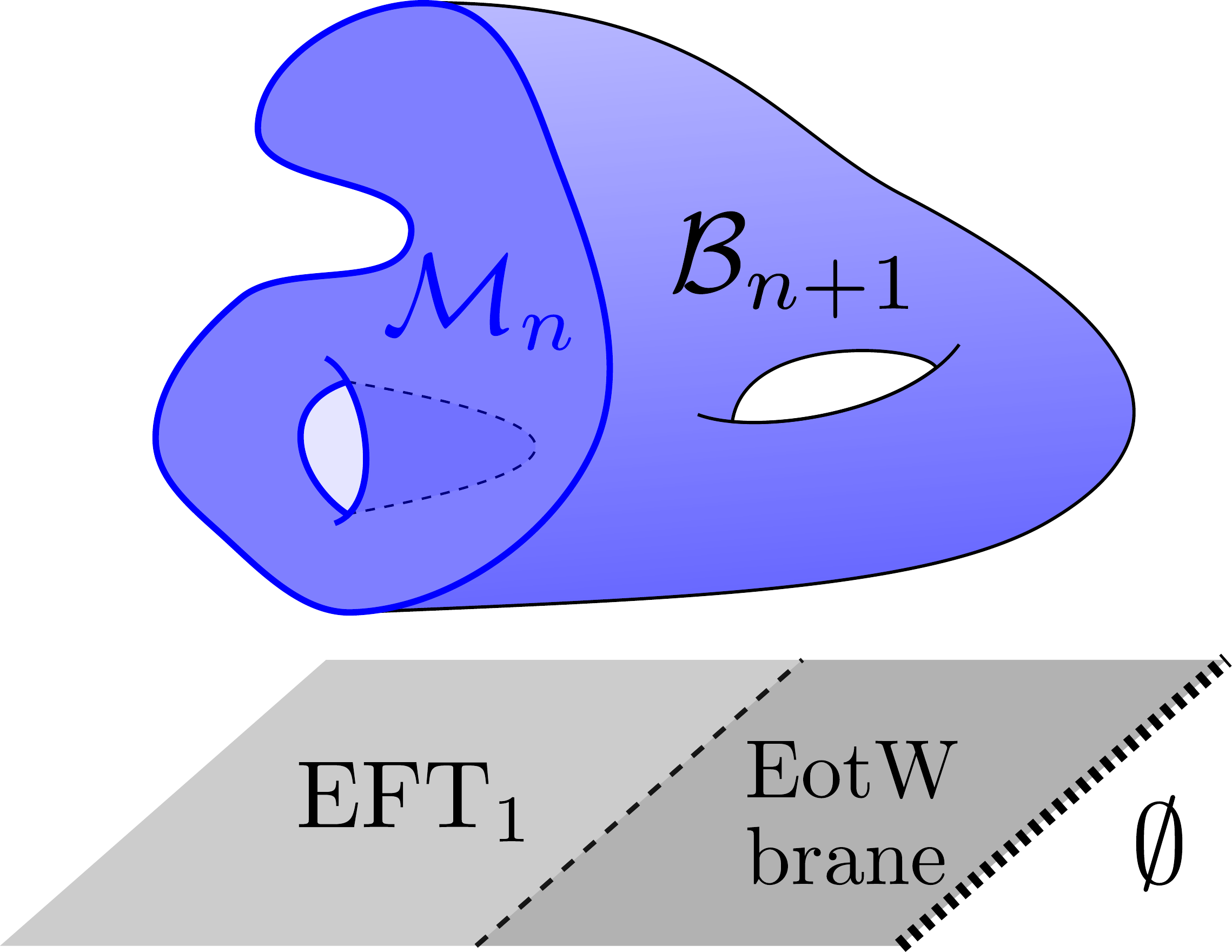}
\caption{\hspace{-0.3em}) Cobordism from a manifold to nothing.} \label{ff.cob2}
\end{subfigure}
\caption{Sketch of different bordisms between $n$-manifolds. In subfigure \ref{ff.cob1} the bordism $\mathcal{B}_{n+1}$ has as components of its boundary $\mathcal{M}_n$ and $\mathcal{N}_n$, and can be interpreted as a domain wall between the two EFTs resulting from compactification on each of the manifolds. In subfigure \ref{ff.cob2} $\mathcal{M}_n=\partial\mathcal{B}_{n+1}$, the bordism connects the EFT to nothing, serving as a boundary of spacetime, which from the lower-dimensional perspective we will call \emph{End of the World brane} in section \ref{ss: EotW}.
\label{f.COB}}
\end{center}
\end{figure}

Bubbles of nothing and other spacetime ending configurations, such as Ho\v rava-Witten walls \cite{Horava:1995qa,Horava:1996ma}, might seem at first a curiosity without further implications. However, as shown in \cite{McNamara:2019rup}, this is not the case at all. The \emph{Swampland program} \cite{Vafa:2005ui,Brennan:2017rbf,Palti:2019pca,vanBeest:2021lhn,Grana:2021zvf,Harlow:2022ich,Agmon:2022thq}, studies the set of constrains that an EFT with a consistent UV completion to Quantum Gravity must fulfill. One of the most established \emph{Swampland Conjectures} is the \textbf{No Global Symmetries} conjecture \cite{Banks:1988yz,Kallosh:1995hi,Banks:2010zn}, which postulates that there are no global charges in quantum gravity, so that any symmetry must be gauged or broken. A specific case of these global symmetries are \emph{topological} ones. Since we expect some topology changes to be dynamically allowed in a theory of Quantum Gravity, this notion of topology change is better encapsulated by that of \textbf{cobordism class}. Two $k$-dimensional compact manifolds $\mathcal{M}_k$ and $\mathcal{N}_k$ are said to be \emph{cobordant} if there exists a $(k+1)$ manifold $\mathcal{B}_{k+1}$ such that $\partial\mathcal{B}_{k+1}=\mathcal{M}\sqcup\overline{\mathcal{N}}$. $\mathcal{M}_k$ and $\mathcal{N}_k$ are then said to be \emph{(co)bordant} and $\mathcal{B}_{k+1}$ is said to be a \emph{bordism} between them. From a physical point, if compactifying our higher-dimensional theory on $\mathcal{M}_k$ and $\mathcal{N}_k$ results two distinct lower-energy descriptions, EFT$_1$ and EFT$_2$, then traversing $\mathcal{B}_{k+1}$ can be thought as crossing the domain wall relating the two. This is illustrated in Figure \ref{f.COB}. 

Two compactifications that can be interpolated through a bordism or, in other words, for which there is a dynamically allowed topology change between them, are part of the same \emph{cobordism class}. The topological charge associated to said classes is global rather than gauged \cite{McNamara:2019rup}, as one can take some spacetime $\mathcal{X}_d$, remove some submanifold $\mathcal{M}_k\subset\mathcal{X}_d$, and glue along it some other $\mathcal{N}_k$ in a different cobordism class then that of $\mathcal{M}_k$. The resulting spacetime is not cobordant to the original one, but far from the region where we have changed the local topology, both look the same, and the topology of this defect introduced cannot be inferred. Per the No Global Symmetries conjecture, this cannot occur in a gravitational theory with a consistent UV completion. The solution to this problem is this topological charge being trivial, or in other words, by requiring that our compactification manifold must be the boundary of an appropriate bordism, $\mathcal{M}_n=\partial\mathcal{B}_{n+1}$. This is known as the \textbf{Swampland Cobordism Conjecture} \cite{McNamara:2019rup}.

While the above might seem like a pretty abstract discussion about compactifications of QG, it has a radical implication: since then all compactifications of QG, say string theory, must be in the same cobordism class, they should be connected through some domain wall to \textbf{nothing}, this is, have spacetime ending boundary such as Witten's BoN.\footnote{As discussed in section \ref{ss: BoN}, an extra requirement will be that such decay is dynamically allowed, this being a general expectation in non-SUSY compactifications.} From the bordism point of view, this corresponds to the internal manifold smoothly capping off to a point, after which spacetime ends. If more stringy, singular objects are allowed, such as O-planes, then this is not necessarily the case, as the compact manifold itself can have boundaries, which themselves serve as domain walls to nothing. More exotic defects might be needed to kill cobordism classes to the trivial one, which has result in the prediction of new, often non-supersymmetric objects \cite{McNamara:2019rup,Montero:2020icj,Dierigl:2022reg}, but their inclusion in the bordism is likely to result in singular solutions.

Controlled solutions, solving the equations of motion, and describing explicit bordisms to nothing are rare in the literature, and usually correspond to simple compactification manifolds such as spheres or tori \cite{Fabinger:2000jd,Dine:2004uw,Horowitz:2007pr,Yang:2009wz,Blanco-Pillado:2010xww,Blanco-Pillado:2010vdp,Brown:2010mf,Blanco-Pillado:2011fcm,Brown:2011gt,deAlwis:2013gka,Brown:2014rka,Blanco-Pillado:2016xvf,Ooguri:2017njy,Dibitetto:2020csn,Bomans:2021ara,Draper:2021ujg,Draper:2021qtc,Petri:2022yhy,Bandos:2023yyo,Ookouchi:2024tfz,Heckman:2024zdo}. The few explicit examples for more complicated cases without the need of stringy defects are quite involved \cite{GarciaEtxebarria:2020xsr}. Smooth solutions are crucial from the phenomenological point of view, as they can be approached from the supergravity approximation. Furthermore, their nucleation rate (per unit volume) as a decay channel for the initial configuration will be dominant, and can be computed semiclassically $\Gamma_{\rm BoN}/{\rm Vol}_d\sim e^{\Delta S^{\rm (E)}}$ \cite{Coleman:1977py}, with $\Delta S^{\rm (E)}$ the difference in Euclidean action between the initial vacua and the BoN solution. Singular bordisms will have divergent terms in their Euclidean action, and thus will be suppressed.

The \textbf{goal of our paper} is to take a \emph{first step} in the construction of \emph{smooth} bordisms into nothing, as well as those which might have singular terms, but for which nonetheless $\mathcal{B}_{n+1}$ is \emph{homeomorphic} to a smooth manifold. We will call said bordisms \emph{smoothable}. We will obtain bounds in the topology of $\mathcal{B}_{n+1}$ from that of $\mathcal{C}_n=\partial\mathcal{B}_{n+1}$, and understand how this changes as we move along the bordism. Precisely for compact manifolds $\mathcal{C}_n$ with an involved internal topology, featuring different internal cycles of various dimensions, these smooth bordisms to nothing $\mathcal{B}_{n+1}$ cannot be the complete manifold $\mathcal{C}_n$ as a whole smoothly shrinking to a point.\footnote{One could simply consider the cone $C(\mathcal{C}_n)=(\mathcal{C}_n\times[0,1])/\sim$, with $(x,1)\sim(y,1)$ for all $x,\,y\in \mathcal{C}_n$, but said cone $C(\mathcal{C}_n)$ is not a topological manifold unless $H_k(\mathcal{C}_n,\mathbb{Z})\simeq \mathbb{Z}$ for $k=0,\,n$ and 0 otherwise, i.e., $\mathcal{C}_n$ is a homology sphere.} Rather, they will have a more involved structure, with the different internal cycles shrinking before the final tip into nothing. This corresponds to intermediate topology changes, which will be associated with a series of domain walls before, or ``coating'', the final boundary to nothing. Our approach will be able to give us information about the internal structure of both Bubbles of Nothing that dynamically nucleate and, as well as more general End of the World branes when interpolating from our EFT theory into nothing.

As it is easy to imagine, explicit solutions featuring these intermediate topology changes are quite involved. We will not attempt to write those, but simply obtain qualitative information about the structure of smooth(able) bordisms into nothing. The tool for this will be \textbf{Morse theory}, interpreting the bordism direction in $\mathcal{B}_{n+1}$ as our Morse function, whose critical points we will understand as the loci where the internal topology changes. As a matter of fact, Morse theory is nothing new in high energy theoretical physics, with the seminal work by Witten \cite{Witten:1982im} giving a supersymmetric system interpretation to it, which allowed him to derive a version of the so-called \textbf{Morse inequalities} we will use in this work, relating the set of critical points of a function with the topology of the manifold it is defined over.\footnote{Morse theory had already been considered in the literature \cite{Sorkin:1989ea,Louko:1995jw,Dowker:1997kc,Dowker:1997hj,Dowker:1999wu,Garcia-Heveling:2022fkf} for spacetime topology changes. Unlike in our case, in which these changes occur as we move spatially towards the end of our spacetime (be it the tip of the BoN or more generally the End of the World brane, see section \ref{sec:Dyn Bord}), there the topology changes occur in the time direction, with the Lorentzian metric vanishing precisely at the critical points of the Morse function with which it is constructed. The similarities between our approach and theirs end here, as we will not need to worry about problems of causality, the main focus of attention of said references.} Our work goes in line with recent approaches trying to infer local information of the EFT from topological global properties \cite{Lust:2024aeg}. We will also be able to use the Morse theory tools to obtain qualitative properties of configurations where more than one bordism to nothing is present.

\vspace{0.25cm}
The structure of the paper is as follows. In section \ref{sec:Dyn Bord} we review basic concepts of Bubbles of Nothing and End of the World branes, using Witten's BoN as the canonical example. In section \ref{sec:AlgTop} we introduce the necessary algebraic topology and Morse theory tools needed to obtain information about the topology of the bordism $\mathcal{B}_{n+1}$ from the compact manifold $\mathcal{C}_n$, as well as the possible intermediate topology changes. We use these results in section \ref{sec:Examples} on some compactification examples, showcasing the changes in the low-energy EFT, while in section \ref{sec:multiple} we consider the possibility that different bordisms to nothing come into play simultaneously, and how Morse theory can help us obtain information about these configurations. Finally, in  section \ref{sec:conc} we conclude
by summarizing our results and highlighting possible future directions.

\section{Dynamical bordisms into nothing}
\label{sec:Dyn Bord}
We will begin with a short review Witten's \textbf{Bubble of Nothing}, as well as its generalization to other compactification manifolds and how the lower EFT ``sees'' this spacetime boundaries. We will then use it as a starting points for generalizations to more involved compact topologies.
    \subsection{Bubbles of Nothing}
    \label{ss: BoN}
    The simplest (and coincidentally first found to appear in the literature) dynamical bordism into nothing is Witten's \textbf{Bubble of Nothing} \cite{Witten:1981gj}, presented here for an arbitrary number of spacetime (non-compact) dimensions. To study this non-perturbative instability of Kaluza-Klein vacuum, consider a circle compactification to a $d$-dimensional Minkowski, with global spacetime $X_{d+1}=\mathbb{R}^{1,d} \times \mathbb{S}^1$, with $\mathbb{S}^1$ of radius $\mathsf{R}$. This instability can be constructed from a Schwarzschild instanton $\hat{\mathcal{S}}_{d+1}$ in the Euclidean continuation of our spacetime, $X_{d+1}^{\rm (E)}=\mathbb{R}^{d+1}\times \mathbb{S}^1$, which will have metric
    \begin{equation}\label{e:BoN euc}
        \dd s_{d+1}^2=r^2\dd \Omega_{d-1}^2+\frac{\dd r^2}{1-\left(\frac{R}{r}\right)^{d-2}}+\mathsf{R}^2\left[1-\left(\frac{R}{r}\right)^{d-2}\right]\dd\theta^2\;,\quad\text{with }\left\{\begin{array}{l}
             \theta\in[0,2\pi)  \\
             r\geq R 
        \end{array}\right.\,,
    \end{equation}
    where $R$ is the nucleation radius of the instanton. It is easy to see that for $r\gg R$, $\dd s_{d+1}^2\approx \dd r^2+r^2\dd\Omega_{d-1}^2+\mathsf{R}\dd\theta^2$ and the space locally looks like $X_{d+1}^{\rm (E)}=\mathbb{R}^{d+1}\times \mathbb{S}^1$. Expanding $\dd \Omega_{d-1}^2=\dd\lambda^2+\sin^2\lambda\dd\Omega_{d-2}^2$, with $\lambda\in[0,\pi)$, we can go back to Lorentzian signature, by identifying $t=0$ with $\lambda\equiv\frac{\pi}{2}$, and Wick-rotating $\lambda\to\frac{\pi}{2}+i\tau$, under which
    \begin{equation}\label{e:WittenBoN}
        \dd s_{d+1}^2=-r^2\dd\tau^2+\frac{\dd r^2}{1-\left(\frac{R}{r}\right)^{d-2}}+r^2\cosh^2\tau\dd\Omega_{d-2}^2+\mathsf{R}^2\left[1-\left(\frac{R}{r}\right)^{d-2}\right]\dd\theta^2\;,
    \end{equation}
    which in the $r\gg R$ limit, and after redefining $x=r\cosh\tau$ and $t=r \sinh\tau$ (see that precisely $t=0$ for $\lambda=\frac{\pi}{2}$), looks locally like $X_{d+1}$,
    \begin{equation}
        \dd s_{d+1}^2\approx-\dd t^2+\dd x^2+x^2\dd\Omega_{d-2}^2+\mathsf{R}^2\dd\theta^2\;.
    \end{equation}
    However, precisely the fascinating observation made in \cite{Witten:1981gj} is that the new time and radial coordinates $t$ and $x$ do not cover the full original $X_{d+1}$, but only those parts with $x^2-t^2=r^2\geq R^2$, with $r=R$ corresponding to the wall of the \emph{bubble} at the time of nucleation $t=0$. The ``inside'' of the $x^2-t^2=R^2$ hyperboloid is now removed from spacetime. The size of the extra dimension
    \begin{equation}\label{eq.sfR}
        \hat{\mathsf{R}}(r)=\mathsf{R}\sqrt{1-\left(\frac{R}{r}\right)^{d-2}}\;,
    \end{equation}
    shrinks as we approach the bubble wall, $x_{\rm BoN}=\sqrt{R^2+t^2}$, which asymptotically expands at the speed of light. In order to avoid a conical singularity at $x_{\rm BoN}$ the condition $R=\mathsf{R}$ (i.e., the bubble nucleates at the KK scale) is required. A smooth solution is required in order to avoid curvature singularities and non-geodesically complete spacetimes, though as we will see in sections \ref{sec:AlgTop} and \ref{ss:defects} for our analysis we will only require that the BoN solution has a topological manifold structure, regardless of possible singularities of stringy origin.
\begin{figure}[t!]
				\centering
				\includegraphics[width=0.9\textwidth]{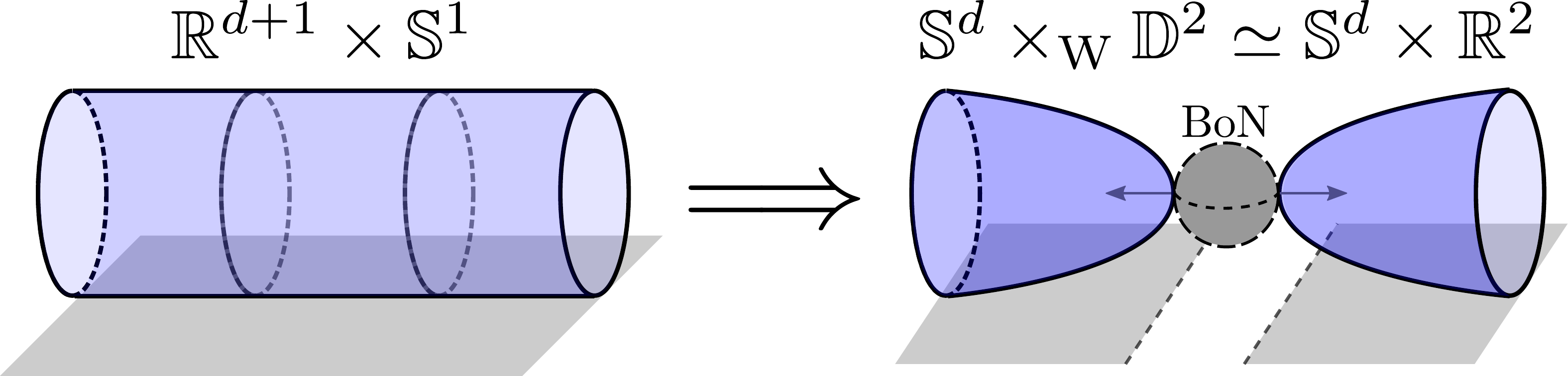}
			\caption{\small Euclidean sketch of the $d$-dimensional version of Witten's BoN (together with the Wick rotated KK Minkowski vacuum), with the macroscopic dimensions being the horizontal gray direction depicted in gray. Notice that far from the BoN surface, both spaces are locally identical.}
			\label{f.WittenBoN}
	\end{figure}
   As shown in Figure \ref{f.WittenBoN}, the Euclidean solution has global topology $\hat{\mathcal{S}}_{d+1}\simeq \mathbb{S}^{d-1}\times_{\rm W}\mathbb{D}^2\simeq \mathbb{S}^{d-1}\times\mathbb{R}^2$, where $\times_{\rm W}$ is a warped product and $\mathbb{D}^n=\{x\in\mathbb{R}^n\,:\,\|x\|\leq 1\}$ is the $n$-disc, with indeed $\partial \mathbb{D}^2=\mathbb{S}^1$.
   
   Already in the original appearance of Witten's BoN \cite{Witten:1981gj} a possible topological obstruction to the KK vacuum decay was postulated in the presence of fermions. Being $\hat{\mathcal{S}}_{d+1}$ simply connected, there is a unique spinor structure, which here it is set by the asymptotics far from the bubble wall, where space locally looks like $\mathbb{R}^{d}\times\mathbb{S}^1$. Under a $2\pi$ rotation along $\mathbb{S}^1$, spinors can have either symmetric or antisymmetric boundary conditions. However, if we want to be able to extend the solution to the complete $\hat{\mathcal{S}}_{d-1}$, the spin structure must be extended to the whole $\mathbb{D}^2$. Since this is contractible, only the antiperiodic spin structure makes sense for the BoN solution to exist. Since this accounts for non-covariantly constant spinors, SUSY must be broken either in the higher-dimensional theory or explicitly through the imposition of Scherk-Schwarz boundary conditions \cite{Scherk:1978ta,Scherk:1979zr,Kounnas:1989dk} when compactifying.

   As for whether the vacuum decay is energetically allowed, the ADM mass computation shows \eqref{e:WittenBoN} to have 0 energy (same as $X_{d+1}=\mathbb{R}^{1,d-1}\times\mathbb{S}^1$). The \emph{Positive Energy Theorem} \cite{Schon:1979rg,Witten:1981mf}, i.e., that every non-flat solution of Einstein's equations asymptoting to Minkowski has positive energy (thus semiclassically preventing Minkowski decay), does not apply here when antiperiodic boundary conditions are used in $\mathbb{S}^1$, as the proof of the theorem relays on the existence of covariantly constant spinors.  The decay rate (per unit volume per unit time) can be computed to be $\Gamma\sim\exp\left\{-2\pi^{d/2}\Gamma(\frac{d}{2})^{-1}(\mathsf{R}M_{{\rm Pl},d})^{d-2}\right\}$, with all contributions coming from the Gibbons–Hawking–York term (the BoN solution being Ricci-flat) and having $\Gamma\to 0$ for large radius.

    As mentioned above, the topology of the Euclidean instanton $\hat{\mathcal{S}}_{d+1}$ \eqref{e:BoN euc} is not $\mathbb{R}^{d+1}\times \mathbb{S}^1$, but rather $\mathbb{S}^{d-1}\times_{\rm W}\mathbb{D}^2$. As explained in \cite{GarciaEtxebarria:2020xsr}, this can be made more explicit by rewriting the Euclidean solution \ref{e:BoN euc} as
     \begin{equation}\label{e:BoN euc2}
        \dd s^2_{d+1}=W(\rho)^2\mathsf{R}^2\dd\Omega_{d-1}^2+\dd \rho^2+\hat{\mathsf{R}}(\rho)^2\dd\theta^2\;,\quad\text{with }\rho\in[0,\infty)\,,\;\theta\in[0,2\pi)\;,
    \end{equation}
    where $W(\rho)$ is some warping function.\footnote{
    The radial directions are related by $r=\mathsf{R}W(\rho)$, with the bubble located at $\rho=0$ and the warping fulfilling
    \begin{equation}
        W'(\rho)=\mathsf{R}^{-1}\sqrt{1-W(\rho)^{-(d-2)}}\,,\quad W(0)=1\;\Longleftrightarrow {}_2F_1\left(\frac{1}{2},\frac{1}{2-d},\frac{3-d}{2-d}W(\rho)^{2-d}\right)W(\rho)=\frac{\rho}{\mathsf{R}}\;,
    \end{equation}
    which for large $\rho$ asymptotics to $W(\rho)\approx \frac{\rho}{\mathsf{R}}$ and $r\approx \rho$, as expected.
    } 
    Precisely for constant angle $\mathbb{S}^{d-1}$ direction in \eqref{e:BoN euc2} the metric simplifies to 
    \begin{equation}
        \dd s_2^2=\dd\rho^2+\hat{\mathsf{R}}(\rho)^2\dd\theta^2\;,
    \end{equation}
    which is precisely the $\mathbb{R}^2\simeq\mathbb{D}^2$ metric in polar coordinates, with the $\rho\to\infty$ limit resulting in $\hat{\mathsf{R}}^2\dd\theta^2=\mathsf{R}^2 \dd\theta^2$ metric for the compact $\mathbb{S}^1$ in the original spacetime.
    
    \vspace{0.25cm}
    While Witten's Bubble of Nothing is in many aspects very simple, it serves as a first step from which we can generalize to more involved bubbles of nothing. In general, starting with a $X_{d+n}=\mathbb{R}^{1,d-1}\times \mathcal{C}_{n}$ Minkowski compactification, the associated Euclidean instanton mediating the bubble of nothing will have $\hat{\mathcal{S}}_{d+n}\simeq \mathbb{S}^{d-1}\times_{\rm W} \mathcal{B}_{n+1}$ topology, with $\mathcal{B}_{n+1}$ the bordism from $\mathcal{C}_n$ to nothing, i.e.,  $\partial \mathcal{B}_{n+1}=\mathcal{C}_{n}$. In the language of \eqref{e:BoN euc2}, this results in the following $SO(d-1)$ symmetric ansatz \cite{GarciaEtxebarria:2020xsr} for the Euclidean solution:
    \begin{equation}\label{e.gen bon}
        \dd s_{d+n}^2=W(y)^2\mathsf{R}^2\dd \Omega_{d-1}^2+h_{\alpha\beta}^\mathcal{B}(y)\dd y^\alpha\dd y^\beta\;,
    \end{equation}
    where $\mathsf{R}$ is the nucleation radius and $h_{\alpha\beta}^\mathcal{B}$ is the metric of the bordism $\mathcal{B}_{n+1}$, parameterized by $\{y^\alpha\}_{\alpha=1}^{n+1}$. Denoting the radial direction from the bubble as $\rho(y)=W(y)\mathsf{R}\in[0,\infty)$, far away $\rho\to\infty$ we can rewrite \eqref{e.gen bon} as
    \begin{equation}\label{e.large rho}
        \dd s_{d+n}^2\approx \rho^2\dd\Omega_{d-1}^2+\underbrace{\dd \rho^2+{h}^{\mathcal{C}}_{\bar\alpha\bar\beta}\dd y^{\bar\alpha}\dd y^{\bar\beta}}_{h_{\alpha\beta}^\mathcal{B}(y)\dd y^\alpha\dd y^\beta}\;,
    \end{equation}
    where now we can separate the bordism coordinates into the radial direction and the internal parametrization of the compact $\mathcal{C}$, $\{y^\alpha\}_{\alpha=1}^{n+1}\to \{\rho\}\cup\{y^{\bar\alpha}\}_{\bar\alpha=1}^n$, and ${h}^{\mathcal{C}}_{\bar\alpha\bar\beta}$ the asymptotic metric of $\mathcal{C}_{n}$ in $X_{d+n}$, independent of the radius $\rho$. This way, far from the bubble wall, the solution locally looks like the initial compactification. The Wick rotation back to Minkowski can be performed in the same way as in Witten's BoN, with no additional difficulties from the more complicated compact $\mathcal{C}_{n}$, as the Wick rotation is performed on the sphere $\mathbb{S}^{d-1}$ to a hyperboloid.

\vspace{0.25cm}

    As explained earlier in this section, for Witten's BoN to exist, it was crucial that the compactification manifold $\mathbb{S}^1$ had antiperiodic (i.e., SUSY-breaking) boundary conditions. Furthermore, one would have to argue that the process is not kinematically obstructed and has a finite decay rate. This translates to more general BoN decays through the following two conditions \cite{GarciaEtxebarria:2020xsr}:
    \begin{itemize}
        \item \textbf{Topological Condition}: This first requirement translates in all compactification manifolds belonging to the trivial cobordism class. When considering $\mathcal{M}_k$ and $\mathcal{N}_k$ to be $\mathsf{g}$-manifolds, with $\mathsf{g}$ some appropiate structure, then the $\mathsf{g}$-bordism is defined such that $\mathcal{B}_{k+1}$ is also a $\mathsf{g}$-manifold, and the associated $\mathsf{g}$-structure restricted to each of $\partial\mathcal{B}_{k+1}$ components corresponds to that of $\mathcal{M}_k$ and $\mathcal{N}_k$. The notion of cobordism defines an equivalence relation between compact manifolds, and $[\mathcal{M}_k]=[\mathcal{N}_k]$. We denote the set of cobordism classes by $\Omega_k^{\mathsf{g}}$, which can be shown to have a Abelian group structure, such that $[\mathcal{M}]+[\mathcal{N}]=[\mathcal{M}\sqcup\overline{\mathcal{N}}]$. Through the \textbf{Swampland Cobordism Conjecture} this condition is automatically satisfied by requiring $\Omega^{\rm QG}_\bullet=0$. As the complete, non-perturbative description of quantum gravity is not yet known, we will approximate this QG structure by simpler $\mathsf{g}$ whose cobordism groups are better understood. In section \ref{ss:defects} we comment on the necessary defects that might be required to trivialize non-zero cobordism groups, $\Omega_k^{\mathsf{g}}\neq 0\to\Omega_k^{\mathsf{g}+{\rm defects}}=0$.

        Translating this topological condition to Witten's BoN, the circle with antiperiodic boundary conditions $\mathbb{S}^1_a$ belongs to the trivial class of $\Omega_1^{\rm Spin}\simeq\mathbb{Z}_2$ \cite{Anderson1967:SPIN}, with the second class given by the periodic circle $\mathbb{S}^1_p$. As in principle the later is an allowed compactification manifold, from the Cobordism Conjecture point of view, there must exist either UV defects that can change the spin structure \cite{McNamara:2019rup,Garcia-Etxebarria:2015ota}, or either the QG structure $\mathsf{g}={\rm Spin}$ must be refined to $\hat{\mathsf{g}}$ such that $\Omega^{\hat{\mathsf{g}}}_1=0$, for instance to $\Omega_1^{\rm Spin^c}=0$, see \cite{Wan:2018bns}.

        \item \textbf{Dynamical Condition}: The Positive Energy Theorem (PET) \cite{Witten:1981mf, Hertog:2003ru, Hertog:2003xg} states that the ADM mass of asymptotically $\mathcal{M}_{d+n}\simeq\mathbb{R}^{1,{d-1}}\times\mathcal{C}_n$ (with $\mathcal{
        C}_n$ compact) spacetimes is bounded from below by 0, with the only spacetime saturating the inequality being precisely $\mathbb{R}^{1,{d-1}}\times\mathcal{C}_n$. This prevents any bubble nucleation, since any BoN spacetime will have more energy, and as such the decay will be obstructed. PET relays on the assumption that \textbf{(1)} $\mathcal{M}_{d+n}$ admits an asymptotically covariantly constant spinor (always the case when SUSY is preserved) and \textbf{(2)} the \emph{Dominant Energy Condition} (DEC), i.e., $-g_{BC}T^{AB}k^C$ is causal and future-pointing for all $k^C$ causal and future-pointing vectors, holds. While \textbf{(1)} is a topological condition which depends on the global structure of spacetime, \textbf{(2)} depends on the matter and higher-derivative ingredients of our EFT. There is a rich literature on how any of these two conditions can be broken, specially \textbf{(2)} \cite{Coleman:1977py,Curiel:2014zba,GarciaEtxebarria:2020xsr}, so that the PET does not apply.
    \end{itemize}

    As for the most part of this paper we will be interested in the qualitative properties of our bordisms to nothing, we will assume that the above conditions hold and thus such bordism can be realizable. In general, the difficulty in building the appropriate BoN solution for a given compactification stems from finding the appropriate bordism solution $h^{\mathcal{B}}_{\alpha\beta}$ to the equations of motion, as well as including the appropriate defects to kill the bordism group $\Omega_n^{\mathsf{g}}$. While much effort has been done in the past years in finding explicit BoN constructions \cite{Witten:1981gj,Fabinger:2000jd,Dine:2004uw,Horowitz:2007pr,Yang:2009wz,Blanco-Pillado:2010xww,Blanco-Pillado:2010vdp,Brown:2010mf,Blanco-Pillado:2011fcm,Brown:2011gt,deAlwis:2013gka,Brown:2014rka,Blanco-Pillado:2016xvf,Ooguri:2017njy,Dibitetto:2020csn,GarciaEtxebarria:2020xsr,Bomans:2021ara,Draper:2021ujg,Draper:2021qtc,Petri:2022yhy,Bandos:2023yyo,Ookouchi:2024tfz,Heckman:2024zdo} (see also \cite{Draper:2023ulp,Blanco-Pillado:2023hog,Blanco-Pillado:2023aom,Delgado:2023uqk,Friedrich:2024aad} for related problems and techniques), most of the solutions feature relatively simple topologies along the bordism direction. One can imagine that as the topology of the internal manifold $\mathcal{C}_n$ becomes more involved, so will that of the bordism $\mathcal{B}_{n+1}$. In section \ref{sec:AlgTop} we will see that indeed we can obtain plenty of information about these simply from the topology of $\mathcal{C}_n$. 

    \subsection{End of the World Branes}
    \label{ss: EotW}

    In the above section, we have kept a higher-dimensional description. One can wonder now how Witten's BoN looks from the $d$-dimensional description once the internal coordinate is integrated out. Dimensionally reducing \eqref{e:BoN euc} down to $d$-dimensions (see \cite{Angius:2022aeq} for more details) results in the following metric in the (Euclidean) Einstein frame:
    \begin{equation}\label{e: BoN down}
        \dd s_d^2=\left(\frac{\hat{\mathsf{R}}(r)}{\mathsf{R}}\right)^{2\frac{3-d}{d-2}}\dd r^2+\left(\frac{\hat{\mathsf{R}}(r)}{\mathsf{R}}\right)^{\frac{2}{d-2}}\dd\Omega_{d-1}^2\;,
    \end{equation}
    with $\hat{\mathsf{R}}(r)$ as defined in \eqref{eq.sfR}, and the $(d+1,d+1)$ component of the higher dimensional metric being promoted to a (massless) scalar, which can be canonically normalized to the following radion profile
    \begin{equation}
        \varrho(r)=-\sqrt{\frac{d-1}{d-2}}\log\left[\frac{\hat{\mathsf{R}}(r)}{\mathsf{R}}\right]\;,
    \end{equation}
        with the overall sign being only a matter of convention. Note that as we approach the BoN tip $r\to\mathsf{R}$, $\hat{\mathsf{R}}(r)\to 0$ and the radion blows up, $\varrho\to\infty$. As the compact manifold is flat, there is no induced curvature potential to the $d$-dimensional action. Note that, even though, being a solution of Einstein's equations in vaccum, \eqref{e:BoN euc} is Ricci-flat,\footnote{The Riemann tensor, as well as some curvature invariants derived from it, are not null for \eqref{e:BoN euc}, though they remain small if $\mathsf{R}$ is large in $M_{\rm Pl,5}$ units, which in turn results in a smaller decay rate.} this is no longer the case once we compactify, with the Ricci scalar of \eqref{e: BoN down} given by
        \begin{equation}
            \mathcal{R}=\frac{(d-2)(d-1)}{4\mathsf{R}^2}\left(\frac{\mathsf{R}}{r}\right)^{2(d-1)}\left(\frac{\hat{\mathsf{R}}(r)}{\mathsf{R}}\right)^{-2\frac{d-1}{d-2}}\;,
        \end{equation}
        which blows up as we approach $r\to\mathsf{R}$. In terms of the actual traversed physical distance in the radial direction, this can be computed from \eqref{e: BoN down} to be
        \begin{equation}
            \Delta r=\frac{2(d-2)^2\mathsf{R}}{d-1} \left(\frac{\hat{\mathsf{R}}(r)}{\mathsf{R}}\right)^{\frac{d-1}{d-2}}\;,
        \end{equation}
        which means that the bubble wall $r=\mathsf{R}$ ($\Delta r=0$) is at finite spacetime distance. Furthermore, as we approach said locus, both the radion and Ricci scalar scale as
        \begin{equation}\label{e: scalings BoN}
            \varrho\sim-\sqrt{\frac{d-2}{d-1}}\log\left(\frac{\Delta r}{\mathsf{R}}\right)\to\infty \;,\qquad\mathcal{R}\sim e^{2\sqrt{\frac{d-1}{d-2}}\varrho}\sim \frac{1}{\Delta r^2}\to\infty\;.
        \end{equation}
        As expected from infinite distance regions in moduli space being proven, $\varrho\to\infty$, there is a break-down of the lower-dimensional EFT from the blowing-up spacetime curvature $\mathcal{R}\to\infty$.

        The above behaviors are not unique to Witten's BoN, but rather universal features of the close distance behavior of \emph{End of the World branes}. As introduced in \cite{Angius:2022aeq}, general solutions to $d$-dimensional Einstein gravity coupled to a canonically normalized scalar $\phi$ (generalizable to more scalars and non-canonical metrics) and a potential $V(\phi)$, in the presence of a real codimension-1 boundary (the End of the World Brane), feature the following scalings
        \begin{equation}\label{e: EotW scaling}
            \phi(\Delta r)\sim-\frac{2}{\delta}\log \Delta r\,,\qquad|\mathcal{R}|\sim \Delta r^{-2}\sim e^{\delta\phi}\;.
        \end{equation}
        Both scalings are expressed in terms of a \emph{critical exponent} $\delta$ given by
        \begin{equation}\label{e. critical exponent}
            \delta=2\sqrt{\frac{d-1}{d-2}(1-a)}\;,
        \end{equation}
        where $a$ is given by
        \begin{equation}
            a=\frac{V(\phi)}{V(\phi)-\frac{1}{2}(\partial_{\Delta r}\phi)^2}\leq 1\;,
        \end{equation}
        as from the e.o.m. $\frac{1}{2}(\partial_{\Delta r}\phi)^2\geq V(\phi)$. Note that for Witten's BoN there is no potential and as such $a=0$, with $\delta=2\sqrt{\frac{d-1}{d-2}}$, see \eqref{e: scalings BoN}. For $a<1$ equations of motion result in a scalar potential of the form $V(\phi)\sim V_0e^{\delta \phi}$, with the potential (if present) blowing up at the EotW brane. Solutions with $a=1$ are more exotic, as they allow for sub-exponential dependence, with expressions \eqref{e: EotW scaling} not applying.

The example above features a smooth bordism into nothing, with the internal cycle quietly shrinking to a point. In \cite{Friedrich:2023tid} a general analysis of non-smooth bordisms is presented, with the angular defect at the tip of the bordism being related to the tension of the EotW brane, \cite{Vilenkin:1984ib,Garfinkle:1985hr}.

\vspace{0.25 cm}

        The EotW brane description has been extensively studied from the dynamical cobordism point of view \cite{Dudas:2000ff,Dudas:2002dg,Dudas:2004nd,Basile:2018irz,Antonelli:2019nar,Basile:2021mkd,Mourad:2021qwf,Mourad:2021roa,Buratti:2021yia,Buratti:2021fiv,Angius:2022mgh,Blumenhagen:2022mqw,Basile:2022ypo,Blumenhagen:2023abk,Angius:2023xtu,Huertas:2023syg,Friedrich:2023tid,Angius:2023uqk,GarciaEtxebarria:2024jfv,Huertas:2024mvy,Angius:2024pqk}, and while there is not a one-to-one correspondence between the critical exponent $\delta$ and the UV resolution of the EotW brane, some steps have been performed towards a classification \cite{Angius:2024zjv}. While in the cobordisms to nothing from Witten's BoN and other instances in the literature (see \cite{Angius:2022aeq} for some simple examples) the scalar $\phi$ denotes the volume of the internal manifold, which shrinks to nothing, as we will see in section \ref{sec:Examples}, more involved bordisms have several (canonically normalized) moduli going to infinite distance as we radially traverse our bordism towards its ``boundary to nothing''. These scalars need not blow up at the same locus, with the EotW brane being ``thick'' and having some internal structure, precisely corresponding with the different topology changes which we will describe shortly in \ref{sec:AlgTop}. This is in no way surprising, as both the EFT description breaking can be seen from the scalars traveling to infinite distance in moduli space, or equivalently from the change in the compactification manifold topology which results in a different lower-dimensional EFT.

\section{Bordism Topology and Morse-Bott Inequalities}
\label{sec:AlgTop}
In the general introduction to BoN and EotW branes, the bordism $\mathcal{B}_{n+1}$ from the compact manifold $\mathcal{C}_n$ to nothing was extensively used. As discussed in sections \ref{sec:intro} and \ref{sec:Dyn Bord}, a necessary condition for $\mathcal{B}_{n+1}$ to exist was that $\Omega_n^{\mathsf{g}}=0$, with $\partial \mathcal{B}_{n+1}=\mathcal{C}_n$ and the structure $\mathsf{g}$ extending from $\mathcal{C}_n$ to $\mathcal{B}_{n+1}$. However, even when such bordism exists, it will not be unique.

\begin{figure}[h]
\begin{center}
\begin{subfigure}[b]{0.45\textwidth}
\center
\includegraphics[width=0.71\textwidth]{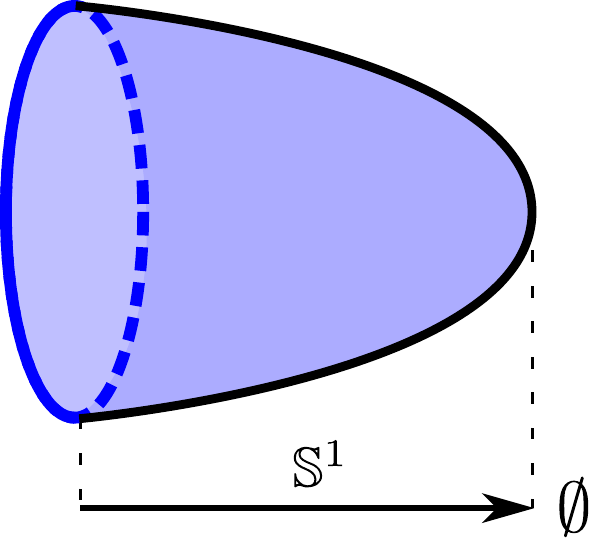}
\caption{\hspace{-0.3em}) $\mathbb{D}^2$} \label{ff.exS11}
\end{subfigure}
\begin{subfigure}[b]{0.54\textwidth}
\center
\includegraphics[width=0.7\textwidth]{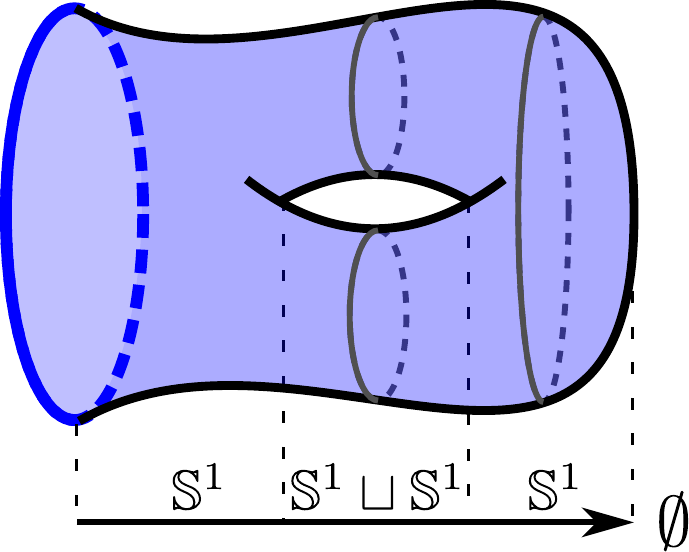}
\caption{\hspace{-0.3em}) $\mathbb{T}^2-\mathbb{D}^2$} \label{ff.exS12}
\end{subfigure}
\caption{Two instances of oriented bordism from $\mathbb{S}^1$ to nothing, with the associated minimal topology changes in each.
\label{f. Disk torus}}
\end{center}
\end{figure}

As an illustrative example, take $\mathbb{S}^1$ together with the oriented bordism. As $\Omega_1^{SO}=0$ \cite{Thom1954QuelquesPG}, there is no obstruction from this side, and actually it is straightforward that $\mathbb{S}^1=\partial\mathbb{D}^2$. However, more generally the circle is the boundary of any genus-$g$ Riemann surface $\Sigma_g$ with a disk removed, $\mathbb{S}^1=\partial(\Sigma_g-\mathbb{D}^2)$, with $\mathbb{D}^2=\mathbb{S}^2-\mathbb{D}^2$, see Figure \ref{f. Disk torus}. More in general, given some compact $\mathcal{C}_n=\partial\mathcal{B}_{n+1}$, one can always define $\mathcal{B}_{n+1}'=\mathcal{B}_{n+1}\#\mathcal{Y}_{n+1}$, with some appropriate $\mathcal{Y}_{n+1}$ not homeomorphic to $\mathbb{S}^{n+1}$ and to which the structure $\mathsf{g}$ also extends. This way $\mathcal{B}_{n+1}'$ and $\mathcal{B}_{n+1}$ will be topologically distinct while both having $\mathcal{C}_n$ as boundary. This means that (at least only from topological data), both $\mathbb{S}^{d-1}\times_{\rm W} \mathcal{B}_{n+1}$ and $\mathbb{S}^{d-1}\times_{\rm W} \mathcal{B}_{n+1}'$ are two different topologies for BoN decay channels of $\mathbb{R}^{d}\times\mathcal{C}_n$, both looking the same far from the bubble wall/EotW brane.

Intuitively, as we move along the radial direction $\rho$ towards the bubble wall $\rho=0$, the difference between the two constructions lies in the \emph{necessary} topology changes 
\begin{equation}\label{eq: top chain}
    \mathcal{C}_n=\mathcal{C}_n^{(0)}\to \mathcal{C}_n^{(1)}\to\dots\to \mathcal{C}_n^{(m)}=\emptyset
\end{equation}
in the local compactification manifold, given by the section or fiber of $\mathcal{B}_{n+1}$. The local theory undergoes a series of changes in the compact manifold, with different internal cycles opening and closing, until we reach the boundary to nothing. As an illustration, return to the bordisms of the circle $\mathbb{S}^1=\partial\mathbb{D}^2=\partial(\mathbb{T}^2-\mathbb{D}^2)$, depicted in Figure \ref{f. Disk torus}. While in the former case the minimal topology change is $\mathbb{S}^1\to\emptyset$ as the circle shrinks to nothing (Witten's BoN), for the later one undergoes $\mathbb{S}^1\to\mathbb{S}^1\sqcup\mathbb{S}^1\to\mathbb{S}^1\to \emptyset$, which involves two extra topology changes.

As a result, the dimensional reduction from the higher-dimensional description results in distinct low-energy EFTs, such that before the final bordism into ``nothing'' one would encounter a series of domain walls separating the different topologies. Whilst from the EFT point of view the EotW brane and the accompanying diverging scalar correspond to the domain wall associated with the first topology change we encounter, there can be a ``more complicated'' structure ``inside'' such brane, dressed by said first domain wall.

Of the two bordisms depicted in Figure \ref{f. Disk torus}, we known that the one that actually describes a known``dynamical'' process is that of Figure \ref{ff.exS11}, Witten's BoN, for which there are no intermediate topology changes. We will encounter that of Figure \ref{ff.exS12} again in section \ref{ss:defects} in the context of more complicated bordism groups. The question that naturally arises now is whether one can always go to nothing in a single step, or if intermediate topology changes are required once the internal manifold $\mathcal{C}_n$ is complicated enough. To answer this, we will need to obtain information about \textbf{(1)} the topology of the possible bordisms $\mathcal{B}_{n+1}$ and \textbf{(2)} how the radial direction $\rho$ interpolates the bordism $\mathcal{B}_{n+1}$ on its way to $\emptyset$, and how this translates into the \emph{necessary topology changes}.

    \subsection{Bounding the topology of the bordism}
    \label{ss:exact seq}
    The first of the two above tasks is the most straightforward, as it will only involve some careful manipulation of long exact sequences. For simplicity, we will restrict to bordisms admitting a continuous deformation to a smooth topological manifold\footnote{\label{fn.String DEF}For example, we will allow for conical singularities, as these could be smoothed-out to a finite curvature region. Some other bordism present in the literature, such as $(\mathbb{Z}/\mathbb{Z}_2)\times \mathbb{S}_p^1$ or $(\mathbb{R}\times\mathbb{S}_p^1)/\mathbb{Z}_2$ for type IIA or IIB string theory on a periodic circle $\mathbb{S}^1_p$, see \cite{McNamara:2019rup}, are not. Other bordisms in the same reference include M-theory on non-orientable manifolds. A description of these non-smooth(able) or non-orientable bordisms is out of the reach and assumptions needed to apply the algebraic topology and Morse theory tools we use, so they will not be considered.}. Consider for this a $(n+1)$-dimensional smooth manifold $\mathcal{B}_{n+1}$ with boundary $\mathcal{C}_n=\partial\mathcal{B}_{n+1}\neq\emptyset$. For further simplicity we will assume $\mathcal{B}_{n+1}$ to be orientable.

    We start by applying the zig-zag lemma, which results in the following long exact sequence:
    \begin{equation}\label{e.les1}
        \cdots\xrightarrow{\partial}  H_k(\mathcal{C}_n;\mathbb{Z})\xrightarrow{i_\ast}  H_k(\mathcal{B}_{n+1};\mathbb{Z})\xrightarrow{j_\ast} H_k(\mathcal{B}_{n+1},\mathcal{C}_n;\mathbb{Z})\xrightarrow{\partial} H_{k-1}(\mathcal{C}_n;\mathbb{Z})\xrightarrow{i_\ast} \cdots\;,
    \end{equation}
    where $H_k(\mathcal{X};\mathbb{Z})$ is the $k$-th homology group with integer coefficients and $H_k(\mathcal{X},\mathcal{Y};\mathbb{Z})$, with $\mathcal{Y}\subseteq\mathcal{X}$, is the relative homology group, whose elements are represented by the $n$-chains $\alpha$ of $\mathcal{X}$ whose boundary $\partial\alpha$ is a $(n-1)$-chain of $\mathcal{Y}\subseteq\mathcal{X}$. Applying now Lefschetz duality,\footnote{In the pairings between homology and cohomology groups through Lefschetz duality, cohomology is to be taken with \emph{compact support}, $H_k(\mathcal{B}_{n+1},\mathcal{C}_n;\mathbb{Z})\simeq H^{n+1-k}_\mathbf{c}(\mathcal{B}_{n+1};\mathbb{Z})$ and $H^k_\mathbf{c}(\mathcal{B}_{n+1},\mathcal{C}_n;\mathbb{Z})\simeq H_{n+1-k}(\mathcal{B}_{n+1};\mathbb{Z})$,
 \cite{maunder1996algebraic}. However, since $\mathcal{B}_{n+1}$ is compact, we do not need to worry about this subtlety, and can consider de Rham cohomology.} $H_k(\mathcal{B}_{n+1},\mathcal{C}_n;\mathbb{Z})\simeq H^{n+1-k}(\mathcal{B}_{n+1};\mathbb{Z})$, we can rewrite \eqref{e.les1} as
    \begin{equation}\label{e.les2}
        \cdots\xrightarrow{j_\ast} H^{n-k}(\mathcal{B}_{n+1};\mathbb{Z})\simeq H_{k+1}(\mathcal{B}_{n+1},\mathcal{C}_n;\mathbb{Z})\xrightarrow{\partial}  H_k(\mathcal{C}_n;\mathbb{Z})\xrightarrow{i_\ast} H_k(\mathcal{B}_{n+1};\mathbb{Z})\xrightarrow{j_\ast}\cdots\;.
    \end{equation}
    On the other hand, given some manifold $\mathcal{M}$, we can separate the torsion and free parts of its homology groups,
    \begin{equation}
        T_k(\mathcal{M}):=\tau H_k(\mathcal{M};\mathbb{Z})\;,\qquad 
        fH_k(\mathcal{M};\mathbb{Z}):=H_k(\mathcal{M};\mathbb{Z})/T_k(\mathcal{M})\;,
    \end{equation}
    which for orientable $n$-manifolds without boundary, such as $\mathcal{C}_n$, obey
    \begin{equation}
        fH_k(\mathcal{M};\mathbb{Z})\simeq fH_{n-k}(\mathcal{M};\mathbb{Z})\;,\qquad T_k(\mathcal{M})\simeq T_{n-k-1}(\mathcal{M})\;,
    \end{equation}
    note the extra $-1$ in the torsion part. Now, for integral (co)homology, the Universal Coefficient Theorem states that
    \begin{equation}
        H_k(\mathcal{M};\mathbb{Z})\simeq \mathbb{Z}^{b_k(\mathcal{M})}\oplus T_k(\mathcal{M})\;\Longleftrightarrow\;H^k(\mathcal{M};\mathbb{Z})\simeq \mathbb{Z}^{b_k(\mathcal{M})}\oplus T_{k-1}(\mathcal{M})\;,
    \end{equation}
    where $b_k(\mathcal{M})$ are the (integral) Betti numbers of $\mathcal{M}$. This allows us to rewrite \eqref{e.les2} as
    \begin{equation}
        \cdots\to \mathbb{Z}^{b_{n-k}(\mathcal{B}_{n+1})}\oplus T_{n-k-1}(\mathcal{B}_{n+1})\to\mathbb{Z}^{b_k(\mathcal{C}_n)}\oplus T_k(\mathcal{C}_n)\to\mathbb{Z}^{b_k(\mathcal{B}_{n+1})}\oplus T_k(\mathcal{B}_{n+1})\to\cdots\;.
    \end{equation}
    Denoting $t_k(\mathcal{M})={\rm rank}\, T_k(\mathcal{M})$, we can conclude\footnote{\label{fn.torsion free}
    We can study the free and torsional ranks separately, as there are no torsional $k$-cycles acting as boundary of torsion-free cycles in $H_{k+1}(\mathcal{B}_{n+1},\mathcal{C}_n;\mathbb{Z})$. To see this, we tensor the \eqref{e.les1} long exact sequence by $\mathbb{Q}$, which acts as an exact functor, getting rid of the torsion parts:
    \newline
    {\begin{center}    
   \begin{tikzcd}[ampersand replacement=\&]
  \cdots\arrow[r,"i_\ast"]\&  H_k(\mathcal{B}_{n+1};\mathbb{Z})\arrow[r,"j_\ast"]\arrow[d,"\otimes\mathbb{Q}"]\&   H_k(\mathcal{B}_{n+1},\mathcal{C}_n;\mathbb{Z})\arrow[r,"\partial"]\arrow[d,"\otimes\mathbb{Q}"]\&   H_{k-1}(\mathcal{C}_n;\mathbb{Z})\arrow[d,"\otimes\mathbb{Q}"]\arrow[r,"i_\ast"]\&\cdots\\
   \cdots\arrow[r,"i_{\ast\otimes\mathbb{Q}}"]\&  H_k(\mathcal{B}_{n+1};\mathbb{Z})\otimes\mathbb{Q}\arrow[r,"j_{\ast\otimes\mathbb{Q}}"]\&   H_k(\mathcal{B}_{n+1},\mathcal{C}_n;\mathbb{Z})\otimes\mathbb{Q}\arrow[r,"\partial_{\otimes\mathbb{Q}}"]\&   H_{k-1}(\mathcal{C}_n;\mathbb{Z})\otimes\mathbb{Q}\arrow[r,"i_{\ast\otimes\mathbb{Q}}"]\&\cdots
\end{tikzcd}
\end{center}
}
Since the torsion elements in the image of a given map $f:X\to Y$ are trivialized after tensoring by $\mathbb{Q}$, we have that $\{x\in X\,:\,f(x)\in\tau Y\}\subseteq\ker(f_{\otimes\mathbb{Q}})$, and thus any free $k$-cycle $\alpha_k$ in $H_k(\mathcal{B}_{n+1},\mathcal{C}_n;\mathbb{Z})$ with torsional boundary, $\partial[\alpha_k]\in\tau H_{k-1}(\mathcal{C}_n;\mathbb{Z})$, will have $\partial_{\otimes\mathbb{Q}}[\alpha_k]=0$. Now, having $\I (j_{\ast\otimes\mathbb{Q}})\simeq \ker (\partial_{\otimes\mathbb{Q}})$, $[\alpha_k]\in\I (j_{\ast\otimes\mathbb{Q}})$. But since $\alpha_k$ is a free cycle and $j_{\ast\otimes\mathbb{Q}}:H_k(\mathcal{B}_{n+1};\mathbb{Z})\otimes\mathbb{Q}\to H_k(\mathcal{B}_{n+1},\mathcal{C}_n;\mathbb{Z})\otimes\mathbb{Q}$, then $[\alpha_k]\in \I(j_\ast)$ and thus is boundary-less. One then concludes that for relative homology, the boundary of a torsion-free cycle of must also be torsion-free.
    }
    \begin{subequations}\label{e.ineq}
    \begin{empheq}[box=\widefbox]{align}
           b_k(\mathcal{B}_{n+1})+ b_{n-k}(\mathcal{B}_{n+1})&\geq b_k(\mathcal{C}_n)=b_{n-k}(\mathcal{C}_n)\label{e.ineq1}\\
           t_k(\mathcal{B}_{n+1})+ t_{n-k-1}(\mathcal{B}_{n+1})&\geq t_k(\mathcal{C}_n)=t_{n-k-1}(\mathcal{C}_n)\;,
        \end{empheq}
    \end{subequations}
    for $k=0,\,...,\,n$.
    %, where $\tilde{t}_{k}(\mathcal{C}_{n})$ denotes the rank of the possible $T_{k}(\mathcal{C}_n)$ subgroup generated by the torsional $k$-cycles acting as boundary of torsion-free cycles in $H_{k+1}(\mathcal{B}_{n+1},\mathcal{C}_n;\mathbb{Z})$. Saturating the first inequality \eqref{e.ineq1}, i.e., $  b_k(\mathcal{B}_{n+1})+ b_{n-k}(\mathcal{B}_{n+1}= b_k(\mathcal{C}_n)$, automatically results in $\tilde{t}_{k}(\mathcal{C}_{n})=0$, with all torsion-free cycles of $H_{k+1}(\mathcal{B}_{n+1},\mathcal{C}_n;\mathbb{Z})$ having also a torsion-free boundary.    
    As a matter of fact the first inequality is always saturated for the middle Betti numbers when $n$ is even, in the so-called \emph{Half lives, half dies} Theorem \cite{Putman:2022},
    \begin{equation}\label{e.halfhalf}
        \boxed{b_{n/2}(\mathcal{B}_{n+1})=\frac{1}{2}b_{n/2}(\mathcal{C}_n)\;,}
    \end{equation}
    irrespective of the choice of bordism realization $\mathcal{B}_{n+1}$. In general we will expect $\mathcal{B}_{n+1}$ to be connected and not enclosing any volume, so that $b_0(\mathcal{B}_{n+1})=1$ and $b_{n+1}(\mathcal{B}_{n+1})=0$, which further simplifies computations.
    
    Unlike the free part, the torsion part of the homology is not determined (up to isomorphisms) by the torsion rank. However, using the Fundamental Theorem of finite abelian groups one can show that $
        T_k(\mathcal{C}_n)\simeq S_k\oplus S'_{n-k-1}$, with  $S_k\leq T_k(\mathcal{B}_{n+1})$ and $S'_{n-k-1}\leq T_{n-k-1}(\mathcal{B}_{n+1})$, though we will not make use of this relation along the paper. Furthermore, for most of the examples considered, the torsion parts will be trivial.

        Although along this paper we will focus on bordisms to nothing, such that $\mathcal{C}_{n}$ has unique component, the above relations hold in generals, so in principle one could apply them for $\partial\mathcal{B}_{n+1}=\mathcal{C}_n\sqcup\overline{\mathcal{C}}_n'$, with bordisms between two compactifications.

\vspace{0.25cm}

    Now, denoting $\vec{b}(\mathcal{X}_n)=(b_0(\mathcal{X}_n),\dots,b_n(\mathcal{X}_n))$, we can illustrate the above relations \eqref{e.ineq} and \eqref{e.halfhalf} with some examples:
    \vspace{-0.25cm}
    \begin{itemize}
	\item $\mathcal{C}_1=\mathbb{S}^1$: $\vec{b}(\mathbb{S}^1)=(1,1)$, resulting in $\vec{b}(\mathcal{B}_2)=(1,\beta,0)$, with $\beta\geq0$. This precisely saturated by the disk, as $\vec{b}(\mathbb{D}^2)=(1,0,0)$, corresponding with Witten's BON. In general $\mathbb{S}^1$ can be seen as the boundary of $\Sigma_g$ with a disk removed, with $\vec{b}(\Sigma_g-\mathbb{D}^2)=(1,2g,0)$, which for $g>0$ can be thought of as topologically ``more complicated'' than $\mathbb{D}^2$.
	\item $\mathcal{C}_2=\Sigma_2=\mathbb{T}^2\#\mathbb{T}^2$: $\vec{b}(\Sigma_2)=(1,4,1)$, so that $\vec{b}(\mathcal{B}_3)=(1,2,\beta,0)$, with $\beta\geq0$, saturated by the solid double-torus or the manifold $\mathcal{B}_3$ sketched in Figure \ref{f.SKETCH}.
	\item $\mathcal{C}_3=\mathbb{T}^3$: $\vec{b}(\mathbb{T}^3)=(1,3,3,1)$, so that $\vec{b}(\mathcal{B}_4)=(1,\beta_1,\beta_2,\beta_3,0)$, with $\beta_1,\,\beta_2\,,\beta_3 \geq0$ and $\beta_1+\beta_2\geq 3$. Note that (though in this case just 2) the number of possibilities saturating the inequalities is not unique. The first one of them, $\vec{b}(\mathcal{B}_4)=(1,2,1,0,0)$, corresponds to the straightforward bordism $\mathcal{B}_4=\mathbb{T}^2\times\mathbb{D}^2$, while $\vec{b}(\mathcal{B}_4)=(1,1,2,0,0)$ does not immediately correspond to any known 4-manifold with boundary. We will come back to this example is section \ref{ss:nothing is certain} with a more involved bordism.
	\item $\mathcal{C}_6={\rm K3}\times \mathbb{T}^2$: $\vec{b}({\rm K3}\times \mathbb{T}^2)=(1,2,23,44,23,2,1)$, resulting in $\vec{b}(\mathcal{B}_7)=(1,\beta_1,\beta_2,22,$ $\beta_4,\beta_5,\beta_6,0)$, with $\beta_i\geq 0$ and $\beta_1+\beta_5\geq 2$ and $\beta_2+\beta_4\geq 23$. Note that again several possible different bordisms of $X_6$ (up to 72 possible solutions for $\vec{b}(\mathcal{B}_7)$) can saturate the inequalities. An evident instance of the bordism is $\vec{b}({\rm K3}\times \mathbb{S}^1\times\mathbb{D}^2)=(1, 1, 22, 22, 1, 1,0,0)$.
        \item  We finally take the lens space $\mathcal{C}_3=L(p;q)$, with homology groups $\mathbb{Z}$, $\mathbb{Z}_p$, 0 and $\mathbb{Z}$, which unlike the previous examples has torsion. We find that $\vec{b}(\mathcal{B}_4)=(1,\beta_1,\beta_2,\beta_3,0)$ with $\beta_i\geq 0$ for $i=1,\,2,\,3$, while $t_1(\mathcal{B}_4)\geq 1$ (the rest of $\mathcal{B}_4$ torsion ranks can be null).
	\end{itemize}
	It must be empathized that inequalities \eqref{e.ineq} and \eqref{e.halfhalf} are only \emph{necessary} conditions, so a vector $\vec{b}$ obeying them does not immediately correspond to an actual bordism.

	\subsection{Morse-Bott inequalities for manifolds with boundary}
    \label{ss:Morse-Bott}
	Once topological information for the bordism $\mathcal{B}_{n+1}$ has been obtained, it is time to relate it with the number and type of necessary topology changes $\mathcal{C}_n\to\cdots\to\emptyset$. An important tool to do this will be the ans\"atze \eqref{e.gen bon} and \eqref{e.large rho} for the BoN solutions, such that we can split the bordism and angular part of our solutions, and the radial direction $\rho:\mathcal{B}_{n+1}\to (-\infty,0]$,\footnote{For our purposes in this section, we will want $\rho$ to grown towards the tip of the bordism, so we will take this definition rather than $\rho:\mathcal{B}_{n+1}\to [0,+\infty)$.} which indicates the distance to the BoN/EotW wall. Note the following two properties of $\rho$:
 \begin{itemize}
     \item As given in \eqref{e.large rho}, for $\rho\to\infty$, we can split $h_{\alpha\beta}^\mathcal{B}(y)\dd y^\alpha\dd y^\beta\approx\dd \rho^2+{h}^{\mathcal{C}}_{\bar\alpha\bar\beta}\dd y^{\bar\alpha}\dd y^{\bar\beta}$, so that in this limit $\mathcal{B}_{n+1}$ locally looks like $\mathbb{R}\times \mathcal{C}_n$. This means that in this regime $\dd\rho\neq 0$, as $h^\mathcal{B}_{\alpha\beta}$ is non-degenerate.
     \item  As it parameterizes the smooth solution $\mathcal{B}_{n+1}$, we expect the critical points of $\rho$, i.e., where $\dd \rho=0$, to be a (not necessarily connected) submanifold of $\mathcal{B}_{n+1}$. For technical reasons, we also ask the Hessian of $\rho$ to be non-degenerate in the normal directions to the critical sub-manifold.\footnote{If this is not the case, we can take a deformation $\rho\to\rho'$ such that the monotonous behavior of this radial direction can be captured by the $\mathcal{O}(x_\perp^2)$ terms, with $x_\perp$ a normal direction. This is to say $\rho\sim\pm x_\perp^{2k}+\mathcal{O}(x_\perp^{2k+1})\to\rho'\sim \pm x_\perp^2+\mathcal{O}(x_\perp^3)$ and $\rho\sim\pm x_\perp^{2k+1}+\mathcal{O}(x\perp^{2k+2})\to\rho'\sim \pm x_\perp+\mathcal{O}(x_\perp^2)$.}
 \end{itemize}
 The above properties mean that $\rho$ is a \emph{Morse-Bott function}\footnote{In more precise words, given $\mathcal{B}_{n+1}$ a $(n+1)$-dimensional compact manifold with boundary $\mathcal{C}_n=\partial \mathcal{B}_{n+1}$, a smooth function $\rho:\mathcal{B}_{n+1}\to \mathbb{R}$ is called \emph{Morse-Bott} if the set of critical points ${\rm Crit}(f)=\{p\in \mathcal{B}_{n+1}:\,\dd f(p)=0\}$ is a disjoint union of connected sub-manifolds of $\mathcal{B}_{n+1}-\mathcal{C}_n$, with each connected component being non-degenerate, and $\left.\rho\right|_{\mathcal{C}_n}:X_n\to\mathbb{R}$ is also Morse-Bott. Note that in principle this would also allow $\mathcal{C}_n$ to be a manifold with boundary.}, which will allow us to use the so-called \textbf{Morse-Bott inequalities (for manifolds with boundary)} \cite{ORITA_2018}. In what follows we introduce the needed mathematical machinery in order to use them.

\begin{figure}[t!]
				\centering
				\includegraphics[width=0.65\textwidth]{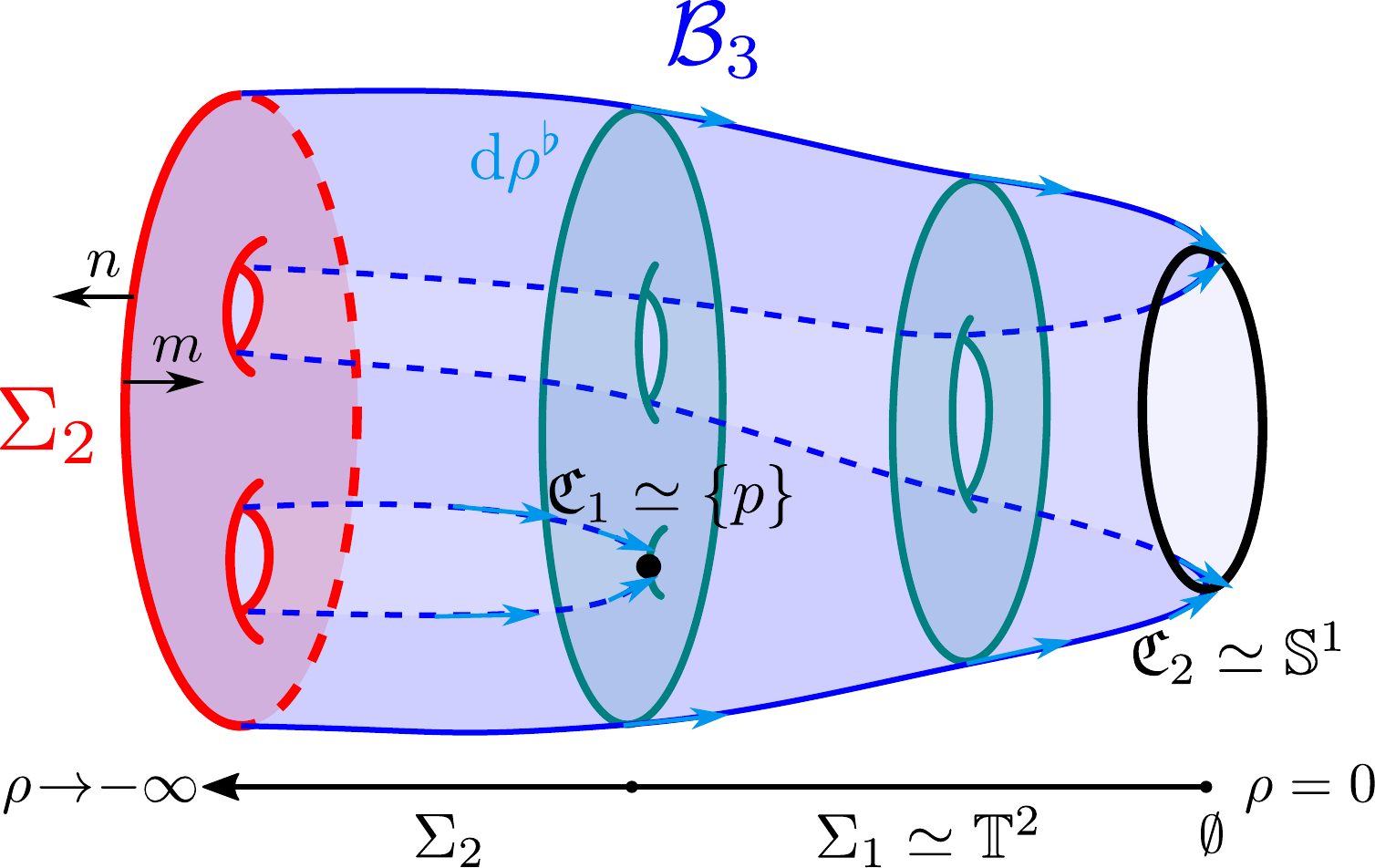}
			\caption{\small Sketch of the bordism $\mathcal{B}_3$ for the genus-2 Riemann surface $\Sigma_2$. The different critical submanifolds of the Morse-Bott function $\rho$ are depicted in black, $\mathfrak{C}_1\simeq\{p\}$ ($c_1=0$) and $\mathfrak{C}_2\simeq\mathbb{S}^1$ ($c_2=1$), both with indices $\lambda_1=\lambda_2=2$. Also depicted are two \emph{outward} and \emph{inward} normal vectors to the boundary, $n$ and $m$, as well as the vector field $\dd \rho^\flat=h^{\mathcal{B}\,\alpha\beta}\partial_\alpha\rho\partial_\beta=\partial^\alpha\rho\partial_\alpha$ associated to $\rho$, in blue. Depicted in green are the internal 2-manifolds at two fixed values of $\rho$.}
			\label{f.SKETCH}
	\end{figure}

	Take for this $\rho$ to be a Morse-Bott function. Then a critical  point of the boundary $p\in{\rm Crit}(\rho|_{\mathcal{C}_n})$ is said to be \emph{type $N$} if $\langle\dd \rho|_{\mathcal{C}_n}(p),n(p)\rangle<0$, where $n(p)\in T_p\mathcal{B}_{n+1}$ is an \emph{outward}\footnote{In the  sense of pointing out of the boundary $\mathcal{C}_n$ in a direction opposite to the interior of $\mathcal{B}_{n+1}$, rather than an \emph{inward} normal vector pointing towards the interior of $\mathcal{B}_{n+1}$. These are respectively represented by $n$ and $m$ in Figure \ref{f.SKETCH}.} normal vector to the boundary at $p$. We then define the following submanifolds and connected components:
	\begin{equation}
	\label{e.defsMB}
	{\rm Crit}(\rho)=\{\dd\rho=0\}=\bigsqcup_ {j=1}^l\mathfrak{C}_j,\quad
	\mathcal{N}(\rho)=\{p\in {\rm Crit}(\rho|_{\mathcal{C}_n}):\, p\text{ is }N\}=\bigsqcup_{s=1}^{l_N} \Gamma_s\;.
	\end{equation}
	We will denote by $c_j$ and $d_s$ the respective dimensions of $\mathfrak{C}_j$ and $\Gamma_s$, and by $\lambda_j$ and $\mu_s$ the Morse index of $\rho$ and $\rho|_{\mathcal{C}_n}$ over them (i.e. the number of negative eigenvalues of its Hessian at each point of the critical submanifold). These different submanifolds and concepts are depicted in Figure \ref{f.SKETCH}.
	
	The \emph{Morse-Bott counting polynomial of type $N$} is defined as
	\begin{equation}
	\label{e.MBpoly}
	\mathsf{MB}_t^{N}(\rho)=\sum_{j=1}^l\mathsf{P}_t(\mathfrak{C}_j)t^{\lambda_j}+\sum_{s=1}^{l_N}\mathsf{P}_t(\Gamma_s)t^{\mu_s}\,,
	\end{equation}
	where $\mathsf{P}_t(\mathcal{M})=\sum_{j=0}^{\dim \mathcal{M}}b_j(\mathcal{M})t^{j}$ is the Poincar\'e polynomial of a given manifold $\mathcal{M}$. Then the \emph{Morse-Bott inequalities for manifolds with boundary} state that for $\mathcal{B}_{n+1}$ a compact manifold with boundary and $\rho$ a Morse-Bott function defined over it, there exists a polynomial $\mathsf{R}\in\mathbb{Z}_{\geq 0}[t]$ with non-negative integer coefficients such that
	\begin{equation}
	\label{e.MorseBottIneq0}
	\boxed{\mathsf{MB}_t^N(\rho)-\mathsf{P}_t(\mathcal{B}_{n+1})=(1+t)\mathsf{R}(t)}
	\end{equation}

 From the behavior of $\rho$, as it is constant in $\mathcal{C}_n$ for $\rho\to\infty$, which we identify with the boundary of $\mathcal{B}_{n+1}$, $\mathcal{N}(\rho)=\mathcal{C}_n$, with $\mu=0$, so that 
 
	\begin{equation}\label{e.MorseBottIneq}	\boxed{\sum_{j=1}^l\mathsf{P}_t(\mathfrak{C}_j)t^{\lambda_j}+\mathsf{P}_t(\mathcal{C}_n)-\mathsf{P}_t(\mathcal{B}_{n+1})=(1+t)\mathsf{R}(t)}
	\end{equation}
	for some $\mathsf{R}(t)\in\mathbb{Z}_{\geq 0}[t]$. Note that evaluating at $t=-1$, we have $
     \chi(\mathcal{B}_{n+1})=\chi(\mathcal{C}_n)+\sum_{j=1}^l(-1)^{\lambda_j}\chi(\mathfrak{C}_j)$,  with $\chi(\mathcal{M})$ the Euler characteristic of $\mathcal{M}$.

\vspace{0.25cm}
 
Topology changes occur when internal cycles $\mathcal{A}\subseteq\mathcal{C}_n$ shrink over critical submanifolds $\mathfrak{C}$ of $\rho$, with $\dd \rho=0$ indicating that such cycle no longer ``propagates'' as we move towards the BoN/EotW brane. The dimensionality of the ``disappearing'' cycle is given by $c+\lambda-1$, where $c=\dim\mathfrak{C}$ and $\lambda$ its critical index, as $\lambda-1$ dimensions shrink (one has to discount the radial direction) over the $c$-dimensional base the cycle is fibered over. Furthermore, apart from this $(c+\lambda-1)$-cycle $\mathcal{A}$ that shrinks (what we could interpret as a cobordism for $\mathcal{A}$ into nothing), the dual $(n-c-\lambda+1)$-cycle $\mathcal{A}'$ also disappears, not shrinking \emph{per se} but being interpreted now as a chain with boundary at $\mathcal{C}$. This results in a reduction of the homology groups $H_{c+\lambda-1}(\rho^{-1}(r);\mathbb{Z})$ and  $H_{n-c-\lambda+1}(\rho^{-1}(r);\mathbb{Z})$ for $r$ smaller and larger than $r_*$, with $r_*$ the location of the critical point. This translates into a lowering (not necessarily by a unit) of the respective Betti numbers and/or torsion ranks. We will revisit this cycle shrinking/transforming into a chain with boundary interpretation when discussing flux potentials and their fate after topology changes in section \ref{ss:IIB Riemann}.

 Analogously, it could be the case that the topology change results in the creation of some cycle (and its dual), as occurs in Figure \ref{ff.exS12}, but for these cases the cobordism into nothing will feature more topology changes than necessary.

	The number $l$ of critical submanifolds will correspond to the number of topology changes (including the ultimate collapse into nothing), so impossibility to satisfy \eqref{e.MorseBottIneq} with $l=1$ means an intermediate topology change must occur. Note that the index of the final critical submanifold is given by $\lambda+c-1=n$, as all dimensions must close at the top of the BoN/EotW brane.

\vspace{0.25cm}

Inequalities \ref{e.MorseBottIneq} only consider the free part of the homology groups (i.e. the Betti numbers), with the torsion ranks not entering into account. While most of the examples considered will not feature torsion, one would expect that bordisms with torsion will probably require additional topology changes in order to get rid of this topological charge. Unfortunately, to our knowledge there are not results concerning Morse-Bott inequalities for manifolds with boundary \emph{and torsion} in the Mathematics literature. The closest result are the \emph{Morse-Pitcher inequalities} \cite{Pitcher} for closed manifolds without boundary (but with torsion). Having $\rho:\mathcal{M}_n\to\mathbb{R}$ a \emph{Morse} function (this is, critical loci will be isolated points rather than general submanifolds, as in Morse-Bott functions), then
\begin{subequations}\label{e.morse torsion}
    \begin{align}
        m_\lambda&\geq b_\lambda(\mathcal{M}_n)+t_\lambda(\mathcal{M}_n)+t_{\lambda-1}(\mathcal{M}_n)\\
        \sum_{k=0}^\lambda(-1)^{\lambda-k}m_k&\geq \sum_{k=0}^\lambda(-1)^{\lambda-k}b_k(\mathcal{M}_n)+t_\lambda(\mathcal{M}_n)\;,
    \end{align}
\end{subequations}
for $\lambda=1,\,\dots,\, n$, where $m_\lambda$ is the number of critical points of $\rho$ with index $\lambda$. While it is clear that having torsion on $\mathcal{B}_{n+1}$ will probably result in more critical points (and topology changes) appearing, it is not clear the way this can be quantified. In section \ref{ss:BoNCollide} we will apply the above \eqref{e.morse torsion} to study collision of bubbles of nothing.

Topology changes associated to an internal cycle shrinking into a point and disappearing are associated, from the EotW brane point of view, with the moduli that controlled its size in the original EFT blowing up to infinity. If the rest of cycles and the whole manifold remain relatively unaltered, which can be considered to be a good approximation in the large volume regime, then the critical exponent $\delta$ from \eqref{e. critical exponent} could be read from the dependence of said blowing up scalar in the potential. If the different topology changes are spaced in the $\rho$ spatial direction, then a EFT description is valid between the different domain walls, and critical exponents can be associated  with distinct topology changes/types of critical points, though not necessarily in a 1-to-1 basis. An example of these are flux potentials from $p$-form fluxes threading different internal cycles, as we will see in section \ref{sec:Examples} in different examples.
 
	\subsubsection{Some easy examples}
	
	 We will now consider some ubiquitous compactification manifolds to test and illustrate this.
\paragraph*{Example 1: Spheres $\mathbb{S}^n$}
Our first example is the $n$-dimensional sphere, which is known to be the boundary of the solid ball $\mathbb{D}^{n+1}$. One way one can imagine the compactification to interpolate to nothing is by shrinking the radial direction to a single point, so that ${\rm Crit}(\rho)=\{p\}$ with index $\lambda=n+1$. As $\mathsf{P}_t(\mathbb{S}^n)=1+t^n$ and $\mathsf{P}_t(\mathbb{D}^{n+1})=1$, one has that \eqref{e.MorseBottIneq} translates to
\begin{equation}
	(1+t)\mathsf{R}(t)=1\times t^{n+1}+(1+t^n)-1=(1+t)t^n\Longrightarrow \mathsf{R}(t)=t^n\in \mathbb{Z}_{\geq 0}[t],
\end{equation}
so indeed the Morse-Bott inequalities are satisfied with a single topology change.
\paragraph*{Example 2: Tori $\mathbb{T}^n$}
Due to their simplicity and control, some of the most studied compactifications in the literature are toroidal manifolds. Surprising nobody, they can also be taken to nothing without intermediate topology changes. Indeed, as $\mathbb{T}^n=\mathbb{T}^{n-1}\times \mathbb{S}^1$, one can interpolate to nothing by shrinking one of the 1-cycles over the $\mathbb{T}^{n-1}$ base. This way, using $\mathsf{P}_t(\mathbb{T}^n)=(1+t)^n$
\begin{equation}\label{e.Tn}
	(1+t)\mathsf{R}(t)=(1+t)^{n-1}t^2+ (1+t)^n-(1+t)^{n-1}\cdot 1=(1+t)^{n}(t-1)=t(1+t)^{n}\;,
\end{equation}
so that one can take $\mathsf{R}(t)=t(1+t)^{n-1}$, which means that in principle intermediate topology changes are not necessary for BoN with compact manifold $\mathbb{T}^n$. We will see in section \ref{ss:nothing is certain} more involved bordisms of toroidal compactifications which include additional topology changes.
 
\paragraph*{Example 3: Product manifolds}
An immediate corollary follows for product manifolds $\mathcal{C}_n=\mathcal{X}_m\times \mathcal{Y}_{n-m}$ (both manifolds without boundary) in which $\mathcal{X}_m$ can go to nothing without intermediate topology changes. In other words, there exists a $(m+1)$-smooth manifold $\mathcal{D}_{m+1}$ with $\partial \mathcal{D}_{m+1}=\mathcal{X}_m$, a Morse-Bott function $\rho:\mathcal{D}_{m+1}\to [0,+\infty)$ with a single critical submanifold $\mathfrak{C}\subset \mathcal{D}_{m+1}$ and polynomial $\mathsf{R}(t)\in\mathbb{Z}_{\geq 0}[t]$ such that
\begin{equation}
	\mathsf{P}_t(\mathcal{C})t^{m-\dim \mathfrak{C}}+\mathsf{P}_t(Y_m)-\mathsf{P}_t(\mathcal{D}_{m+1})=(1+t)\mathsf{R}(t)\;.
\end{equation}
Now, defining $\mathcal{B}_{n+1}=\mathcal{D}_{m+1}\times \mathcal{Y}_{n-m}$, we have $\partial \mathcal{B}_{n+1}=(\partial\mathcal{D}_{m+1}\times \mathcal{Y}_{n-m})\cup (\mathcal{D}_{m+1}\times \partial \mathcal{Y}_{n-m})=\mathcal{X}_m\times \mathcal{Y}_{n-m}=\mathcal{C}_n$, thus providing a bordism realization for $X_n$. We can then extend the Morse-Bott function
\begin{equation}
	\begin{array}{rccl}
	\hat{\rho}:\mathcal{B}_{n+1}=&\mathcal{D}_{m+1}\times \mathcal{Y}_{n-m}&\longrightarrow&[0,+\infty)\\
	&(y,z)&\longmapsto&\rho(y)
	\end{array}\;,
\end{equation}
such that the only critical submanifold of $\hat{\rho}$ is $\hat{\mathfrak{C}}=\mathfrak{C}\times \mathcal{Y}_{n-m}$. Finally, using $\mathsf{P}_t(A\times B)=\mathsf{P}_t(A)\mathsf{P}_t(B)$, we obtain
\begin{equation}
	\mathsf{P}_t(\hat{\mathfrak{C}})t^{m-\dim \mathfrak{C}}+\mathsf{P}_t(\mathcal{C}_n)-\mathsf{P}_t(\mathcal{B}_{n+1})=(1+t)\mathsf{R}(t)\mathsf{P}_t(\mathcal{Y}_{n-m})\;.
\end{equation}
As the Poincar\'e polynomial always has non-negative coefficients, $\mathsf{R}(t)\mathsf{P}_t(\mathcal{Y}_{n-m})\in\mathbb{Z}_{\geq 0}[t]$, so that in principle there is no need for intermediate topology changes in $\mathcal{C}_n=\mathcal{X}_M\times \mathcal{Y}_{n-m}$ going to nothing. As we will see in section \ref{ss:K3}, this does not extend to non-trivial fibrations.

    \subsection{Non-minimal bordisms and defects}
    \label{ss:defects}

In the above sections only the global topology (actually only the homology groups) of the bordism $\mathcal{B}_{n+1}$ was considered, assuming from the start that the relevant bordism groups $\Omega^{\mathsf{g}}_n$ vanished. However, as explained in section \ref{sec:intro}, in order for this to happen there might be extra ingredients or \emph{defects} that must be added to our theory, such that $\Omega_n^{\mathsf{g}}\neq0\to\Omega^{\mathsf{g}+{\rm def.}}_n=0$. In a general way, $\Omega_k^{\mathsf{g}}\neq0$ implies the existence of a $D-k-1$-global symmetry in our theory, with associated conserved topological charge/invariant \cite{McNamara:2019rup}
\begin{equation}
    \begin{array}{cccl}
         Q:&\Omega_k^{\mathsf{g}}&\longrightarrow&\mathbb{Z}\\
         &[\mathcal{M}_k]&\longmapsto&\int_{\mathcal{M}_k}G_{k}
    \end{array}\;,
\end{equation}
where $G_{k}$ is some associated closed $k$-form. Note that by definition $Q$ is independent of the bordism representative. Now, in order to kill this bordism invariant, a $D-k-1$ defect is needed. For the case of oriented bordisms in the presence of $A_p$ forms, $\Omega_{p+1}^{SO,\,A_p}\simeq \mathbb{Z}$, represented by the magnetic charge $Q$ above. The $D-(p+1)-1=D-p-2$ defects are given by the D$(D-p-3)$-branes on whose world-volume the magnetic dual gauge potential $\tilde{A}_{D-p-2}$ lives (see \cite{McNamara:2019rup} for more details on this).

    \begin{figure}[h]
\begin{center}
\begin{subfigure}[b]{0.45\textwidth}
\center
\includegraphics[width=0.75\textwidth]{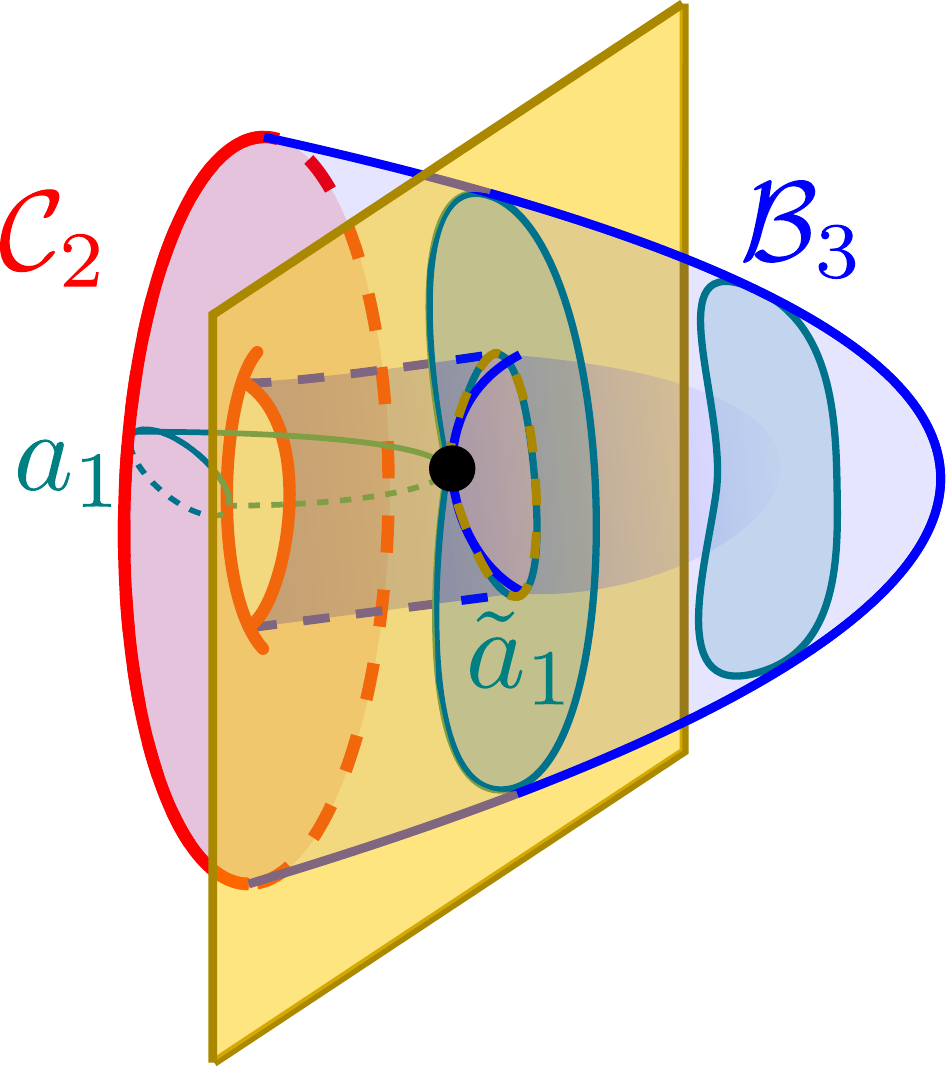}
\caption{\hspace{-0.3em}) Thin wall on bordism for $\mathcal{C}_2=\mathbb{T}^2$.} \label{ff.thinwall}
\end{subfigure}
\begin{subfigure}[b]{0.54\textwidth}
\center
\includegraphics[width=0.75\textwidth]{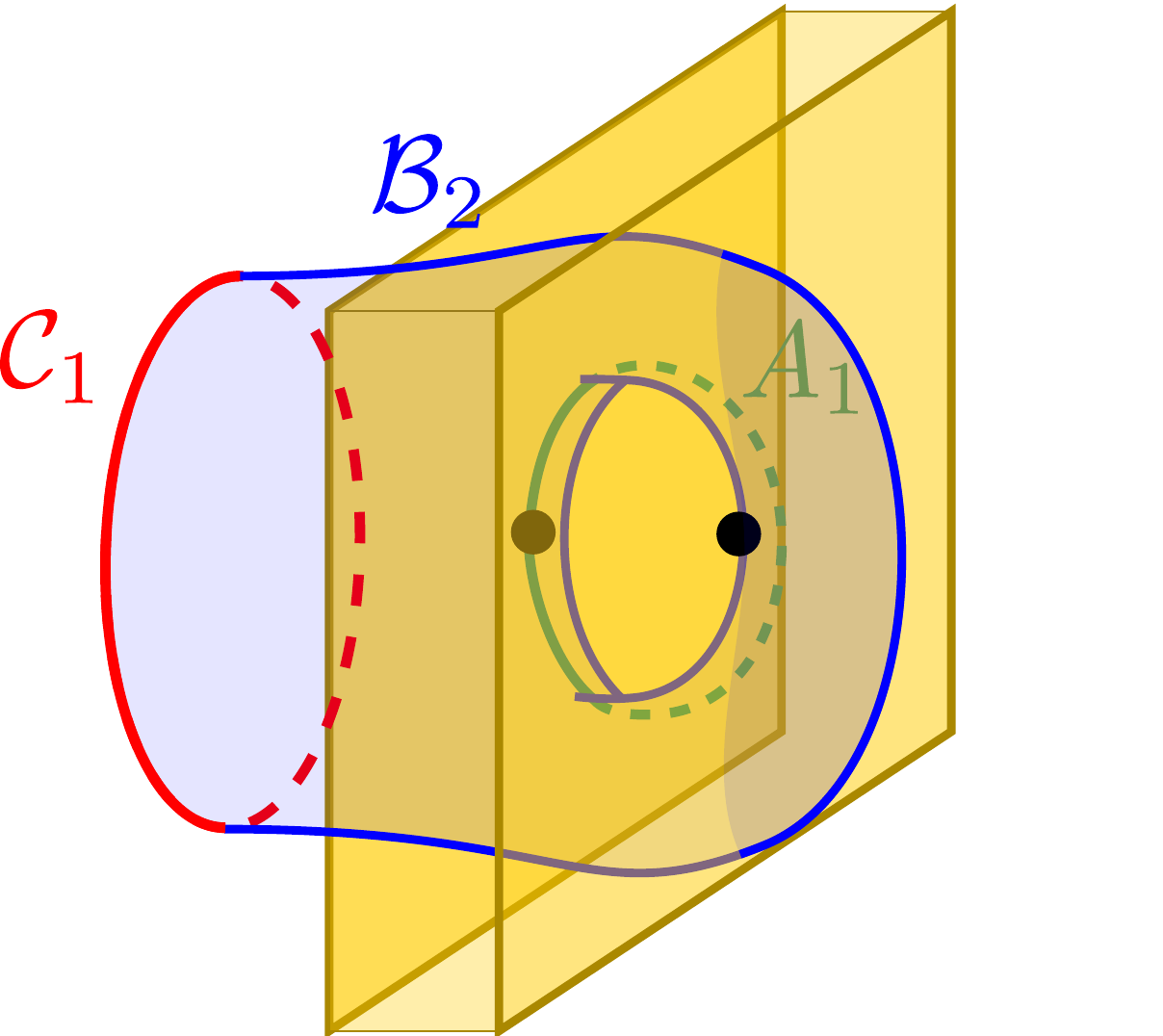}
\caption{\hspace{-0.3em}) Thick wall on non-trivial bordism for $\mathcal{C}_1=\mathbb{S}^1$.} \label{ff.thickwall}
\end{subfigure}
\caption{Illustrations of thin and thick walls in different bordisms, respectively for $\mathcal{C}_2=\mathbb{T}^2$ (subfigure \ref{ff.thinwall}) and $\mathcal{C}_1=\mathbb{S}^1$ (subfigure \ref{ff.thickwall}). In the thin wall case the $(d+n-k-1=d)$-defect simply extends over $d-1$ spacetime directions and wrapping along the $(n-k=2-1=1)$-cycle $\tilde{a}_1$ dual to the 1-cycle $a_1$ that is shrinking. In the thick wall case the analogous $d$-defect extends over $d-1$ spacetime directions and wraps a 1-cycle $A_1$ of $\mathcal{B}_2\simeq \mathbb{T}^2-\mathbb{D}^2$. This second example can be found in section 7.3 from \cite{Debray:2023yrs}.
\label{f.walls}}
\end{center}
\end{figure}

In general, when compactifying on $\mathcal{C}_n$ to $d=D-n$ dimensions, we will also have internal $k$-cycles over which topological charges can be defined. The $d+n-k-1$-defects associated with $\{\Omega_k^{\mathsf{g}}\}_{k=0}^n$ will allow the associated $k$-cycles to go to nothing in the intermediate topology changes described in section \ref{ss:Morse-Bott}. As at the different sides of the topology change \eqref{eq: top chain} the compactification manifold $\mathcal{C}_n^{(k)}\to\mathcal{C}_n^{(k+1)}$ will be different, the lower $d$-dimensional EFT will change, with the $d+n-k-1$-defects being seen from the macroscopic point of view as codimension-1 $d-1$-domain walls, which are localized at the topology change loci. Depending on how the additional $n-k$ directions are wrapped, we can have the following two types of domain walls:
\begin{itemize}
    \item \textbf{Thin domain walls}: The $d+n-k-1$-defect extends over $d-1$ of the macroscopic directions, while wrapping the $n-k$-cycle dual to the $k$-cycle in $\mathcal{C}_n$ supporting the topological charge, and that the defect allows to shrink to nothing. Note that after the topology change the dual $n-k$-cycle no longer exists either, so the domain wall also acts as a boundary of both cycles. An illustration of this is depicted in Figure \ref{ff.thinwall}, and will be found in explicit examples in section \ref{ss:IIB Riemann}.
    \item \textbf{Thick domain walls}: The $d+n-k-1$-defect extends over $d-1$ macroscopic directions (perpendicular to the bordism radial direction $\rho$), while wrapping a $n-k$-cycle in the bordism $\mathcal{B}_{n+1}$, in such a way that the cycle also extends in the $\rho$ direction (for a finite interval), with the $n-k$-cycle not corresponding to any specific cycle in any of the $\mathcal{C}_n^{(k)}$ slices. Note that in this case the localization occurs around said cycle, which by definition is non-contractible, unlike those of $\mathcal{C}_n^{(k)}
    $ once embedded in $\mathcal{B}_{n+1}$.
    
    This is illustrated in Figure \ref{ff.thickwall}, with an example for this disposition being found in section 7.3 of \cite{Debray:2023yrs} (albeit not in the context of an explicit BoN construction), the non-trivial 1-cycle $A_1\subset \mathcal{B}_2$ has a monodromy that kills the $\mathbb{Z}_3$ factor of $\Omega_1^{{\rm Spin-Mp}(2,\mathbb{Z})}\simeq\mathbb{Z}_8\oplus\mathbb{Z}_3$ by performing a reflection $g\to g^{-1}$ on the generator of $\mathbb{Z}_3=\langle g\rangle$ (we refer to the original, \cite{Debray:2023yrs} for more details). This is equivalent to having a 8-defect spanning between the two critical points (in black in Figure \ref{ff.thickwall}) associated to $A_1$, resulting in a \emph{thick} (7+1)-domain wall.\footnote{We thank Markus Dierigl and Miguel Montero for referencing a specific example of this class to us.}
\end{itemize}

The defects considered here in general are of stringy origin and will backreact on the spacetime geometry, resulting in a loss of the smooth geometry (e.g. conical defects and angular deficits). We expect that, in some of these cases, the topological description of our bordism holds in a way that can be smoothed-out and the Morse-Bott inequalities hold.

With respect to thick domain walls, in order for the defects or non-trivial geometry (such as monodromies in the above example) to wrap internal (non-contractible) cycles of the bordism, it will be required that the topology of $\mathcal{B}_{n+1}$ is more involved than the minimal structure required from \ref{e.ineq}. As we will see in section \ref{ss:nothing is certain}, this can also be the case in order to include the necessary (in this case Spin) defects while holding a smooth description of the bordism.

\section{Some more physical examples}
\label{sec:Examples}
After the theoretical introduction and mathematical results from the previous sections, we will now consider some more realistic constructions, as well as the qualitative properties of the cobordisms to nothing/end of the world branes.

     \subsection{Spheres with fluxes}
    We first begin by considering a pretty straightforward set of examples. Take for this the 10d Einstein-dilaton-RR sector of the type II bosonic supergravity action in Einstein frame \cite{Ibanez:2012zz,VanRiet:2023pnx},
    \begin{equation}\label{e.II action}
        S_{\rm II}\supseteq \frac{1}{2\kappa_{10}^2}\int\dd^{10}x\sqrt{-G_{10}}\left\{\mathcal{R}_{10}-(\partial\hat\Phi)^2-\frac{1}{2n!}e^{\frac{5-n}{\sqrt{2}}\hat\Phi}|F_n|^2\right\}\;,
    \end{equation}
    where $\Hat{\Phi}$ is the canonically normalized 10 dimensional dilaton, and for simplicity we have considered a single RR field strength. One can compactify the above theory on a $n$-sphere $\mathbb{S}^{n}$ with volume $\mathcal{V}$ in $M_{\rm Pl,10}$ units and $N$ units of flux, which, assuming that $F_p$ only has internal legs, results in a scalar potential in the $(d=10-n)$-dimensional EFT from the curvature and flux contributions:
    \begin{equation}\label{eq.sphere pot}
        \kappa_d^{2}V(\hat\Phi,\rho)=-\frac{\pi}{2}(10-d)(9-d)\Gamma\left(6-\frac{d}{2}\right)^{-\frac{2}{10-d}}e^{-2\sqrt{\frac{8}{(10-d)(d-2)}}\rho}+\frac{N^2}{4}e^{\frac{d-5}{\sqrt{2}}\hat\Phi-(d-1)\sqrt{\frac{10-d}{2(d-2)}}\rho}\;,
    \end{equation}
    where $\rho=\sqrt{\frac{8}{(10-d)(d-2)}}\log\frac{\mathcal{V}}{\mathcal{V}_0}$ is the canonically normalized radion, with $\mathcal{V}_0=\left(\frac{\kappa_{10}}{\kappa_d}\right)^2$ some fixed volume. While for $d=5$ (precisely when $\Phi$ is not sourced by the RR field strength, see \eqref{e.II action}) \eqref{eq.sphere pot} only depends on the volume $\mathcal{V}\sim N^{\frac{5}{4}}$ and has an AdS minimum $V_{\rm AdS}\sim -N^{-\frac{4}{3}}$ which becomes deeper as $N$ decreases, for $d\neq 5$ the potential gradient flow pushes towards small volume configurations and large/small coupling (depending on the sign of $5-d$), with $\hat\Phi=\sqrt{\frac{d-2}{10-d}}\frac{d-9}{5-d}\rho$.

    Now, as explained in section \ref{ss:defects}, the inclusion of $(p-1)$-forms in our theory (which act as gauge potentials for the RR field strength) results in a conserved topological charge (the flux number) upon compactification on oriented surfaces such as $\mathbb{S}^p$. In order to kill the $\Omega_{p}^{SO,\, A_{p-1}}$ invariant, D$(8-p)$-brane defects must be included. Precisely this D$(8-p)$-brane will act as the EotW brane, with the $\mathbb{S}^p$ sphere shrinking as we approach it. This was studied in more generality in \cite{Angius:2022aeq}, to which we refer for more detail. In general the inclusion of a D-brane results in a singular term in the EFT action, and as such the bordism $\mathcal{B}_{p+1}$ will have a conical singularity at its tip, where the Dp-brane is located. However, we still have $\mathcal{B}_{p+1}\simeq\mathbb{D}^{p+1}$, and so we can still consider a generic smooth structure on the bordism and the results from section \ref{sec:AlgTop}.

    In the presence of $F_n$ RR flux, a D$(8-n)$ brane (with $n=2,\dots,7$)\footnote{We do not consider D8- and D7-branes as their backreaction prevents asymptotically flat spacetime, complicating the analysis.} solution to \eqref{e.II action} has the following 10d Einstein metric and $\hat{\Phi}$ profiles \cite{Ibanez:2012zz}:
    \begin{subequations}
        \begin{align}
            \dd s^2_{10}&=Z(r)^{\frac{1-n}{8}}\underbrace{\eta_{\mu\nu}\dd x^\mu\dd x^\nu}_{\dd x^2_{\parallel}}+Z(r)^{\frac{7-n}{8}}\underbrace{(\dd r^2+r^2\dd\Omega^2_{8-n})}_{\dd x^2_\perp}\\
            \hat{\Phi}&=\frac{n-5}{4\sqrt{2}}\log Z(r)+\hat{\Phi}_{\infty}\;,\quad\text{where}\quad Z(r)=1+\left(\frac{r_0}{r}\right)^{n-1}\;,
        \end{align}
    \end{subequations}
    where $r$ is a radial direction perpendicular to the brane and $r_0$ is a length scale given by
    \begin{equation}
        r_0=\sqrt{\alpha'}\left[(4\pi)^{\frac{n-3}{2}}\Gamma\left(\frac{n-1}{2}\right)e^{\sqrt{2}\hat{\Phi}_\infty}N\right]^{\frac{1}{n-1}}\;,
    \end{equation}
    explicitly depending on the flux number $N$. For $n=5$ (i.e. D3 branes), the dilaton does not blow up at $r=0$ (and as commented above results in an AdS minimum with a flat direction for $\hat{\Phi}$ upon compactification), and the D3 brane will not correspond to the EotW brane description from section \ref{ss: EotW}, as there is neither a cycle shrinking or a scalar going to infinity.\footnote{In this $d=5$ the cobordism to nothing is more subtle \cite{Brown:2011gt}, occurring through nucleation of D3-branes that eat up units of flux charge, $N\to N-1$, in such a way that the minimum of the potential $V_{\rm AdS}\sim-N^{-4/3}$ becomes deeper and the $\mathbb{S}^5$ shrinks, $\mathcal{V}\sim N^{5/4}$. In the $N\to 0$ limit the AdS curvature radius goes to zero and the 5-sphere shrinks to a point. Though this does not happen in a remotely smooth way either (the integer flux jumps when crossing each brane), from the topological point of view the cobordism to nothing still corresponds in the sphere shrinking to a point without intermediate topology changes. See also \cite{Horowitz:2007pr}, where a (non-smooth) BoN solution for the non-supersymmetric AdS$_5\times \mathbb{S}^5/\mathbb{Z}_k$ is built, based in the Hopf fibration structure $\mathbb{S}^1\hookrightarrow\mathbb{S}^5/\mathbb{Z}_k\to\mathbb{CP}^2$, with the $\mathbb{S}^1$ fiber shrinking over the AdS$_5\times \mathbb{CP}^2$ similarly to Witten's BoN.} We will thus not consider this case and remain with $n\neq 5$.

    Upon compactification on $\mathbb{S}^{n}$ (transversal to the brane) down to $d=10-n$ dimensions, the D$(8-n=d-2)$-brane has now spatial codimension one (as expected from EotW branes), with the Einstein frame metric and the radion profile now given by \cite{Angius:2022aeq}
    \begin{subequations}
        \begin{align}
            \dd s^2_d&=\left[\left(\frac{r}{r_0}\right)^2Z(r)^{\frac{d-1}{8}}\right]^{\frac{10-d}{d-2}}\left\{Z(r)^{\frac{d-9}{8}}\eta^{\perp}_{\mu\nu}\dd x^\mu\dd x^\nu+Z(r)^{\frac{d-1}{8}}\dd r^2\right\}\\
            \rho(r)&=\sqrt{\frac{2(10-d)}{d-2}}\log\left[\left(\frac{r}{r_0}\right)^2Z(r)^{\frac{d-1}{8}}\right]\;,
        \end{align}
    \end{subequations}
    where now the volume scales as
    \begin{equation}
        \frac{\mathcal{V}(r)}{\mathcal{V}_0}=e^{\frac{1}{2}\sqrt{\frac{(d-2)(10-d)}{2}}\rho(r)}=\left(\frac{r}{r_0}\right)^{10-d}Z(r)^{\frac{(10-d)(d-1)}{2}}\;,
    \end{equation}
    with $V(r)\to 0$ as $r\to 0$ except for $d=5$ (equivalently $n=5$ or D3 branes), where the volume remains finite, and indeed the D$(d-2)$-branes serve as EotW branes, coinciding with the tip of the $\mathbb{S}^{10-d}$ bordism to nothing, topologically given by the solid disk $\mathbb{D}^{11-d}$. The $\Omega_{d-2}^{SO,\, A_{10-d}}$ defect given by the D-brane is precisely located in the topology change (in this case into nothing).

    For completion, we refer to \cite{Angius:2022aeq}, where these solutions were studied with the EotW brane formalism, where the following critical exponent\footnote{This critical exponent can also be immediately computed by obtaining the exponential rate of potential \eqref{eq.sphere pot} along the trajectory $\hat{\Phi}=\sqrt{\frac{d-2}{10-d}}\frac{d-9}{5-d}\rho$ the scalars follow for $d\neq 5$.}
    \begin{equation}
        \delta=\frac{2|d-5|}{\sqrt{(d-2)(11-d)}}=\frac{2|5-n|}{\sqrt{(8-n)(n+1)}}
    \end{equation}
    was computed, with again $\delta=0$ for $d=n=5$. Note that for these examples there is a 1-to-1 correspondence between the dimensionality of the sphere shrinking to a point and the critical exponent. 

\vspace{0.25cm}

    The above description only discusses the local EotW behavior of the scalars as they approach the brane/cobordism wall and the sphere threaded by fluxes shrinks to a point, which as we have argued for and seen, can happen without intermediate topology changes. Actual vacuum instability solutions corresponding to BoN are much more involved, and we will not detail them. For this we refer to the literature, such as \cite{Yang:2009wz,Blanco-Pillado:2010xww,Blanco-Pillado:2010vdp,Brown:2010mf,Blanco-Pillado:2016xvf,Draper:2023ulp,Blanco-Pillado:2023aom}.

    \subsection{Type IIB string theory on compact Riemann surfaces}
    \label{ss:IIB Riemann}
    We will now consider a more involved example with actual intermediate topology changes being needed in order to have a smooth(able) bordism to nothing, and study how this intermediate changes in the internal dimension affect the lower-dimensional EFT. For this we will consider Type IIB string theory compactified on a Riemann surface $\Sigma_g$ of genus $g\geq 2$.

     For simplicity, we will not consider the full ${\rm Spin}-GL^+(2,\mathbb{Z})$ bordism of type IIB string theory\footnote{Type IIB string theory has fermions and a duality group $SL(2,\mathbb{Z})$, which after including the F-theory torus reflections (and their action on fermions) is promoted to $GL^+(2,\mathbb{Z})$ \cite{Tachikawa:2018njr,Pantev:2016nze}. This way, the actual structure our bordism must take into consideration is ${\rm Spin}-GL^+(2,\mathbb{Z})$ (see \cite{Debray:2021vob,Dierigl:2022reg,Debray:2023yrs} for more on this, including the rest of cobordism groups and generators, as well as properties of the required defects), with 
     \begin{equation}
         \Omega^{{\rm Spin}-GL^+(2,\mathbb{Z})}_1=\langle\{\mathbb{S}^1_R,L^1_4\}\rangle \simeq \mathbb{Z}_2\oplus\mathbb{Z}_2\;,\qquad \Omega^{{\rm Spin}-GL^+(2,\mathbb{Z})}_2=\langle\mathbb{S}^1_p\times\mathbb{S}^1_R\rangle\simeq \mathbb{Z}_2\;,
     \end{equation}
     where $\mathbb{S}^1_p$ is the circle with periodic boundary conditions, $\mathbb{S}^1_R$ is the circle with a non-trivial $GL^+(2,\mathbb{Z})$ reflection bundle, and $L^1_4\simeq\mathbb{S}^1/\mathbb{Z}_4$ a Lens space, naturally endowed with a $\mathbb{Z}_4$ bundle.} and simply restrict to Spin bordism, with \cite{Anderson1967:SPIN,McNamara:2019rup}
     \begin{equation}\label{e.spin bord}
         \Omega^{\rm Spin}_1=\langle\mathbb{S}_p^1 \rangle\simeq \mathbb{Z}_2\;,\qquad \Omega^{\rm Spin}_2=\langle\mathbb{S}_p^1 \times \mathbb{S}_p^1 \rangle\simeq \mathbb{Z}_2\;.
     \end{equation}
     If we further take $\Sigma_g$ to be oriented and include the $p$-fields in the 10d SUGRA description, as explained in section \ref{ss:defects}, D-branes will have to be include in order to ``eat'' the magnetic charges of the fluxes.

     Before studying the 8d EFT description after compactifying on the Riemann surface, let's study how many and how the topology changes occur.

     \subsubsection*{Morse-Bott inequalities for $\Sigma_g$}
     For our purposes, a Riemann surface $\Sigma_g$ is determined by its genus $g$ and its Spin structure. Having no torsion, homology is set by the Betti numbers $\vec{b}(\Sigma_g)=(1,2g,1)$, so there are $b_1(\Sigma_g)=2g$ 1-cycles, paired in intersecting dual cycles. By using \eqref{e.ineq} and \eqref{e.halfhalf}, the Poincar\'e polynomial of the bordism $\mathcal{B}_3$ is $\mathsf{P}_t(\mathcal{B}_3)=1+gt$.\footnote{Riemann surfaces automatically set the Betti numbers of their bordism to nothing (assuming it has a single connected component), as the \emph{Half dies, half lives} theorem implies $b_1(\mathcal{B}_3)=\frac{1}{2}b_1(\Sigma_g)=g$.} Because of the low dimensionality we are working with, the only possible critical submanifolds are $\{p\}$ and $\mathbb{S}^1$, of dimension $c=0$ and $c=1$, respectively. Now, as at each critical locus with index $\lambda$ we have a $(\lambda+c-1)$-cycle shrinking to nothing, this only allows us to have \textbf{(a)} a 1-cycle collapsing on a point, \textbf{(b)} a 1-cycle collapsing on $\mathbb{S}^1$, or \textbf{(c)} a 2-cycle without internal cycles\footnote{If torsion has not been induced on $\mathcal{B}_3$, this is then $\mathbb{S}^2$ shrinking.} collapsing on a point. As the two last possibilities saturate the $\lambda+c-1\leq 2=\dim \Sigma_g$ inequality, and thus correspond to the final topology change into nothing, there can only be one of them.

     Considering $(a,b,c)$ topology changes of the above types (i.e., a total $N=a+b+c$ topology changes), the Morse-Bott inequalities will be given by
     \begin{equation}
         at^2+b(1+t)t^2+ct^3+(1+2gt+t^2)-(1+gt)=(1+t)\mathsf{R}(t)\;,
     \end{equation}
     with $\mathsf{R}(t)\in\mathbb{Z}_{\geq 0}[t]$ and $b+c=1$. In order to have $1+t$ being a factor of the LHS of the above equation, it is needed that $a=g+c-1$. This results in the following options, depending on the genus $g$:
     \begin{itemize}
         \item $g=0$: The sphere $\Sigma_0=\mathbb{S}^2$ shrinks to nothing in a single topology change, with $(a,b,c)=(0,0,1)$.
         \item $g=1$: For the torus $\Sigma_1=\mathbb{T}^2$ it is possible to go to nothing via one of the 1-cycles shrinking over its dual, so $(a,b,c)=(0,1,0)$. It is also possible to have two topology changes, with a 1-cycle pinching over a point, leading to a sphere that later shrinks, $\mathbb{T}^2\to\mathbb{S}^2\to\emptyset$, $(a,b,c)=(1,0,1)$. 
         \item $g\geq 2$: The contribution of $a=g+c-1$ to the total number of topology changes is minimized by $(a,b,c)=(g-1,1,0)$. As each \textbf{(a)}-type topology change ``kills'' a pair of dual 1-cycles, this results in a final $\Sigma_1=\mathbb{T}^2$ which shrinks through a type-\textbf{(b)} topology change, with a total $N=g$. An extra \textbf{(a)} and \textbf{(c)} can be in place of the last \textbf{(b)}, amounting for $N=g+1$.
     \end{itemize}

     This way, a bordism (that does not increment the number of internal components or 1-cycles) between $\Sigma_g$ and nothing features either $g$ ($g-1$ type \textbf{(a)} and a final \textbf{(b)}) or $g+1$ ($g$ type \textbf{(a)} and a final \textbf{(c)}) topology changes. At each of these, the compact manifold will change as we move in the $\rho$ direction of the bordism, resulting in different low-energy EFTs behind each domain wall.  

     In Appendix \ref{app:RiemannianSurfaces} some basics about Riemann surfaces and their shape moduli are presented. While the final shrinking of $\mathbb{T}^2$ or $\mathbb{S}^2$ (type \textbf{(b)} or \textbf{(c)} topology changes from the above description) is well understood, (accounting simply for a compactification limit in the EotW brane description), type \textbf{(a)} changes, with a 1-cycle pinching on a point and decaying, are not as simple. This can be well understood through tachyon condensation from a world-sheet perspective \cite{Saltman:2004jh,Adams:2005rb}. In Appendix \ref{app:tachyon} we give a brief overview of how 1-cycles of Riemann surfaces with anti-periodic boundary conditions decay once their thickness at some point of the handle becomes small enough, with tachyon condensation effectively preventing spacetime degrees of freedom from propagating along the handle direction.

     \subsubsection*{The low-energy EFT(s) under topology changes}
     After the above introductory remarks, we can proceed to study the changes in the lower-dimensional EFT under the topology changes. To do so, we first take the 10D Type IIB SUGRA bosonic effective action in Einstein frame \cite{Ibanez:2012zz,VanRiet:2023pnx}
    \begin{align}\label{e: IIB eff action}
        S_{\rm IIB}\supseteq\frac{1}{2\kappa_{10}^2}&\left\{\int\dd^{10}x\sqrt{-G}\left[\mathcal{R}_G-\frac{1}{2}(\partial\Phi)^2-\frac{e^{-\Phi}}{2}|H_3|-\frac{e^\Phi}{2}|F_1|^2-\frac{e^{\frac{\Phi}{2}}}{2}|\tilde F_3|^2-\frac{1}{2}|\tilde F_5|^2\right]\right.-\notag\\
        &\left.\quad-\frac{1}{2}\int C_4\wedge H_3\wedge\dd C_2\right\}\;,
    \end{align}
    where $|\omega_p|^2=\omega_p\wedge\star\omega_p$, $\Phi$ is the 10-dimensional dilaton and
    \begin{equation}
        H_3=\dd B_2\,,\quad F_1=\dd C_0\;,\quad \tilde F_3=\dd C_2-C_0H_3\,,\quad \tilde F_5=\dd C_4-\frac{1}{2}C_2\wedge\dd  B_2+\frac{1}{2}B_2\wedge\dd C_2\;.
    \end{equation}
    For simplicity, we will set $C_2,\,C_4\equiv 0$, under which \eqref{e: IIB eff action} simplifies to
    \begin{equation}
        S_{\rm IIB}'\supseteq\frac{1}{2\kappa_{10}^2}\int\dd^{10}x\sqrt{-G}\left\{\mathcal{R}_G-\frac{1}{2}(\partial\Phi)^2-\frac{e^{-\Phi}+C_0^2e^{\frac{\Phi}{2}}}{2}|H_3|^2-\frac{e^{\Phi}}{2}|F_1|^2\right\}\;,
    \end{equation}
    again considering only the bosonic part. We now compactify our theory on a Riemann surface $\Sigma_g$, with volume $\mathcal{V}$ in 10d Planck units. Taking the legs of $F_1$ and one of $H_3$ laying on $\Sigma_g$, the following 8d effective action results:
    \begin{equation}\label{e:eft action}
        S_{\Sigma_g}=\frac{1}{2\kappa_8^2}\int\left\{\star\left[\mathcal{R}-2V(\vec\varphi)\right]-\mathsf{G}_{ab}(\vec\varphi)\dd\varphi^a\wedge\star\dd\varphi^b-\frac{1}{2}\mathcal{A}_{AB}(\vec\varphi) H_2^A\wedge\star H_2^B\right\}\;,
    \end{equation}
    where here $\vec\varphi=(\Phi,\mathcal{V},\vec\tau)$ are the moduli corresponding to the 10d dilaton, $\Sigma_g$ volume and complex structure (respectively 0, 1 and $3(g-1)$ for $g=0,\,1$ or more), and 
    \begin{equation}
        H_2^A\equiv\dd B_1^A=\int_{a_A}H_3\;,\qquad A=1,\dots,\,2g
    \end{equation}
    are the field strengths of some effective 1-forms obtained from integrating $H_3$ along the $2g$ 1-cycles of $\Sigma_g$, with gauge kinetic matrix
    \begin{equation}
        \mathcal{A}_{AB}(\vec\varphi)=\left(\frac{\mathcal{V}}{\mathcal{V}_0}\right)^{\frac{1}{3}}(e^{-\Phi}+C_0^2e^{\frac{\Phi}{2}})\mathbb{A}_{AB}(\vec\tau)\;,
    \end{equation}
    where $\mathbb{A}_{AB}(\vec\tau)$ is a positive definite symmetric matrix with $\det \mathbb{A}=1$, given in terms of the period matrix $\Omega$, depending only on the complex structure moduli $\vec\chi$. It is given, as derived in Appendix \ref{app:RiemannianSurfaces}, by \eqref{e:mathbbA}, \cite{Saltman:2004jh,Adams:2005rb}
    \begin{equation}\label{e.AA g}
    \mathbb{A}=i\begin{pmatrix}
        2\Omega(\Omega-\bar\Omega)^{-1}\bar\Omega&-(\Omega+\bar\Omega)(\Omega-\bar\Omega)^{-1}\\
        -(\Omega+\bar\Omega)(\Omega-\bar\Omega)^{-1}&2(\Omega-\bar\Omega)^{-1}
    \end{pmatrix}\;,    
\end{equation}
which for $g=1$ simplify to the following $2\times 2$ real matrix
\begin{equation}\label{e.AA 1}
    \mathbb{A}=\tau_2^{-1}\begin{pmatrix}
        \tau_1^2+\tau_2^2&-\tau_1\\
        -\tau_1&1
    \end{pmatrix}\;.
\end{equation}
Regarding the scalar potential, it has contributions both from the $\Sigma_g$ internal curvature and the different fluxes, with
\begin{align}\label{e. IIb potential}
    V(\varphi)&=-\frac{1}{2}\left(\frac{\mathcal{V}}{\mathcal{V}_0}\right)^{-\frac{4}{3}}\int_{\Sigma_g}\dd ^2 y\sqrt{g_\Sigma}\mathcal{R}_{\Sigma}+\frac{1}{2}\left(\frac{\mathcal{V}}{\mathcal{V}_0}\right)^{-\frac{4}{3}}e^{\Phi}\int_{\Sigma_g}F_1\wedge\star F_1\notag\\
    &=\left(\frac{\mathcal{V}}{\mathcal{V}_0}\right)^{-\frac{4}{3}}\left[\pi(g-1)+\frac{1}{2}e^{\Phi}\sum_{I=1}^{2g}Q^{I\intercal} \mathbb{A}(\vec\chi)Q^I\right]\;,
\end{align}
where Gauss-Bonnet theorem has been used, and $Q^{Ii}=\int_{a_i}F_I$ are the charges vectors of the different $F_1$ components with respect to the homology basis. It is easy to see that for $g\geq 1$ $V(\varphi)\geq 0$ and negative for $g=0$ (note that for $g=0$ there is no flux potential), with gradient flows respectively pointing to $\mathcal{V\to\infty},\,\Phi\to 0$ and $\mathcal{V}\to 0$.

As for the complex structure moduli, these are stabilized for a generic flux configuration of $F_1$ \cite{Saltman:2004jh}.\footnote{\label{fn:stab}To see this, notice that all the eigenvalues $\mathcal{A}_{AB}\propto\mathbb{A}_{AB}$ are positive. From the definition \eqref{eq.holo basis} of the period matrix $\Omega^{ij}$, in a topology change $\Sigma_g\to\Sigma_{g-1}$ a 1-cycle pinches and becomes small, and as such some eigenvalue of ${\rm Im}\Omega=\frac{1}{2i}(\Omega-\bar\Omega)$ goes to 0, or equivalently, one of $(\Omega-\bar\Omega)^{-1}$ blows up. Then in such a limit $\mathbb{A}(\tau)$ has some (positive) eigenvalue diverging. An arbitrary flux configuration is expected to have $F_1$ components along the complete $\mathbb{A}(\tau)$ eigenbasis, so that for any diverging eigenvalue the flux part of $V(\varphi)$ diverges. As this happens for any 1-cycle pinching and the flux potential is non-negative, there must be some minimum in the bulk of the complex structure moduli space, resulting in their stabilization.} Note that, as explained in Appendix \ref{app:tachyon}, for the topology change to occur it is not needed that the cycle radius actually pinches to zero, as for $\mathsf{R}\leq \sqrt{2\alpha'}$ (with $\alpha'$ the string tension) a tachyonic instability appears. This means that the $\Sigma_g\to\Sigma_{g-1}$ transition does not encounter an infinite potential barrier, but rather a finite-height slope up-hill until the instability region is reached and tachyon condensation occurs. 

    \begin{figure}[t!]
				\centering
				\includegraphics[width=0.9\textwidth]{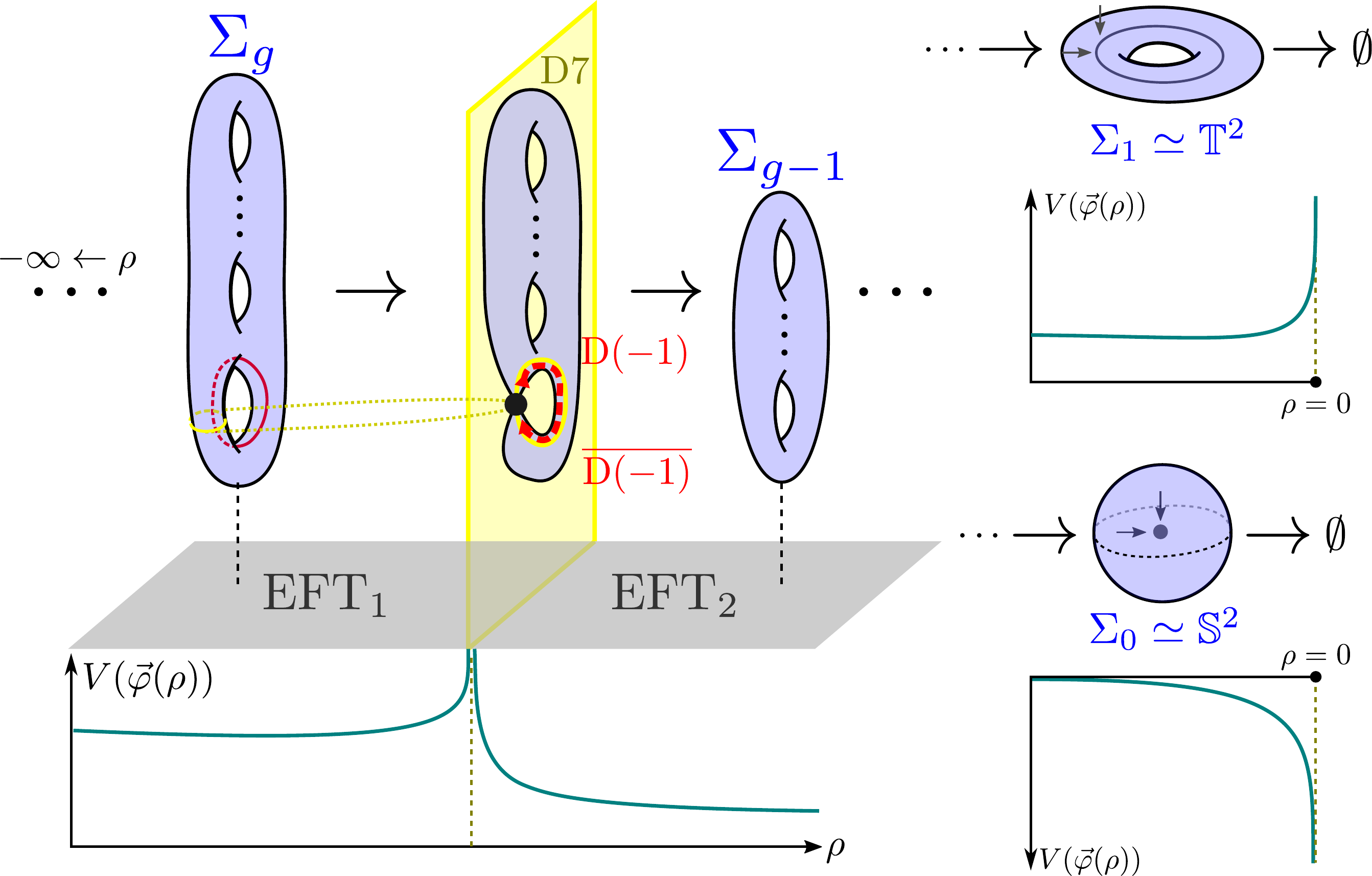}
			\caption{\small Sketch of different topology changes along the cobordism to nothing, depicting the intermediate losing of a handle for $g\geq 2$ (including the arrangement of the D7-domain walls, in yellow, and the D$(-1)$-$\overline{\text{D}(-1)}$ bounded on it, in red with a dashed line presenting the 1-chain joining them) and the two possible final topology changes into nothing, $\mathbb{T}^2\to\emptyset$ and $\mathbb{S}^2\to\emptyset$. The qualitative behavior of the scalar potential is also depicted.}
			\label{f.IIBonRieamann}
	\end{figure}

\vspace{0.25cm}
By inspection of action \eqref{e:eft action}, it is evident that compactifying on $\Sigma_g$ results in distinct low energy EFTs depending on the value of $g$, with the topology changes loci acting as domain walls between different theories. We will now comment on how these changes occur as we move in the radial direction towards the end of the bordism to nothing. Assuming that the different topology changes are sufficiently separated from one another such that a lower $d=8$ description is valid between them, 
\begin{itemize}
    \item The most evident feature of $\Sigma_g\to\Sigma_{g-1}$ is the change in the \textbf{scalar potential} $V(\vec\varphi)$, as depicted in Figure \ref{f.IIBonRieamann}. We first take the \textbf{internal curvature term}. Gauss-Bonnet theorem applied to compact Riemann surfaces result in this being proportional to $(g-1)$ times an extra $\mathcal{V}^{-4/3}$ factor coming from the rescaling to 8d Einstein frame. For $g\geq 2$ $\Sigma_g$ are negatively curved, resulting in a runaway \emph{positive} potential to large volume. At the topology change locus a cycle pinches, with the overall volume of the internal manifold not being dramatically affected. Assuming that said volume remains approximately constant as we cross the domain wall (if our bordism $\mathcal{B}_3$ is smooth then this is the case), then the internal curvature contribution to the vacuum energy experiences a change of $\Delta V_{\mathcal{R}}\approx -\pi\left(\mathcal{V}/\mathcal{V}_0\right)^{-4/3}<0$, with the change being energetically favored. For $g=1$ (i.e. a  torus compactification), there is no internal contribution to the curvature. If our bordism features an additional topology change to the sphere $\Sigma_0=\mathbb{S}^2$, the curvature potential becomes negative. In this case the $F_1$ flux potential is 0, as there are no more 1-cycles, so there is no extra contributions to the potential \eqref{e. IIb potential}. This drives $\mathcal{V}\to 0$, corresponding precisely to the $\mathbb{S}^2\to \emptyset$ final topology change into nothing.

    As for the \textbf{flux} contributions to the potential, apart from the overall $\mathcal{V}^{-4/3}e^{\Phi}$ factor driving towards a weak coupling and large volume limit, there is the $\sum_{I=1}^{2g}Q^{I\intercal} \mathbb{A}(\vec\chi)Q^I$ factor. We can take the factorization approximation so that from \eqref{e.AA 1},
    \begin{equation}
        \sum_{I=1}^{2g}Q^{I\intercal} \mathbb{A}(\vec\chi)Q^I\approx \frac{|\hat\tau|^2}{\hat\tau_2}Q^2+\frac{1}{\hat\tau_2}Q'{}^2+\sum_{I=3}^{2g}Q^{I\intercal} \mathbb{A}'(\vec\chi)Q^I\;,
    \end{equation}
    where $\tau=\tau_1+i\tau_2$ is the complex structure associated to cycle $c$ pinching, $Q$ and $Q'$ the flux charges of  that cycle and its dual ($c'$), and $\mathbb{A}'$ the \eqref{e.AA g} matrix from the rest of factorized cycles. For $\tau_2\to 0$ then the flux contribution to the potential grows, until the tachyon instability regime is reached and the topology changes occurs. 

    As for the units of flux lost in the topology change process, we must differentiate between those associated with the shrinking cycle $c$ ($Q$) and those with its dual $c$ ($Q'$). As was explained in sections \ref{ss:Morse-Bott} and \ref{ss:defects}, the shrinking of the cycle can be understood as a cobordism into nothing, with $\mathbb{S}^1=\partial\mathbb{D}^2$. In order for this to be possible, we need a defect at the tip of the disk which eats up the charge. As reasoned in section \ref{ss:defects}, this corresponds to a D7-brane, electrically charged under the $C_8$ form dual to $C_0$. The (6+1)-dimensional topology change locus/domain wall (precisely where the flux potential naively would blow up) corresponds to the D7 brane, with the extra direction wrapped along the dual 1-cycle. This fixes the position of this intermediate D7 to be precisely at the topology change locus.

    As for the charges associated to the dual 1-cycle, the topology change does not shrink it to nothing, but rather transforms it into a 1-chain with boundary. The total charge then must remain the same, with
    \begin{equation}
        Q'=\int_{c'}F_1=\int_{\partial c'}C_0=C_0|_{(\partial c')^+}-C_0|_{(\partial c')^-}\,.
    \end{equation}
    We argue that this corresponds to a pair of $\rm D(-1)-\overline{D(-1)}$ branes (D-instantons) localized on the D7 \cite{Billo:2021xzh,Reymond:2024mwe}, that are unstable and annihilate each other, along with their charges. As a result these charges are not present at the other side of the domain wall, and do not contribute to the potential anymore.

    Under the factorization approximation the vacuum energy is lowered by \begin{equation}
        \Delta V_F\approx -\left(\frac{|\hat\tau^{(0)}|^2}{\hat\tau_2^{(0)}}Q^2+\frac{1}{\hat\tau_2^{(0)}}Q'{}^2\right)\,,
    \end{equation} with $\hat{\tau}^{(0)}$ the value of the pinching cycle complex structure before the topology change process starts. Again this is positive, so that the  topology change process is energetically favored, both from the curvature and flux contributions to the vacuum energy.

    \item We now turn to the other term in the \eqref{e:eft action} action, corresponding to the gauge kinetic term for the resulting 2-from fields in the 8d EFT, $\mathcal{A}_{AB}(\vec\varphi) H_2^A\wedge\star H_2^B$, with $A=1,\,\dots,\, 2g$. As explained in \cite{Adams:2005rb},  for a 1-cycle $c$ shrinking, with dual $c'$ , both taken to be decoupled from the rest, we have that\footnote{Note that there is a different sign and $\tau_2$ factor in the expression when compared to that in \cite{Adams:2005rb}, though the conclusions are the same.}
    \begin{align}\label{e.effective gauge kin}
        \mathbb{A}_{AB}H_2^A\wedge\star H_2^B\supseteq\frac{\star 1}{\tau_2}|H_2^c-\tau H_2^{c'}|^2=\frac{1}{\tau_2}|H_2^-|^2+\tau_2|H_2^+|^2\,,
    \end{align}
    where $\tau=\tau_1+i\tau_2$ is the factorized complex structure associated to $c$ and $c'$ and where we assume that the kinetic terms for the additional forms are not affected by the topology change. In the shrinking cycle limit $\tau_2\to 0^+$, the combination $H_2^+=H_2^c-\tau_1H_2^{c'}$ becomes weakly coupled, while $H_2^-=H_2^{c'}$ becomes strongly coupled.

    As the 1-cycle $c$ shrinks along the bordism direction $\rho$ in the topology change process, the 1-forms associated with the angular coordinate along $c$, $\dd \theta$, ceases to be harmonic, as the internal metric is warped with respect to $\rho$ (i.e., $\mathcal{B}_3$ is no longer locally $\Sigma_g$ times an interval). When decomposing $B_2=B_1\wedge\dd\theta$, close to the topology change locus $\Delta\theta\neq 0$, in such a way that the 8d field $B_1$ is massed-up and the $U(1)$ 1-form symmetry is removed from the EFT. At the same time, the angular profile $\theta$ of the condensing tachyon \eqref{e:tachyon}, which is wound along it, acts as a Goldstone boson of the broken $U(1)$ symmetry, see \cite{Delgado:2023uqk}. As for the gauge field associated to the dual $c'$, fundamental strings charged under the $B_2$ field and winding around $c'$ in opposite directions are seen as charge-anticharge pairs from the (7+1)-dimensional point of view, charged under $B_1^+$, where $H_2^+=\dd B_1^+$. As $\tau_2\to 0^+$, from \eqref{e.effective gauge kin} the gauge coupling of these grows stronger, with these pairs experimenting a strong attractive force (think of tension full flux string in 8d), until they encounter and annihilate each other. This results in the winding charge along $c'$ to be effectively 0 as $\tau_2\to 0$ and the $U(1)$ 1-form gauge symmetry is confined, thus not obstructing the handle decay. This is illustrated in Figure \ref{f.confinement}.

    In the new EFT resulting on the other side of the topology change domain wall, the rank of the gauge group is two units lower, as the two $U(1)$ 1-forms associated to the two no-longer-present cycles have either been massed-up or classically confined \cite{Adams:2005rb}, thus changing the gauge symmetry of the theory from $U(1)^{2g}$ to $U(1)^{2g-2}$.

    \begin{figure}[t!]
				\centering
				\includegraphics[width=\textwidth]{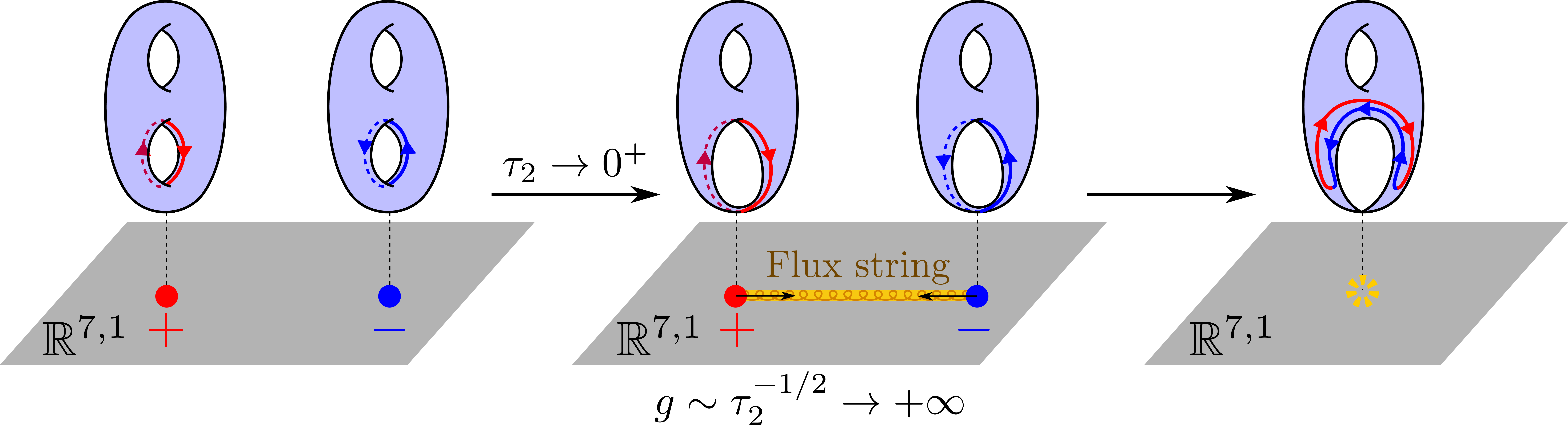}
			\caption{\small Sketch of the decay process of a handle in a $\Sigma_g\to \Sigma_{g-1}$ topology change, and the fate of the gauge $U(1)$ gauge form associated to the dual cycle, through confinement as the gauge coupling grows.}
			\label{f.confinement}
	\end{figure}
    
    While for simplicity we have considered only the dynamics of the $U(1)$ 1-form fields obtained from the integration of the $B_2$ field with one leg on the internal dimensions, analogous arguments could be done about 1- or 3-forms from doing the same with the $C_2$ and $C_4$ fields, respectively. Through each topology change, similar scenarios of massing-up/confining would arise.

\end{itemize}
Summing up, we find that after any of the $g$ or $g+1$ necessary topology changes in the bordism from Type IIB string theory on $\Sigma_g$ to nothing, the vacuum energy lowers and the rank of the gauge group becomes smaller, with the different changes being energetically favored.

Before finishing this example, we briefly comment on the EotW brane picture of the above topology changes. These can be associated to global volume changing, parameterized by a canonically normalized volume modulus $\hat{\mathcal{V}}=\sqrt{2/3}\log\frac{\mathcal{V}}{\mathcal{V}_0}$, the canonically normalized 10d dilaton $\hat{\Phi}=\frac{1}{\sqrt{2}}\Phi$ appearing in the flux term, and the complex structure moduli, from which only the imaginary part of that corresponding to the shrinking cycle is of interest to us, $\hat{\tau}_2=\frac{1}{\sqrt{2}}\log\tau_2$, where we have taken the factorization limit, see \eqref{e. complex structure} in Appendix \ref{app:RiemannianSurfaces}. This allows us to approximate the scaling of the potential \eqref{e. IIb potential} as
\begin{equation}\label{e. approx pot IIB}
    V(\varphi)\sim e^{-2\sqrt{\frac{2}{3}}\hat{\mathcal{V}}+\sqrt{2}\Hat{\Phi}+\sqrt{2}|\hat{\tau}_2|}\to\infty\;,
\end{equation}
where we do not distinguish between the $\hat{\tau}_2\to+\infty$ and $-\infty$ limits, which are equivalent for the potential, and as we will see now not necessarily the three scalars simultaneously blow up. From our analysis, there are the following (qualitatively) different topology changes:
\begin{itemize}
    \item $\Sigma_g\to \Sigma_{g-1}$: Here one of the 1-cycles shrinks to a point, i.e., $|\hat{\tau}_2|\to\infty$, while the overall volume is remains approximately constant. On the other hand, as we approach the D7-brane, the dilaton goes to a weak coupling limit, $\hat{\Phi}\to-\infty$, see section 18.5 from \cite{Blumenhagen:2013fgp}. This results in $V(\varphi)\sim \exp\left(\sqrt{2}\Hat{\Phi}+\sqrt{2}|\hat{\tau}_2|\right)\;$, with critical exponent $\delta=1$.
    \item $\Sigma_1\simeq\mathbb{T}^2\to\emptyset$: This case is analogous as the previous one, with also a D7-brane as a cobordism defect, but now the shrinking of a 1-cycle over its dual is also associated to the total volume going to 0, with the potential being \eqref{e. approx pot IIB}, and $\delta=\sqrt{\frac{8}{11}}$.
    \item $\Sigma_0\simeq\mathbb{S}^2\to\emptyset$: For this final case the sphere shrinks to a point, with the 1-fluxes no longer contributing to the potential, so that $V(\varphi)\sim e^{-2\sqrt{\frac{2}{3}}\hat{\varphi}}$. It is then straightforward that in this case $\delta=\sqrt{\frac{8}{3}}$.
\end{itemize}
This way, the different topology changes are associated with distinct critical exponents, so from the EFT point of view are associated with different EotW brane solutions. As all $\Sigma_g\to \Sigma_{g-1}$ topology changes qualitative look the same, it is not possible, at least with this approach, to learn about the possible additional topology changes that the internal manifold must undergo after the first domain wall is crossed.

    \subsection{An example of non-minimal bordism}
    \label{ss:nothing is certain}
    In the previous examples studied, the bordisms considered saturated the Morse-Bott \eqref{e.MorseBottIneq} and $\mathcal{B}_{n+1}$ homology \eqref{e.ineq} inequalities, in such a way that the number of topology changes was minimized to only those unavoidable. There are, however, some results in the literature in which smooth bordisms which are non-minimal, i.e., those inequalities are fulfilled but not saturated. One of such examples was commented upon in Figure \ref{ff.thickwall} and section \ref{ss:defects}, \cite{Debray:2023yrs}. For a more detailed instance, we will comment on the BoN construction described in \cite{GarciaEtxebarria:2020xsr}, whose relevant ingredients we review now. The following $D$-dimensional (bosonic) effective action is considered, which for simplicity we present in string frame\footnote{The Gauss-Bonnet invariant $\mathsf{R}_{\rm GB}^2$ transforms in a pretty involved way under the conformal rescaling required to move to Einstein frame, see eq. (III.4) from \cite{Dabrowski:2008kx}. As we will not make use of the actual equations of motion, we simply present the effective action in order to present the field content.}
    \begin{equation}
        S_D^{(s)}=\frac{1}{2\kappa_D^2}\int\dd^Dx\sqrt{-G}e^{-2\Phi}\left\{\mathcal{R}+4(\partial\Phi)^2-\frac{1}{2\cdot 3!}|H_3|^2+\frac{1}{8}\alpha \mathsf{R}_{\rm GB}^2\right\}\;,
    \end{equation}
    where $\Phi$ is the $D$-dimensional dilaton, $H_3=\dd B_2$ and $ \mathsf{R}_{\rm GB}^2=\mathcal{R}^2-4R_{MN}R^{MN}+R_{MNPQ}R^{MNPQ}$ is the Gauss-Bonnet invariant, which is topological for $D=4$ and gives second-order e.o.m. for higher $D$. For $\alpha>0$ both SUSY and the Dominant Energy Condition are broken. See \cite{GarciaEtxebarria:2020xsr} for more details on the embedding of the above action in $D=6$ heterotic and the dual type I/IIB descriptions.

    \begin{figure}[t!]
				\centering
				\includegraphics[width=0.8\textwidth]{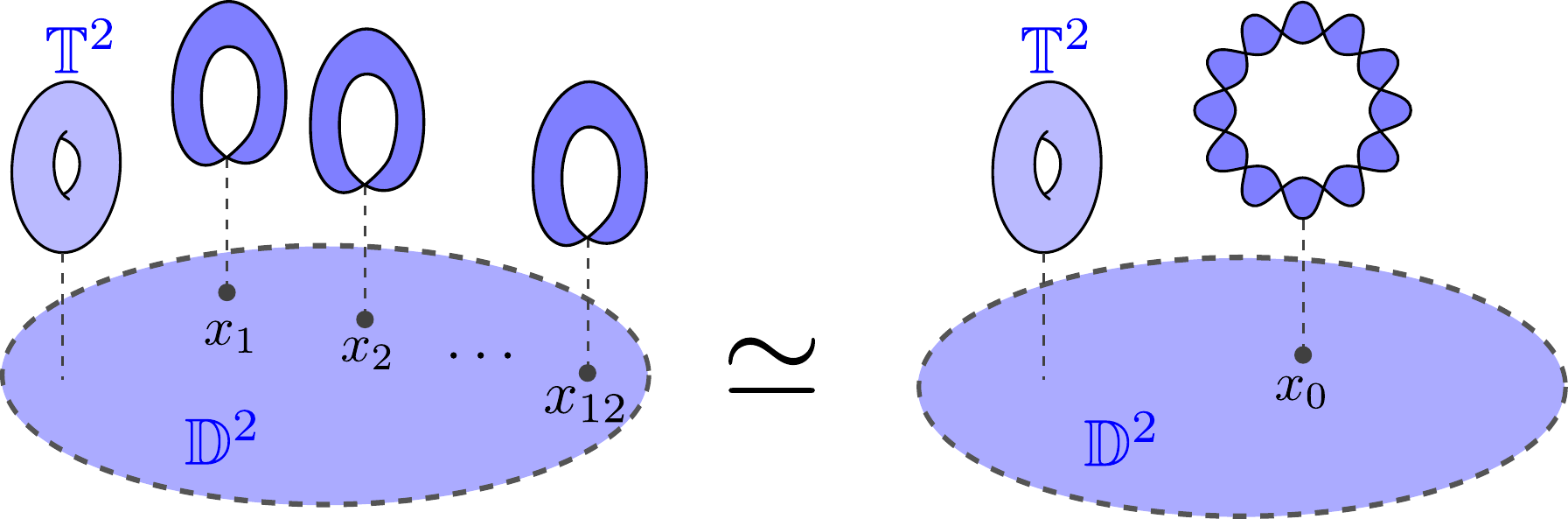}
			\caption{\small Sketch of the smooth fibration structure of $\mathbb{T}^2\hookrightarrow\mathcal{B}_4\to\mathbb{D}^2$, with the $\mathbb{T}^2$ fiber pinching at 12 points $\{x_i\}_{i=1}^{12}$. Considering the topology, the manifold can be continuously deformed such that the fibration is everywhere $\mathbb{T}^2$ except for a point $x_0$ of the $\mathbb{D}^2$ base where it pinches 12 times.}
			\label{f.B4}
	\end{figure}
    
    The above effective action can be further compactified to $d=3$ on a 3-torus $\mathcal{C}_3=\mathbb{T}^3$ with SUSY preserving (i.e., periodic) boundary conditions.    
    The bordism structure considered is $\mathsf{g}={\rm Spin}$, and since $\Omega_{3}^{\rm Spin}\simeq 0$ \cite{Anderson1967:SPIN,McNamara:2019rup}, there exists a Spin manifold $\mathcal{B}_4$ with $\mathcal{C}_3=\partial\mathcal{B}_4$, and thus from the discussion in section \ref{ss: BoN} BoN constructions should exist. Due to the Spin structure on $\mathcal{C}_3$, $\mathcal{B}_4$ cannot simply be $\mathbb{D}^2\times\mathbb{T}^2$, as we cannot just contract one of the 1-cycles to a point. Without the need of including stringy Spin-defects, a smooth solution is precisely given in \cite{Scorpan2005TheWW,GarciaEtxebarria:2020xsr}, with $\mathcal{B}_4$ given by elliptic fibration over $\mathbb{C}\simeq\mathbb{D}^2$ such that the $\mathbb{T}^2$-fiber pinches in 12 points over the disk. If these degeneration loci are isolated, then $\mathcal{B}_4$ is smooth. In Figure \ref{f.B4} a sketch of the $\mathcal{B}_4$ manifold is presented. By continuously deforming it such that the 12 pinching points occurs on the same point of the base $\mathbb{D}^2$, we can easily obtain $\vec b(\mathcal{B}_4)=(1,1,12,0,0)$ (rather than $\vec b(\mathbb{D}^2\times \mathbb{T}^2)=(1,2,1,0,0)$), with no torsion. Note that these Betti numbers fulfill but do not saturate the inequalities  \eqref{e.ineq}. In any case, Morse-Bott inequalities \eqref{e.MorseBottIneq} are satisfied as
    \begin{equation}
        (1+t)\mathsf{R}(t)=(1+t)^2t^2+12t^2+(1+t)^3-(1+t+12t^2)\Longrightarrow \mathsf{R}(t)=2r+2t^2+t^3\in\mathbb{R}_{\geq 0}[t]\;,
    \end{equation}
    where both the 12 pinching points have index 2 and the center of $\mathbb{D}^2$, where the 1-cycle closes over the $\mathbb{T}^2$ fiber, has index 2. Note that after each critical point, where a 1-cycle pinches and then grows again, the local topology of the solution is actually the same, $\mathbb{T}^3$, so that no actual topology change occurs. In any case, these types of critical points also enter in the Morse-Bott \eqref{e.MorseBottIneq} and $\mathcal{B}_{n+1}$ homology \eqref{e.ineq} inequalities, with certainly the bordism into question being different than $\mathbb{T}^2\times \mathbb{D}^2$.   

    As explained in great detail in \cite{GarciaEtxebarria:2020xsr}, each of the 12 degeneration points can be locally seen as a Taub-NUT/KK monopole solution \cite{Gross:1983hb,Sorkin:1983ns}, with the compact space smoothly pinching, so that could also consider this solution as a 12-centered BoN solution. Close to this degeneration points, there is a partial decompatification, with the elliptic fiber becoming large while the $\mathbb{S}^1$ basis shrinks (so that the overall volume remains finite). This results in an infinite distance limit in moduli space being proven.\footnote{As explained in section 5.2.2 from \cite{GarciaEtxebarria:2020xsr}, the associated KK tower is vital to cancel an \emph{a priori} divergence in the metric around the degeneration points, such that the BoN solution remains smooth.} From the EotW brane approach from section \ref{ss: EotW}, there is no potential here for it to blow-up, so that $a=0$ and $V_0=0$, and the critical exponent is $\delta=2\sqrt{\frac{d-1}{d-2}}$. Again we refer to the extensive work of \cite{GarciaEtxebarria:2020xsr} for the detailed expressions.

    \vspace{0.25cm}

    This example might feel a bit underwhelming, as it involves a topology for $\mathcal{B}_4$ much more complicated than $\mathbb{T}^2\times \mathbb{D}^2$, motivated by having a smooth bordism rather than using stringy defects such as those mentioned in footnote \ref{fn.String DEF}. Given the topology of $\mathcal{B}_4$, which as described in \cite{GarciaEtxebarria:2020xsr} is determined by requiring 12 degeneration points of the elliptic fiber over the $\mathbb{D}^2$ base, one can use Morse-Bott inequalities to show that
    \begin{equation*}
        \#\{\text{1-cycle $\twoheadrightarrow\{\rm pt.\}$}\}-\#\{\text{2-cycle $\twoheadrightarrow\{\rm pt.\}$}\}+\left\{\begin{array}{ll}
             2(1-g)&\text{if $\mathbb{S}^1\twoheadrightarrow\Sigma_g$ end}\\
             0&\text{otherwise}
        \end{array}\right.=1-b_1(\mathcal{B}_4)+b_2(\mathcal{B}_4)\;,
    \end{equation*}
    where we denote $A\twoheadrightarrow B$ as a cycle $A$ shrinking over a critical locus $B$. For our example we have $1-b_1(\mathcal{B}_4)+b_2(\mathcal{B}_4)=12$, with only 12 1-cycles shrinking over points and a final shrinking  $\mathbb{S}^1\twoheadrightarrow \mathbb{T}^2=\Sigma_1$, with no 2-cycles shrinking over a point. Note that, in  this aspect, the number of critical points is the minimum needed, and other solutions to the above equations would result in a higher amount of them.

    \subsection{A small step towards smooth bordisms for K3 compactifications.}
    \label{ss:K3}
    We end this section with a comment on the bordism structure for a more complicated internal manifold than the Riemann surfaces from section \ref{ss:IIB Riemann}: the 4-dimensional K3 surface. Considering the $\mathsf{g}={\rm Spin}$ structure, we have that $\Omega_4^{\rm Spin}\simeq\mathbb{Z}\simeq \langle {\rm K3}\rangle$ \cite{Anderson1967:SPIN}, this is, precisely generated by K3, which carries negative unit charge under the conserved current $J_{p_1}=\frac{1}{48}p_1(R)=\frac{1}{48\cdot 8\pi^2}{\rm tr}(R\wedge R)$, with $p_1$ the first Pontriaguin class of the curvature 2-form $R$. How this $U(1)$ symmetry is killed in different string theories is discussed in detail in \cite{McNamara:2019rup}, to which we refer for more details. For our purposes, type IIA/M-theory can be compactified on pin$^+$-manifolds \cite{Witten:2016cio}, on which the bordism is broken to $\Omega_4^{\rm Spin}\simeq\mathbb{Z}\to \Omega^{\rm pin+}_4\simeq \mathbb{Z}_{16}\simeq\langle\mathbb{RP}^4\rangle$, which is then broken by MO5-planes \cite{Dasgupta:1995zm,Witten:1995em}. As for type IIB string theory, this is partly broken by non-trivial elliptic fibrations in F-theory, while, as argued in \cite{McNamara:2019rup}, there must be some new, non-supersymmetric defect with spatial dimensionality 6 that kills the remaining the remaining class. In order to obtain \emph{smooth bordisms}, we will assume that defects are located far from the tip of the bordism, in such a way that our K3 surface can be taken belong to the trivial class of the spin bordism. As in section \ref{ss:IIB Riemann}, we will only care about flux potentials and the location of the branes needed to eat-up the associated charges.

    Before this, it is necessary to obtain information about the topology of the bordism $\mathcal{B}_5$ with ${\rm K3}=\partial\mathcal{B}_5$. As K3 has Poincar\'e polynomial $\mathsf{P}_t({\rm K3})=1+22t^2+t^4$ (i.e. we only have 2-cycles and the whole 4-manifold to shrink). Using \eqref{e.ineq} and \eqref{e.halfhalf} we find that any bordism $\mathcal{B}_5$ for it must have Betti numbers $\vec{b}(\mathcal{B}_5)=(1,\alpha,11,\beta,0,0)$, with $\alpha,\,\beta\in\mathbb{Z}_{\geq 0}$. Without creating new cycles, we have that the following possible topology changes to \textbf{(A)} 2-cycles shrinking to a point, \textbf{(B)} a 2-cycle shrinking over a compact 2-cycle of genus $g$, and \textbf{(C)} a 4-cycle shrinking to a point. The last two possibilities make 4 internal dimensions disappear, and as such correspond to the final topology change. Respectively denoting the number of each type by $a\in\mathbb{Z}_{\geq 0}$ and $b,\; c\in\{0,1\}$, we will then have that the LHS in \eqref{e.MorseBottIneq0} reads as 
\begin{align}
\mathsf{MB}_t^N(\rho)-\mathsf{P}_t(\mathcal{B}_5)
&=(c+b)t^5+(1+2bg)t^4+(a+b-\beta)t^3+11t^2-\alpha t\notag\\&=(1+t)\mathsf{R}(t)\;.
\end{align}
Now, in order for the above polynomial to be divisible by $1+t$, we need $a=12-c+2b(g-1)+\alpha+\beta$, which results in
\begin{equation}
\mathsf{R}(t)=(b+c)t^4+(1-b-c+2bg)t^3+(11+\alpha)t^2-\alpha t\;.
\end{equation}
As $\mathsf{R}\in\mathbb{Z}_{\geq 0}[t]$, then $\alpha=0$ (i.e. $\mathcal{B}_5$ does not have non-trivial 1-cycles), and as we need a single final change into nothing, $b+c=1$, so that $\mathsf{R}(t)=t^4+2bgt^3+11t^2$, with $a=12-c+2b(g-1)+\beta$, resulting in a total number of $N=a+b+c=12+b(2g-1)+\beta\geq 11 $
topology changes in our cobordism into nothing. The above inequality is saturated for $(b,c,g,\beta)=(1,0,0,0)$. We thus conclude that it is not possible to  build a smooth BoN/EotW brane solution that interpolates into nothing without intermediate topology changes.

There are at least 11 topology changes, of which all but the final one (i.e. at least 10) correspond to 2-cycles shrinking to a point, thus resulting in its disappearance and that of its Poincar\'e dual 2-cycle. As for the final topology change, there are two possibilities, as commented before:
	\begin{enumerate}
	\item $(b,c)=(1,0)$: Assume that after the intermediate $a$ topology changes ,in each of which two (dual) 2-cycles disappear, we end up with a 4-manifold $\mathcal{X}_4$ with $\vec{b}(\mathcal{X}_4)=(1,0,2\gamma,0,1)$ before the final topology change into nothing. By similar arguments as above, this final bordism $\mathcal{M}'_5$ will have $\vec{b}(\mathcal{M}'_5)=(1,0,\gamma,\beta',0,0)$ and by Morse-Bott inequalities $\gamma=1-2g-\beta'$, with $\beta'\in\{0,1\}$. This results in $g=0$ and two additional cases:
	\begin{enumerate}
		\item $\beta'=0$: Then $\gamma=1$ and $\mathcal{P}_t(\mathcal{X}_4)=1+2t^2+t^4=(1+t^2)^2$, and as such can be seen as a fibration $\mathbb{S}^2\hookrightarrow \mathcal{X}_4\to \mathbb{S}^2$, such as one of the Hirzebruch surfaces $H_n$ \cite{Hirzebruch1951}. Note that these have different topological invariants for distinct values of $n$, so on general they would correspond to different topologies of $\mathcal{B}_5$. We also have that $a=10+2\hat{a}$, with the extra $2\hat{a}=\beta\geq 0$ topology changes corresponding to creation and disappearance of pairs of 2-cycles.
		\item $\beta'=1$: Then $\gamma=0$ and $\mathcal{P}_t(\mathcal{X}_4)=1+t^4$, so $\mathcal{X}_4$ is an homology sphere. Again these are not unique \cite{SUCIU1987103}, and could give rise to different bordism topologies. There will be  $a=11+2\hat{a}$, with additional $2\hat{a}=\beta-1\geq 0$ topology changes corresponding to additional creation and disappearance of pairs of 2-cycles.
	\end{enumerate}
	\item  $(b,c)=(0,1)$: As in the previous case, after the $a$ topology changes, our final 4-manifold before going to $\emptyset$ has $\vec{b}(\mathcal{X}_4)=(1,0,2\gamma,0,1)$. Now in this case we have that $\mathcal{X}_4$ shrinks to a point as the boundary of $\mathcal{M}'_5$ with $\vec{b}(\mathcal{M}'_5)=(1,0,\gamma,\beta',0,0)$. From Morse-Bott inequalities we have $\gamma+\beta'=0$, and both numbers must be non-negative, we have $\gamma=\beta'=0$, so that $\mathcal{X}_4$ is a homology 4-sphere that shrinks to a point. As before, we will have $a=11+2\hat{a}$, with the extra $2\hat{a}=\beta\geq 0$ again corresponding to creation and disappearance of pairs of 2-cycles.
	\end{enumerate}
	 To sum up, the minimal number of topology changes to go from a K3 compactification into nothing through a smooth bordism is either 11 or 12, respectively ending on a $\mathbb{S}^2\hookrightarrow \mathcal{X}_4\to \mathbb{S}^2$ fibration or a homology 4-sphere. 

  One could perform a similar analysis to that of section \ref{ss:IIB Riemann} to compactifications on K3. In this case, as $b_k({\rm K3})=0$ for odd $k$, only even fluxes contribute to the flux potential, so type IIA compactifications will be more interesting than type IIB, with D4- and wrapped D6-branes (with bound D2-branes from the dual 2-cycle) serving as domain walls between  topology changes in the presence of flux along the whole K3 or  the different 2-cycles. The analysis of the lower-dimensional gauge fields fate after topology change is analogous to those in section \ref{ss:IIB Riemann}, so we will not comment on it further. 

\vspace{0.25cm}

  In section 7.2 of \cite{Debray:2023yrs} a different bordism from K3 to nothing was argued for. F-theory on K3 (seen as a $\mathbb{T}^2\hookrightarrow{\rm K3}\to\mathbb{CP}^2$ fibration) is considered, which in Sen's weak coupling limit is dual to perturbative type IIB string theory on a $\mathbb{T}^2/\mathbb{Z}_2$ orbifold. The four resulting singularities, which are specified by the monodromy of a linking circle, are argued to be equivalent to four $\mathbb{T}^2$ with the same monodromy. This motivates the substitution of $\mathbb{T}^2/\mathbb{Z}_2$ by these genus-4 surface, which now allows for a smooth bordism to nothing as in section \ref{ss:IIB Riemann}, and over which the F-theory elliptic fiber does not degenerate. The problem comes in that the process $(\mathbb{T}^2\hookrightarrow{\rm K3}\to\mathbb{CP}^2)\to(\mathbb{T}^2\hookrightarrow \mathcal{X}_4\to\Sigma_4)$ is not necessarily smooth, with this $\mathcal{X}_4$, which contains non-trivial 1-cycles, not being required as an intermediate step of the analysis considered previously in this subsection. Since the construction of a completely smooth bordism to nothing for K3 compactifications (which for F-theory constructions need to keep the elliptic fibration until the final topology change into nothing) is not yet a solved matter, we hope this short digression serves to bring some light into the matter.

\section{Stories with more than one ending}
\label{sec:multiple}
In the above sections, only a single BoN/EotW brane was considered for each case. However, instances of multi-centered BoN solutions eventually colliding or networks of EotW branes intersecting are the natural generalization. These cases have been previously studied in the literature (see \cite{Horowitz:2002cx,Freivogel:2007fx,Kleban:2011pg,BoNColTBA} and \cite{Angius:2023uqk,Angius:2024zjv,GarciaEtxebarria:2024jfv}, respectively), and as we will see, we can learn important qualitative properties through our approach.
\subsection{\textit{Seeing double}: Collisions of BoN's}
    \label{ss:BoNCollide}
\begin{figure}[t!]
				\centering
				\includegraphics[width=0.75\textwidth]{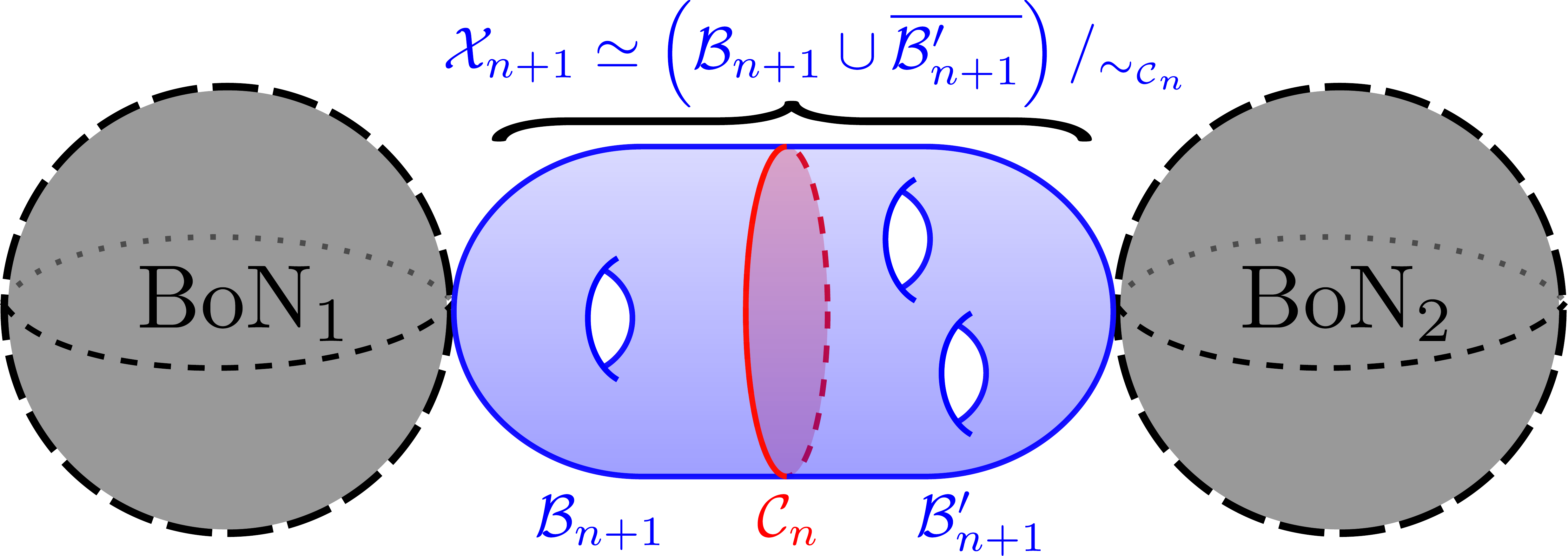}
			\caption{\small Illustration of two BoN approaching in a $\mathcal{C}_n$. The bordism associated to each of them is $\mathcal{B}_{n+1}$ and $\mathcal{B}_{n+1}$, while the radial direction joining both corresponds to the compact manifold $\mathcal{X}_{n+1}$ as defined in \eqref{e.glue}.}
			\label{f. BoN Collision}
	\end{figure}

    As explained in section \ref{ss: BoN}, bubbles of nothing have some non-zero probability of nucleating per unit volume. It is then expected, that, in a large macroscopic region and for a long enough period of time, several BoN nucleate and start growing, eventually approaching each other's surfaces. Considering our lower-dimensional theory to result from compactification on a compact surfaces $\mathcal{C}_n$, a bubble solution will be associated to a bordism and Morse function choice, $(\mathcal{B}_{n+1},\rho)$. Two different BoN can in principle be associated to $\mathcal{B}_{n+1}$ and $\mathcal{B}_{n+1}'$, such that $\mathcal{C}_n=\partial \mathcal{B}_{n+1}=\partial\mathcal{B}_{n+1}'$, without necessarily the two bordisms being homeomorphic. Moving then radially from the surface of one BoN to the other, this can be seen as traveling along some compact $(n+1)$-manifold $\mathcal{X}_{n+1}$ given by
    \begin{equation}\label{e.glue}
        \mathcal{X}_{n+1}\simeq \left(\mathcal{B}_{n+1}\cup \overline{\mathcal{B}'_{n+1}} \right)/_{\sim_{\mathcal{C}_n}}\;,
    \end{equation}
    this is, gluing the two bordisms along their boundary, with the orientation of one of them reversed. This is illustrated in Figure \ref{f. BoN Collision}.
    As the surfaces of the two bubbles approach, the volume of $\mathcal{X}_{n+1}$ shrinks. Whilst the eventual collision of the two tips can reach Planckian curvatures, as well as being affected by the backreaction of large matter densities being dragged by the bubble, one can also have these problems well before the BoN surface if $\mathcal{X}_n$ cannot be contracted to a point in a single topology change, with the curvatures of internal cycles of $\mathcal{X}_{n+1}$ blowing up before that of the whole manifold. Having a single topology change translates in the Morse function interpolating between the tips of the two bubbles having only critical \emph{ two points}, respectively with index $0$ and $n+1$, associated with the tip of each BoN. From \eqref{e.morse torsion}, this translates in 
    \begin{equation}\label{e.homsph}
        b_0(\mathcal{X}_{n+1})=b_{n+1}(\mathcal{X}_{n+1})=1\;,\quad\text{and   }b_k(\mathcal{X}_{n+1})=0\;\text{ otherwise}\;,
    \end{equation}
     and no torsion, or in other words, $\mathcal{X}_{n+1}$ being an \emph{homology} sphere. Through Mayer-Vietoris sequence, one has
     \begin{equation}\label{e.MayerVietoris}
     \adjustbox{scale=0.95,left}{
\begin{tikzcd}
  0 \rar & H_{n+1}(\mathcal{X}_{n+1}) \rar
    & H_n(\mathcal{C}_{n}) \rar\ar[draw=none]{d}[name=X, anchor=center]{}& H_n(\mathcal{B}_{n+1})\oplus H_n(\mathcal{B}'_{n+1})\rar &H_n(\mathcal{X}_{n+1}) \ar[rounded corners,
            to path={ -- ([xshift=2ex]\tikztostart.east)
                      |- (X.center) \tikztonodes
                      -| ([xshift=-3ex]\tikztotarget.west)
                      -- (\tikztotarget)}]{dlll}[at end]{}&\\   
    & \dots \rar&H_{0}(\mathcal{C}_{n}) \rar& H_0(\mathcal{B}_{n+1})\oplus H_0(\mathcal{B}'_{n+1})\rar &H_0(\mathcal{X}_{n+1})\rar&0\;,  
\end{tikzcd}
}
\end{equation}
which implies (splitting between free and torsion parts, see footnote \ref{fn.torsion free})
\begin{subequations}\label{e.double}
\begin{align}
    \sum_{k=0}^{n+1}(-1)^k\left[b_k(\mathcal{X}_{n+1})-b_k(\mathcal{B}_{n+1})-b_k(\mathcal{B}'_{n+1})+b_k(\mathcal{C}_n)\right]&=0\\
    \sum_{k=0}^{n+1}(-1)^k\left[t_k(\mathcal{X}_{n+1})-t_k(\mathcal{B}_{n+1})-t_k(\mathcal{B}'_{n+1})+t_k(\mathcal{C}_n)\right]&=0\;.
    \end{align}
\end{subequations}

\begin{figure}[t!]
				\centering
				\includegraphics[width=0.8\textwidth]{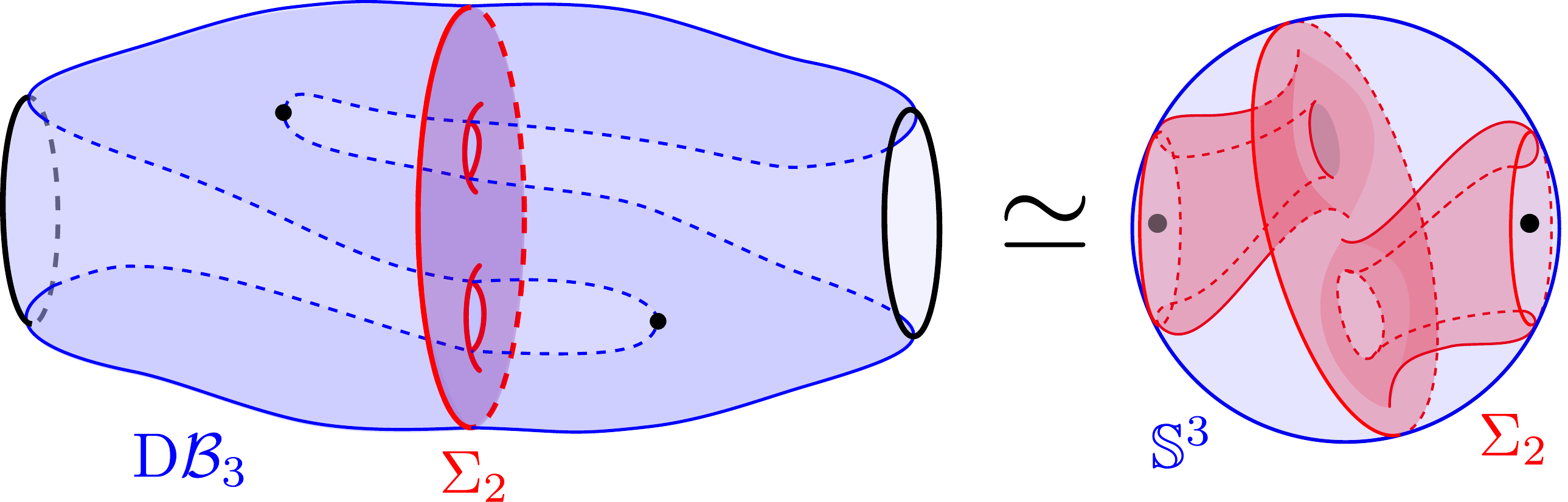}
			\caption{\small Illustration on how the double $\rm D\mathcal{B}_3$ of the minimal bordism $\mathcal{B}_3$ of the genus-2 Riemann surface $\Sigma_2$ is diffeomorphic to the 3-sphere $\mathbb{S}^3$, where $\Sigma_2$ is now seen as a submanifold. While each of the individual bordisms have two critical point loci, the  resulting closed manifold has simply two critical points, required to be able to shrink it to a point. It is easy to see that the same type of deformations apply to arbitrary genus Riemann surfaces.}
			\label{f.homo sphere}
	\end{figure}

    In the absence of torsion (i.e., all $t_k(\mathcal{M})\equiv 0$), a homology sphere solution $\vec{b}(\mathcal{X}_{n+1})=\vec{b}(\mathbb{S}^{n+1})$ is compatible\footnote{Note that this does not imply such solution \emph{exists}, but rather that there is no immediate obstruction to its existence.} with the above \eqref{e.homsph} and \eqref{e.double} equations, by choosing\footnote{For this choice, the resulting ${\mathrm{D}}\mathcal{B}_{n+1}\simeq \left(\mathcal{B}_{n+1}\cup \overline{\mathcal{B}_{n+1}} \right)/_{\sim_{\partial\mathcal{B}_{n+1}}}$ is called \emph{double of} $\mathcal{B}_{n+1}$.} $\mathcal{B}_{n+1}'\simeq \mathcal{B}_{n+1}$ and saturating \eqref{e.ineq}. This is illustrated for $\mathcal{C}_2\simeq \Sigma_2$ in Figure \ref{f.homo sphere}. This means that in principle there should not be a topological obstruction to the collision of two BoN with the compact manifold $\mathcal{X}_{n+1}$ between them shrinking to a point.

    However, in the presence of torsion, its coefficients and Poincar\'e duality do not align, and through \eqref{e.morse torsion} additional critical points in which topology change occurs are needed. As an example, consider the Lens space $\mathcal{C}_3=L(p,q)$ from section \ref{ss:exact seq}, with $\vec{b}(\mathcal{C}_3)=(1,0,0,1)$ and $\vec{t}(\mathcal{C}_3)=(0,1,0,0)$. The bordism saturating \eqref{e.ineq} has $\vec{b}(\mathcal{B}_4)=(1,0,0,0,0)$ and $\vec{t}(\mathcal{B}_4)=(0,1,0,0,0)$. 
    Applying then \eqref{e.MayerVietoris} results in $\vec{b}(\mathcal{B}'_4)=(1,0,0,0,0)$ and $\vec{t}(\mathcal{X}_4)=(0,1,1,0,0)$. From \eqref{e.morse torsion} a total of at least 6 critical points are needed, with $m_4\geq 1$, $m_3\geq 1$, $m_2\geq 2$, $m_1\geq 1$ and $m_0\geq 1$, which means that the $\mathcal{X}_4$ manifold cannot shrink into a point without intermediate topology changes, which as previously stated can result in Planckian curvatures.

\vspace{0.25cm}

    As a final comment, we note that, unlike in the case of a single BoN, which required the presence of D$(8-p)$-branes as cobordisms defects when considering $p$-form fluxes, here their presence is much more mild than the intermediate topology changes commented above. D$(8-p)$-branes, located along along $\mathcal{B}_{n+1}$ and $\mathcal{B}'_{n+1}$, needed to kill the $p$-form charges, have opposite orientations, and thus will annihilate each other when they collide as $\mathcal{X}_{n+1}$ closes. Another way to see this is by considering that the (co)homology groups of $\mathcal{X}_{n+1}$ and $\mathcal{C}_{n}$ are different, and as commented above, $H_k(\mathcal{X}_{n+1},\mathbb{Z})\simeq H^k(\mathcal{X}_{n+1},\mathbb{Z})$ can be trivial for $k=1,\,\dots,\, n$ in the absence of torsion, meaning that the number of $p$-form fields/charges is no longer conserved on the compact manifold connecting the two BoN.

\vspace{0.25cm}

    Going back to the non-minimal $\mathbb{T}^3$ bordism in section \ref{ss:nothing is certain}, as explained in \cite{GarciaEtxebarria:2020xsr}, $\partial \mathcal{B}_4$ can be thought as ``half a K3'' manifold, understood as an elliptical fibration over a $\mathbb{P}^1$ base, with 24 degenerations. This way, the double of our bordism is indeed ${\rm D}\mathcal{B}_4={\rm K3}$, which is not a homology 4-sphere, and as such, cannot be shrunk to a point without intermediate topology changes. The collision of two BoN such as those described in \cite{GarciaEtxebarria:2020xsr} would require Planckian curvatures and stringy effects, even if these were not needed for the single BoN construction.

    \subsection{\textit{Going a bordism further}: Intersection of EotW branes}
    \label{ss:EotW intersect}
    We finally consider the cases with more than one bordism to nothing, in the language of EotW branes. Intersection of these was studied in \cite{Angius:2023uqk,Angius:2024zjv,GarciaEtxebarria:2024jfv}, and in this section we will study them from the point of view of topology changes between ``cobordant'' bordisms to nothing.

    From the EFT point of view, the single EotW brane from section \ref{ss: EotW} was generalized to $k$ intersecting EotW branes in \cite{Angius:2023uqk}. For this the following lower-$d$ EFT action is considered,
    \begin{equation}
        S=\frac{1}{2}\int\dd^dx\sqrt{-g}\left\{\kappa_d^2\left[\mathcal{R}_g-\mathsf{G}_{ij}\partial_\mu\phi^i\partial^\mu\phi^j\right]-2V(\vec\phi)\right\}\;,
    \end{equation}
    where the moduli space metric has unit diagonal, $\mathsf{G}_{ii}=1$, as well as the following ansatz for a conformally flat metric (see \cite{Angius:2024zjv} for EotW solutions with non-conformally flat backgrounds):
    \begin{equation}
        \dd s^2_d=e^{-2\sum_{i=1}^k \sigma_i}\dd s^2_{d-k}+\sum_{i=1}^ke^{-2\sum_{j\neq i}\sigma_ j}(\dd y^i)^2+f_{ij}e^{-\sigma_i-\sigma_j}\dd y^i\dd y^j\;.
    \end{equation}
    with $\phi^i=\phi^i(y^i)$ and $\sigma_i=\sigma^i(y^i)$ and each EotW brane located at $y^i=0$. The $f_{ij}$ metric term is taken to be constant and such that no pair of branes become (anti)parallel (what would result in a degenerate metric). In the same way as in section \ref{ss: EotW}, the potential, metric and scalars have a universal behavior,
    \begin{subequations}
        \begin{align}
            V(\vec\phi)&=\sum_{i=1}^kV_{i,0}e^{\delta_ i\phi^i+\sum_{j\neq i}a_j\delta_j\phi^j}\;,\quad\text{with }\;\mathsf{G}_{ij}=\sqrt{a_ia_j}\\
            \phi^i(y^i)&=-\frac{2}{\delta_i}\log y^i\;,\quad \sigma_i(y^i)=-\frac{4}{(d-2)\delta_i^2}\log y^i\;.
        \end{align}
    \end{subequations}
When moving towards $y^i\to 0$ with the other coordinates fixed, $\phi^i$ scales as if approaching a single EotW brane, eventually exploring a specific limit of moduli space. Different spacetime trajectories when approaching the EotW brane network result in other trajectories along the asymptotic regimes in moduli space. The off-diagonal values of $\mathsf{G}_{ij}$ and $f_{ij}$ depend on the specific construction of the EotW brane intersection.

As explained in \cite{Angius:2023uqk}, the intersection of two EotW branes can be understood as providing a boundary to the theory living on each of the (possibly different) branes, or, more in line with the interpretation we will give now, as domain walls between the different boundaries of the theory.

From the point of view of Morse-Bott theory and the possible topologies of bordisms introduced in section \ref{sec:AlgTop}, intersecting EotW branes have  a very natural reading. We can understand different EotW solutions as having either distinct Morse functions $\rho:\mathcal{B}_{n+1}\to[0,+\infty)$ on the same $\mathcal{B}_{n+1}$ (note that $\rho$ can be directly read from the metric and thus is determined by a solution to the equations of motion), or simply by bordisms to nothing with $\mathcal{B}_{n+1}\not\simeq\mathcal{B}_{n+1}'$. These different bordisms feature different topology changes, either in nature or order, and thus from the EFT point of view have different scalars blowing up, with unrelated critical exponent, determined by the first topology change.

\begin{figure}[t!]
				\centering
				\includegraphics[width=0.65\textwidth]{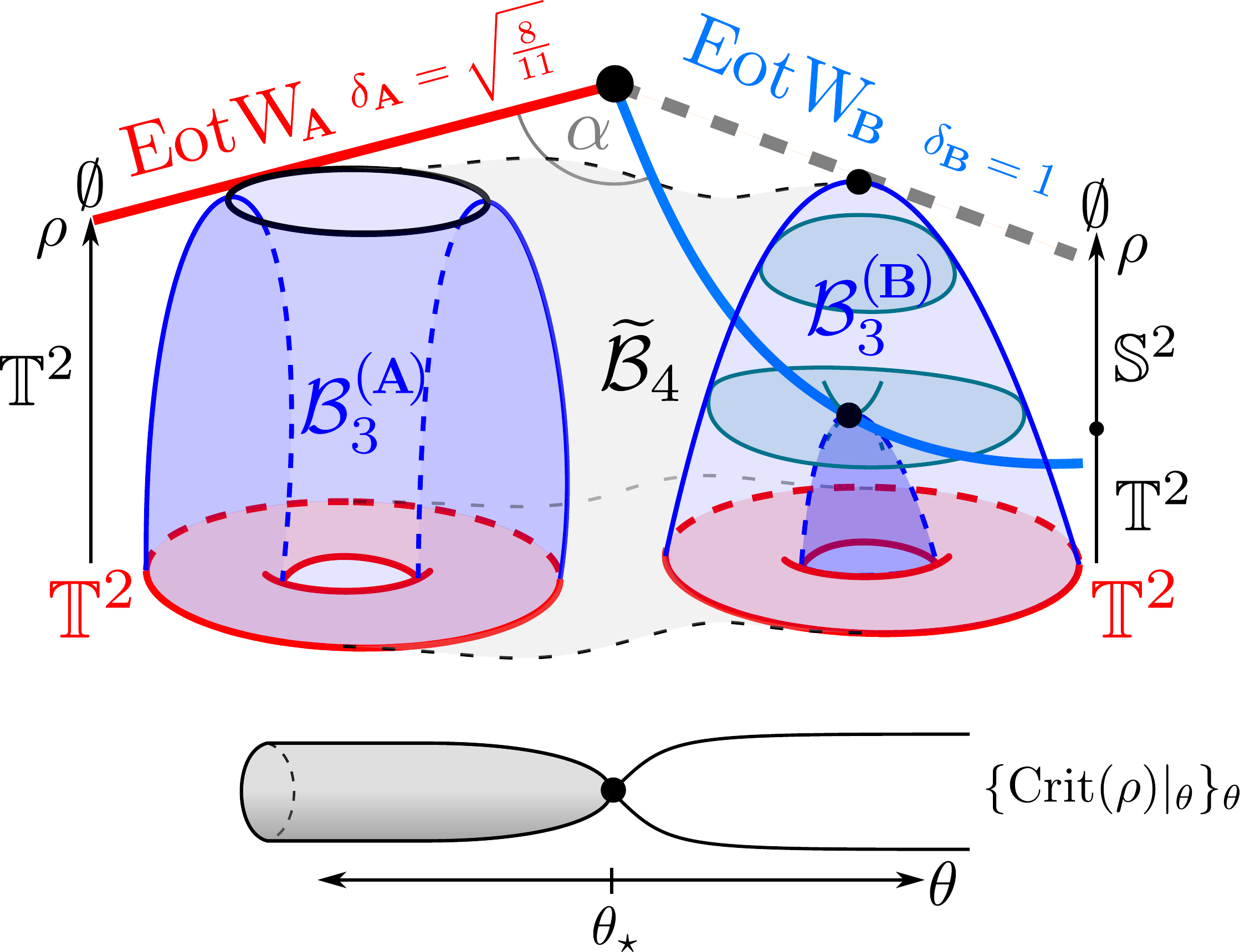}
			\caption{\small Illustration of the intersection of two EotW branes of a $\mathcal{C}_2=\mathbb{T}^2$ compactification, respectively associated to shrinking of a 1-cycle over its dual (\textbf{A}), and of said 1-cycle over a point (\textbf{B}), resulting in a topology change $\mathbb{T}^2\to\mathbb{S}^2$ before the final shrinking of this to nothing. These two possibilities are described in section \ref{ss:IIB Riemann}, and up to the EotW brane surface, they both share the same EFT description. The set $\{{\rm Crit}(\rho)|_\theta\}_\theta$ is also depicted, being topologically $\mathbb{S}^1$ for EotW$_\textbf{A}$ and two points for EotW$_\textbf{B}$. The transition between the two corresponds to the EotW branes intersection. The associated critical exponents are also depicted.}
			\label{f. intersecting t2s2}
	\end{figure}

In the presence of $n$ compact dimensions, we can understand the intersection of an arbitrary number $k<d$ of EotW branes from the perspective of $k$-categories  \cite{Atiyah:1989vu,lurie2009}. From a categorical point of view, the bordisms on $n$-manifolds define a category $\mathbf{Cob}(n+1)$ in the following way:
\begin{itemize}
    \item Elements of $\mathbf{Cob}(n+1)$ are closed and oriented $n$-manifolds.
    \item Given $\mathcal{M}_n,\,\mathcal{N}_{n}\in \mathbf{Cob}(n+1)$, a bordism $\mathcal{B}_{n+1}$ for them defines a morphism $\mathcal{B}_{n+1}:\mathcal{M}_n\to\mathcal{N}_{n}$ in $\mathbf{Cob}(n+1)$. Two bordisms $\mathcal{B}_{n+1}$ and $\mathcal{B}_{n+1}'$ define the same morphism in  $\mathbf{Cob}(n+1)$ if there is an orientation preserving diffeomorphism $\mathcal{B}_{n+1}\simeq\mathcal{B}_{n+1}'$ extending that of their boundaries, $\partial\mathcal{B}_{n+1}\simeq\partial\mathcal{B}_{n+1}'\simeq \mathcal{M}_n\sqcup\overline{\mathcal{N}}_n$.
    \item Given $\mathcal{M}_{n}\in \mathbf{Cob}(n+1)$, the identity morphism ${\rm id}_{\mathcal{M}}$ is given by $\mathcal{I}_{n+1}=\mathcal{M}_n\times[0,1]$.
    \item Composition of morphisms is given by gluing bordisms together by their intermediate boundary.
\end{itemize}
We can use this notion to define the \emph{$k$-category} $\textbf{Cob}_k(n+k)$ recursively, by setting $\textbf{Cob}(n+1)=\textbf{Cob}_1(n+1)$, and defining $\textbf{Cob}_2(n+2)$ as follows:
\begin{itemize}
    \item The elements of $\textbf{Cob}_2(n+2)$ are closed oriented manifolds of dimension $n$.
    \item Given $\mathcal{M}_n,\,\mathcal{N}_n\in\textbf{Cob}_2(n+2)$, the category $\mathscr{C}={\rm Map}_{\textbf{Cob}_2(n+2)}(\mathcal{M}_n,\mathcal{N}_n)$ is given by the bordisms from $\mathcal{M}_n$ to $\mathcal{N}_n$, this is, the oriented $(n+1)$-manifolds $\mathcal{B}_{n+1}$ such that $\partial\mathcal{B}_{n+1}=\mathcal{M}_n\sqcup\overline{\mathcal{N}_n}$. Now, given $\mathcal{B}_{n+1},\,\mathcal{B}'_{n+1}\in\mathscr{C}$, ${\rm Hom}_\mathscr{C}(\mathcal{B}_{n+1},\mathcal{B}'_{n+1})$ is the collections of bordisms $\mathcal{X}_{n+2}$ from $\mathcal{B}_{n+1}$ to $\mathcal{B}'_{n+1}$, reducing to the trivial bordism along their common boundary $\partial\mathcal{B}_{n+1}=\partial\mathcal{B}'_{n+1}=\mathcal{M}_n\sqcup\overline{\mathcal{N}}_n$, in such a way that
    \begin{equation}
        \partial\mathcal{X}_{n+2}\simeq\mathcal{B}_{n+1}\cup\left[(\mathcal{M}_{n}\sqcup\overline{\mathcal{N}_{n}})\times [0,1]\right]\cup\overline{\mathcal{B}'_{n+1}}\;.
    \end{equation}
    In other words, we consider bordisms between bordisms. It can be shown (see \cite{lurie2009} for more details and some nuances in the definitions) that the above structures are well-defined, and as such ${\rm Hom}_\mathscr{C}(\mathcal{B},\mathcal{B}')\neq\emptyset$, so that two bordisms of the same manifolds are always cobordant.
   \end{itemize} 
For our purposes, we will restrict $\textbf{Cob}_2(n+2)$ to consist in $\{\mathcal{C}_n,\emptyset\}$, with the empty set seen as a $n$-dimensional manifold, such that $\mathscr{C}={\rm Map}_{\textbf{Cob}_2(n+2)}(\mathcal{C}_n,\emptyset)$ are the possible bordisms from our compact manifold to nothing. Each of this will correspond to a different EotW brane/BoN construction. Finally, the elements in ${\rm Hom}_\mathscr{C}(\mathcal{B}_{n+1},\mathcal{B}'_{n+1})$ will be the bordisms between this, interpolating between intersecting EotW branes. An ``angular'' coordinate $\theta$ parallel to the two branes and perpendicular to the radial one $\rho$ can act as a new Morse-Bott function over $\mathcal{X}_{n+2}\in {\rm Hom}_\mathscr{C}(\mathcal{B}_{n+1},\mathcal{B}'_{n+1})$. Following an analogous procedure to that of section \ref{sec:AlgTop}, one could bound the topology of $\mathcal{X}_{n+2}$, and from that, the topology changes in going from $\mathcal{B}_{n+1}$ to $\mathcal{B}'_{n+1}$, which can be then understood as the intersecting EotW branes. As the necessary defects to kill topological/cobordism charges should be the same, in principle no additional defects would be needed to be located at these intersections/topology changes (a feature also observed in the universal scalings studied in \cite{Angius:2023uqk}), with $\mathcal{B}_{n+1}$ and $\mathcal{B}'_{n+1}$ differing only on their arrangement. See Figure \ref{f. intersecting t2s2} for a simple example in a toroidal compactification. This analysis could in principle allow us to obtain the qualitative properties and topological structure of intersecting EotW branes, which so far have not been thoroughly explored, though explicit examples appear in \cite{Angius:2023uqk}.

To illustrate the above reasoning, we consider a network of EotW branes in the type IIB on $\Sigma_g$ from section \ref{ss:IIB Riemann}. For simplicity, we take $\Sigma_1\simeq\mathbb{T}^2$. There are two possible bordisms that do not imply the creation of new cycles, namely \textbf{(A)} the shrinking of one cycle over its dual such that $\mathcal{V}\to 0$ and the torus goes to nothing in a single step, and \textbf{(B)} the pinching of a cycle, such that there is a topology change to $\mathbb{S}^2$, which later shrinks to a point. As discussed below \eqref{e. approx pot IIB}, these two bordisms are associated to distinct critical exponents, $\delta_\textbf{A}=\sqrt{\frac{8}{11}}$ and $\delta_\textbf{B}=1$. If the original $\mathbb{T}^2$ is threaded by $F_1$ flux, then this domain wall (which in \textbf{(A)} ends spacetime) corresponds to a D7-brane wrapped along the dual 1-cycle. This is illustrated in Figure \ref{f. intersecting t2s2}. The angle $\alpha>0$ between the different EotW branes should be set by the tension of the unknown object/defect that causes the Morse-Bott function to have different critical points as we move in the parallel direction $\theta$. Understanding its nature would allow us to compute the value of said angle.

Seeing $\{{\rm Crit}(\rho)|_\theta\}_\theta$ as a set parameterized by $\theta$ (for the case shown in Figure \ref{f. intersecting t2s2} this is not even a manifold), the point $\theta_\star$ where the fiber changes topology correspond to the location of the EotW branes intersection.

From Figure \ref{f. intersecting t2s2}, it is easy to visualize analogous EotW branes to \textbf{A} and \textbf{B} given by the bordisms associated to the shrinking/pinching of cycles duals to those presented in Figure \ref{f. intersecting t2s2}. Connecting the different possibilities, one could find a network $$\dots \longleftrightarrow\text{EotW}_\textbf{A}\longleftrightarrow\text{EotW}_\textbf{B}\longleftrightarrow\text{EotW}_\textbf{A'}\longleftrightarrow\text{EotW}_\textbf{B'}\longleftrightarrow\dots\;$$
as we move in the $\theta$ direction. A single type of defect would be enough to explain the same angle $\alpha$ between the branes at each intersection. Identifying its nature and the value of $\alpha$ could allow us to put an upper bound on the number of possible EotW branes, as these could not span more than the celestial sphere. See \cite{Angius:2023uqk} for other explicit computations of the critical exponents in EotW brane intersection networks for compactifications on Calabi-Yau threefolds.

\vspace{0.25cm}

The above reasoning can be extended to intersections of more than two EotW branes. For this we recursively define $\textbf{Cob}_k(n+k)$ (restricted to $\{\mathcal{C}_n,\emptyset\}$), progressively trading spatial directions with coordinates (acting as Morse-Bott functions) in the higher-dimensional bordism. The end point of these sequence of $k$-categories is given by $\textbf{Cob}_{d-1}(d+n-1)$, where every spatial macroscopic direction is capped by an EotW brane, resulting in a (spatially) finite-volume universe. This might be of interest in the context of ``Bubbles of something'' and the ``boundary proposal'' \cite{Friedrich:2023tid,Friedrich:2024aad}, see also \cite{Hawking:1998bn,Turok:1998he,Garriga:1998ri,Bousso:1998pk}, where the universe is nucleated by the reverse process of a bubble of nothing, in such a way that the boundary of these gravitational instantons could be described by local patches of different topology. In any case, we will not comment more on this and leave it for a future work.

\section{Conclusions and Outlook}
\label{sec:conc}

In this paper, we have given a new interpretation of spacetime bordisms to nothing through Morse-Bott theory. The radial distance to the end of spacetime is interpreted as a Morse function over the bordism, and its fibers or level curves correspond to the internal manifold at each spacetime point. We are thus able to obtain properties of the internal topology changes for bordisms to nothing admitting a topological manifold description, learning about which of these changes  must always occur, and where defects, such as D$p$-branes, must be located. Though qualitative in nature, these results serve as a first step towards a piece-wise construction of spacetime ending configurations, such as BoN solutions, for more involved internal geometries. Knowing their topology, it might be possible to proceed analogously to \cite{GarciaEtxebarria:2020xsr}, dividing $\mathcal{B}_{n+1}$ into different regions whose metric can be given an analytic approximate expression, and later gluing them together. As happens with the analysis in section \ref{ss:IIB Riemann}, for large volume regimes in $\mathcal{C}_n$ it is consistent to expect that topology changes on the compact manifold, consisting in the shrinking of some internal cycles, do not immediately trigger topology changes for the rest of cycles. This way it is possible to have an EFT description between the different domain walls.

Furthermore, we are able to extend our results and interpretation to more involved settings such as collision of Bubbles of Nothing, where we identify the possible existence (or lack) of topological obstructions resulting in curvature singularities before the tips of the BoNs come into contact. We also provide an interpretation of the intersection of EotW branes through higher bordisms, which might help in the understanding of the junction structure.

\vspace{0.3cm}
Our results are only a first step, however. The new findings are mostly in the topological, more qualitative, side of the bordisms to nothing. While in section \ref{ss:IIB Riemann} we commented on how, for type IIB compactifications on $\Sigma_g$ each topology change lowered the value of the vacuum energy, in general one would like to also obtain the decay rates associated to them, in line with the examples in \cite{GarciaEtxebarria:2020xsr}, and the general considerations on decay rates/nucleation probabilities on bubbles of nothing/something from \cite{Friedrich:2023tid,Friedrich:2024aad}. Smooth bordisms are interesting precisely for this purpose, as their action can be computed directly with EFT ingredients, without the need to include UV contributions whose behavior might not be well understood, and also can diverge from the EFT perspective. One could expect that the different types of topology changes could be studied separately, in such a way that by considering the different chains \eqref{eq: top chain} $\mathcal{C}_n=\mathcal{C}_n^{(0)}\to \mathcal{C}_n^{(1)}\to\dots\to \mathcal{C}_n^{(m)}=\emptyset$, with the total decay rate given as a sum of blocks associated to each of the changes, in such a way that different bordisms could be sorted by their decay rate. This would be interesting for obvious phenomenological reasons, but goes well over the scope of the paper, and it is left for future work.

Another matter is that of the defects considered. As commented above, for our approach to apply, we need to have smooth bordisms, or at most mild defects such as conical singularities which can be ``smoothed out''. While cobordism defects such as D$p$-branes fulfill this, and indeed we have been able to infer their arrangement along the bordism, other more singular defects, such as O-planes, or more exotic branes \cite{McNamara:2019rup,Dierigl:2023jdp}, might not guarantee that. It would be nice to understand which types of defects allow for some \emph{smooth(able) manifold} description and which do not. As evidenced by the example from \cite{Dierigl:2023jdp} shown in Figure \ref{ff.thickwall}, some of these defects or configurations require for non-trivial cycles along $\mathcal{B}_{n+1}$ (resulting in \emph{thick} domain walls), and thus cannot be located along minimal bordisms. A general understanding of when more involved bordisms are required would also have important implications for the construction of realistic BoN.

Apart from topological constrains, we have not considered the possible dynamical restrictions to vacuum decay. It is possible that, in the process of breaking some of them, specific decay channels or bordism topologies are singled-out or excluded. We leave for a future project the better understanding of the relation between these dynamical constrains and processes through which the topology/geometry of $\mathcal{B}_{n+1}$ changes. On the other hand, as explained in Appendix \ref{app:tachyon} and \cite{Saltman:2004jh,Adams:2005rb}, topology changes consisting in the decay of a 1-cycle can be explained through tachyon condensation. One expects that tachyon condensation is also involved in decay of higher dimensional cycles, but the explicit mechanism does not directly translate from the 1-cycle case. The worldsheet interpretation of $n$-cycle closing is an interesting problem we would like to return to in the future.

Other type of restrictions would be those in which a fibration structure is required along the complete bordism direction. For example, F-theory compactifications need to keep the elliptic fiber for all values of $\rho$, and thus $\mathcal{B}_{n+1}$, with topology changes in which said fiber disappears not being considered. A possible solution would be trying to extend the $\mathbb{T}^2\hookrightarrow\mathcal{C}_n\to \mathcal{D}_{n-2}$ fibration to $\mathbb{T}^2\hookrightarrow\mathcal{B}_{n+1}\to \hat{\mathcal{D}}_{n-1}$, in such a way that $\partial \hat{\mathcal{D}}_{n-1}=\mathcal{D}_{n-2}$ and the bordism fibration restricting to that of $\mathcal{C}_n$ along the boundary. How to properly perform this is beyond the scope of this paper.

\vspace{0.3cm}
As commented in section \ref{sec:AlgTop}, we have made use of two sets of inequalities relating the topology of the bordism $\mathcal{B}_{n+1}$ with the number and type of topology changes/critical points experienced, namely \emph{Morse-Bott inequalities for manifolds with boundary} (not including torsion) \eqref{e.MorseBottIneq0} and \emph{Morse inequalities (including torsion) for manifolds without boundary} \eqref{e.morse torsion}. As the former does not include torsion, while the later requires $\partial \mathcal{M}=\emptyset$ and only consider critical points rather than general submanifolds, their range of applicability is different. Ideally one would like to incorporate torsion into \eqref{e.MorseBottIneq0}, but as far as we know this is not present in the mathematics literature. On the other hand, while \eqref{e.MorseBottIneq0} apply also when the compactification manifold $\mathcal{C}_n$ has a non-zero boundary, it would be interesting to obtain a generalization of the bounds \eqref{e.ineq} on the topology of the bordism for cases other than $\partial \mathcal{C}_n=\emptyset$, as this was an assumption that was needed in their derivation. 

It has been conjectured \cite{Strominger:1996it,chan2014strominger} that Calabi-Yau threefolds always admit a $\mathbb{T}^3$ fibration, what motivated the authors in \cite{GarciaEtxebarria:2020xsr} to propose that the BoN solutions constructed for spin bordisms on this manifold (see also \cite{Acharya:2019mcu} for the additional bordisms of $\mathbb{T}^3$ to nothing) could be used as decay channels for realistic CY$_3$ compactifications. However, as we have shown with examples such as precisely $\mathbb{T}^2\hookrightarrow\mathcal{B}_4\to\mathbb{D}^2$ bordism for $\mathbb{T}^3$ in section \ref{ss:nothing is certain}, bordisms given by non trivial fibrations have different homology groups than their trivial counterparts, resulting in a different topology change sequence. While the claim of \cite{GarciaEtxebarria:2020xsr} probably holds in the sense that an analogous decay channel exists, it is not immediate that the constructions known in the literature can be straightforwardly used, as the $\mathbb{T}^3$ fibration might not be trivial.

Although we have simply considered the $\rho$ component of the bordism metric as our Morse function, it is easy to see that any $p$-form field $\omega$ defined on $\mathcal{C}_n$ can be extended into the whole bordism $\mathcal{B}_{n+1}$ as a $(p+1)$-form by taking $\omega'=\omega\wedge\dd\rho$, being zero at critical points of $\rho$. It might be interesting to relate general properties of these fields with those of the bordism. Furthermore, we would like to point out that only (co)homology coefficients such as Betti numbers and torsion ranks have been considered. We leave for future work the possibility of recovering other topological invariants of the bordism from those of $\mathcal{C}_n$, as well as understanding to which dynamical properties of the bordism to nothing they are related to. A first step for this would be the derivation of generalizations to the Morse-Bott inequalities \eqref{e.MorseBottIneq0} which also take into account, e.g., the spin structure along the boundary, not only its homology. This could result in higher lower bounds on the number of topology changes, maybe finding that the 13 topology changes for the $\mathbb{T}^3$ with periodic boundary conditions from section \ref{ss:nothing is certain} saturate said new bounds.

\vspace{0.3cm}
There is still much to be understood and computed about the possible channels through which our universe can decay to nothing. Computations are usually quite involved, so that apart from specific cases \cite{GarciaEtxebarria:2020xsr}, most constructions in the literature feature relatively simple compactifications. We believe that studying more complicated cases, leading to a better understanding of how our universe can be created and disappear into nothing, is within reach, and we expect that this paper serves as a small step in this direction.

\bigskip
{\bf Acknowledgments:}
We are very grateful to Matilda Delgado, Miguel Montero and Irene Valenzuela for their invaluable contributions through extensive discussions and comments on the manuscript. We would like to also thank Alberto Castellano, Markus Dierigl, Bj\"orn Friedrich, Arthur Hebecker, Camilo las Heras, Jes\'us Huertas, Andriana Makridou, Luca Melotti, Jacob McNamara, Jakob Moritz, Michelangelo Tartaglia, Eduardo Velasco and Matteo Zatti for illuminating discussions and comments along the past two years. I.R. wishes to acknowledge the hospitality of the Department of Theoretical Physics at CERN and the Department of Physics of Harvard University during different stages of this work. I.R. acknowledges the support of the Spanish FPI grant No. PRE2020-094163 and the ERC Starting Grant QGuide101042568 - StG 2021, as well as the Spanish Agencia Estatal de Investigaci\'on through the grant ``IFT Centro de Excelencia Severo Ochoa'' CEX2020-001007-S and the grant PID2021-123017NB-I00, funded by MCIN/AEI/10.13039/ 501100011033 and by ERDF ``A way of making Europe''.

\appendix
\section{Basics of Compact Riemann Surfaces}
\label{app:RiemannianSurfaces}

In this appendix we review the basic results about Riemann surfaces that are used in section \ref{ss:IIB Riemann}. For our purposes, a (compact) genus-$g$ Riemann surface $\Sigma_g$ can be seen as the connected sum of $g$ 2-torus $\mathbb{T}^2=\mathbb{S}^1\times\mathbb{S}^1$, $\Sigma_g:=\#^g\mathbb{T}^2$, with $\Sigma_0=\#^0\mathbb{T}^2\simeq\mathbb{S}^2$. Regardless of its genus, every compact Riemann surface $\Sigma$ admits a conformal Riemann metric \cite{Jost:2006}, i.e., $\dd s_\Sigma^2=\lambda(z)^2\dd z \dd\bar{z}$, with $\lambda\in\mathcal{C}^2(\Sigma)$ and $z=x+i y$. A simple computation yields that the Ricci scalar is given by $\mathcal{R}_\Sigma=-2\Delta\log\lambda$,
with $\Delta=\lambda^{-2}(\partial_x^2+\partial_y^2)$. Given $\Gamma$ a discrete subgroup of
\begin{equation}
    PSL(2,\mathbb{R})=\left\{\begin{pmatrix}
    a&b\\
    c&d
    \end{pmatrix}:\, ad-bc=1\right\}\,,
\end{equation}
acting discontinuously and without fixed points on the upper half plan $\mathbb{H}=\{z=x+iy\in\mathbb{C}:\, y>0\}$, then $\Sigma=\mathbb{H}/\Gamma$ is a compact Riemann surface. Then, for an arbitrary $z_0\in\mathbb{H}$, the set
\begin{equation}
    F=\left\{z\in\mathbb{C}:\,d(z,z_0)\leq d(z,\gamma z_0))\forall \gamma\in \Gamma\right\}\;,
\end{equation}
where $d$ is the geodesic distance associated to the natural hyperbolic metric $\dd s_\mathbb{H}^2=\frac{\dd z\dd \bar z}{2{\rm Im}(z)^2}$, is a convex polygon with finitely many sides, such that all the vertices are equivalent. It can then be shown \cite{Jost:2006} that edges of $F$ can be paired $a\equiv a'$ uniquely, in such a way that the sides can be ordered as either $
    a_0a_0'$ or $a_1b_1a'_1b'_1\dots a_gb_ga'_gb'_g$
in such a way that $F$ has $2$ or $4g$ sides. The fundamental group of $\Sigma$ (which is obtained from $F$ after sides are identified) is then 
\begin{equation}
   \pi_1(\Sigma)=0\,,\qquad\text{or}\qquad\pi_1(\Sigma)=\left\{\langle\{a_i,b_i\}_{i=1}^g\rangle\,:\,a_1b_1a_1^{-1}b_1^{-1}\dots a_pb_pa_p^{-1}b_p^{-1}=1\right\}\;,
\end{equation}
respectively. Being the abelianization of the fundamental group, the first homology groups are then
\begin{equation}
    H_1(\Sigma;\mathbb{Z})=0\,,\qquad\text{or}\qquad H_1(\Sigma;\mathbb{Z})=\mathbb{Z}^{2g}\,,
\end{equation}
respectively corresponding with $\mathbb{S}^2=\Sigma_0$ or $\Sigma_g$.

For $g\geq 1$ we can take dual canonical homology and cohomology basis $\{a_i,b_i\}_{i=1}^g$ and $\{\alpha^i,\beta^i\}_{i=1}^g$ such that we have the following intersections and integrals \cite{Saltman:2004jh}:
\begin{subequations}\label{eq.normal basis}
\begin{align}
	a_i\cdot a_j=b_i\cdot b_j=0,\qquad a_i\cdot b_j=-b_i\cdot a_j=\delta_{ij}\\
	\int_{a_i}\beta^j=\int_{\beta_i}\alpha^j=0,\qquad \int_{a_i}\alpha^j=\int_{b_i}\beta^j=\delta^{ij}\;,
\end{align}
\end{subequations}
such that for any 1-form $\eta\in\Omega^1(\Sigma_g)$
\begin{equation}
\int_{a_i}\eta=\int_{\Sigma_g}\eta\wedge\beta^i,\qquad\int_{b_i}\eta=\int_{\Sigma_g}\alpha^i\wedge\eta\;
\end{equation}
in such a way that one can expand any 1-form as $\eta=m_i\alpha^i+n_j\beta^j$. On the other hand, we can define a holomorphic basis $\{\omega^i\}_{i=1}^g$ with
\begin{equation}\label{eq.holo basis}
\int_{a_i}\omega^j=\delta^{ij},\qquad \int_{b_j}\omega^j=\Omega^{ij}\;,
\end{equation}
where $\Omega$ is a symmetric matrix with positive definite imaginary part known as \emph{period matrix}. One can show that $\Omega$ only depends on the homology classes of the 1-cycles, with the space of period matrices, $\mathbb{H}^g$, having complex dimension $\frac{1}{2}g(g+1)$.

\vspace{0.25cm}

On the other hand, from the Riemann-Roch theorem \cite{Jost:2006,Blumenhagen:2013fgp}, one has that
\begin{equation}
    \#\text{complex moduli parameters}-\#\text{conformal Killing vectors}=-\tfrac{3}{2}\chi(\Sigma_g)=3(g-1)\;,
\end{equation}
which we can study separately:
\begin{itemize}
    \item For $\Sigma_0=\mathbb{S}^2$, we have as metric $\dd s^2=\frac{4\dd z \dd\bar z}{(1+|z|^2)^2}$, with 3 conformal Killing vectors, $\{z^{a}\partial_z\}_{a=0}^2$, resulting in no complex structure moduli (one must still consider the volume).
    \item For $\Sigma_1=\mathbb{T}^2$, we have a flat metric $\dd s^2=\dd z\dd \bar z$. The translations along the torus are given by $U(1)\times U(1)$, with a single complex generator, resulting in one conformal Killing vector and one complex moduli parameter, $\tau=\tau_1+i\tau_2\in\mathbb{H}$. As one considers the $SL(2,\mathbb{Z})$ redundancies, the actual fundamental domain is smaller and corresponding with the famous $\mathcal{F}=\left\{-\frac{1}{2}\leq\tau_1\leq 0\,,\,|\tau|^2\geq 1\right\}\cup\left\{0<\tau\leq \frac{1}{2}\,,\,|\tau|^2>1\right\}$.
    \item For $g\geq 2$ there are no conformal isometries, with the number of complex moduli parameters being $3g-3$. While for $g=2,\,3$ this is the same as $\frac{1}{2}g(g+1)$, for $g\geq 4 $ the Teichm\"uller space $\mathcal{T}_g$ is embedded in $\mathbb{H}^g$ in a non trivial way \cite{Jost:2006}, so that no all choices of $\tau\in \mathbb{H}^g$ actually correspond to a complex structure in $\Sigma_g$. In any case, given $\Sigma_g$ and its homology classes, $\tau$ is uniquely determined. In general, we will have the same problem of defining the fundamental region $\mathcal{F}_g$ from $\mathcal{T}_g$, with \cite{Blumenhagen:2013fgp}
    \begin{equation}
        \mathcal{F}_g=\frac{\mathcal{T}_g}{{\rm Diff}/{\rm Diff}_0}\;,
    \end{equation}
    where $\rm Diff$ are the diffeomorphisms of $\Sigma_g$ and $\rm Diff_0$ those connected to the identity. We will not develop further in this direction.
    
\end{itemize}

\begin{figure}[h!]
				\centering
				\includegraphics[width=0.9\textwidth]{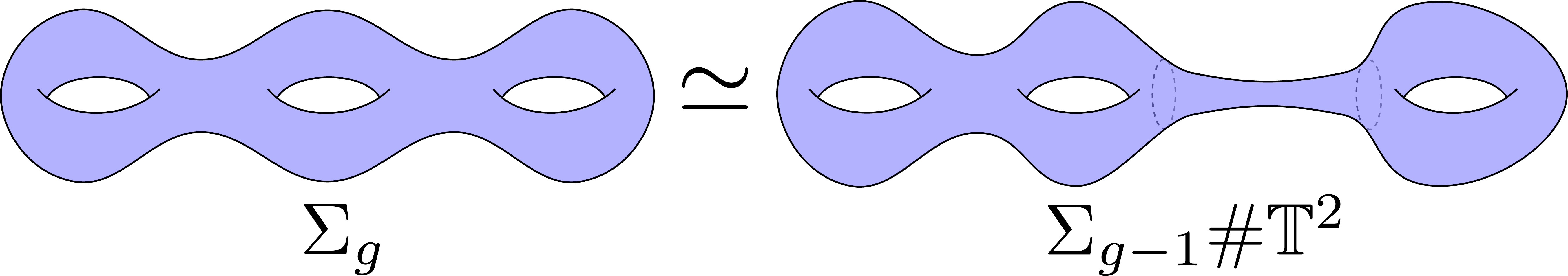}
			\caption{\small Illustration of how a handle of a Riemann surface $\Sigma_g$ can be factorized from the main body by considering the connected sum of a torus $\mathbb{T}^2$ and a lower genus $\Sigma_g$, joined together by a thin tube. For large $g$ the volume of the whole surface is approximately constant, and the dynamics of the handle decay should not affect those of the rest of the manifold. In this limit the factorized handle can be effectively parameterized as $\mathbb{T}^2$.}
			\label{f.deg handle}
	\end{figure}

    The different limits in which a handle degenerates and topology changes occurs are expected to be associated with one of the eigenvalues of $\Omega$ going to infinity or 0. In general, the explicit expression of the complex structure moduli space metric are quite involved and depend on the specific realization and global metric of $\Sigma_g$. However, as shown in Figure \ref{f.deg handle}, we can approximate this by considering a handle of $\Sigma_g\simeq \Sigma_{g-1}\#\mathbb{T}^2$ effectively factorized from the rest, so that it can be seen as $\mathbb{T}^2$, whose (constant volume) metric and complex structure moduli space are given by
    \begin{equation}\label{e. complex structure}
        g_{ab}=\tau_2^{-1}\begin{pmatrix}
            1&\tau_1\\
            \tau_1&|\tau|^2
        \end{pmatrix}\,,\quad\mathsf{G}_{ij}\partial_\mu\varphi^i\partial^\mu\varphi^j\supseteq\frac{\partial_\mu\tau\partial^\mu\bar\tau}{2\tau_2^2}=\frac{(\partial \tau_1)^2}{2\tau_2^2}+\left[\partial\left(\frac{1}{\sqrt{2}}\log\tau_2\right)\right]^2
    \end{equation}
    The decay of said handle can be interpreted as a $\tau_2\to 0,\,+\infty$ limit, which indeed are at infinite distance.

After the previous machinery has been introduced, we will now comment on two specific results from the literature that are used in section \ref{ss:IIB Riemann} and Appendix \ref{app:tachyon}   . First of all, from the \emph{Uniformization theorem} of Schwarz and Klein (see section 4.4 from \cite{Jost:2006} for proof in modern language), every compact Riemann surface admits constant curvature metric. This is indeed useful, as from Gauss-Bonnet theorem we have
\begin{equation}
    \int_{\sigma_g}\sqrt{g_\Sigma}\mathcal{R}_\Sigma=\pi\chi(\Sigma_g)=2\pi(1-g)\,,
\end{equation}
in such cases we would have $\mathcal{R}_\Sigma=2\pi(1-g)\volume(\Sigma_g)^{-1}$. The question is now whether such metric (which results in an Einstein manifold, so solves the equation of motion in the vacuum) is dynamically reached. Indeed, Einstein-dilaton actions,
\begin{equation}
    \int_{\Sigma_g}\dd^2 x\sqrt{g_{\Sigma}}e^{-2\Phi}\left\{\mathcal{R}_\Sigma+4(\partial\Phi)^2\right\}\;,
\end{equation}
feature a Ricci flow (see \cite{Perelman:2006un,Velazquez:2022eco} and references therein) 
\begin{equation}
    \partial_\sigma g_{mn}=-2(\mathcal{R}_{mn}+2\nabla_m\partial_n\Phi)\,,\qquad\partial_\sigma \Phi=-\frac{1}{2}\mathcal{R}_\Sigma-\Delta \Phi\;,
\end{equation}
with $\sigma$ some flow parameter, such that they eventually reach a constant curvature and dilaton configuration. This result no longer holds once extra ingredients, such as fluxes or backreacting branes are introducing, but it is safe to assume that if these are diluted or separated enough, most of the manifold will have an approximately constant scalar curvature.

Finally, we will introduce some machinery in order that we will use for flux potentials. Given a 1-form $F\in \Omega^1(\Sigma_g)$, we can expand it in both the dual and holomorphic basis as \cite{Saltman:2004jh}
\begin{equation}
    F=m^i\alpha_i+n^i\beta_i=u^iw_i+\bar{u i}\bar{w_i}\;,
\end{equation}
modulo an exact form which is not relevant here. From \eqref{eq.normal basis} and \eqref{eq.holo basis} we find that the different coefficients are related by
\begin{equation}
    \begin{pmatrix}
        m\\n
    \end{pmatrix}=\begin{pmatrix}
        \mathbb{I}&\mathbb{I}\\
        \Omega&\bar{\Omega}
    \end{pmatrix}
    \begin{pmatrix}
        u\\\bar{u}
    \end{pmatrix}\;.
\end{equation}
On the other hand, on the $w$ basis the Hodge star acts as
\begin{equation}
    \left.\begin{array}{rl}
         \star w&=-iw  \\
         \star\bar{w}&=i\bar w 
    \end{array}\right\}\Longrightarrow \begin{pmatrix}
        \star u\\\star\bar u
    \end{pmatrix}=\begin{pmatrix}
        -i\mathbb{I}&0\\
        0&i \mathbb{I}
    \end{pmatrix}
    \begin{pmatrix}
        u\\\bar{u}
    \end{pmatrix}\;.
\end{equation}
Finally, from \eqref{eq.normal basis}, we are finally able to write the Hodge norm (which is precisely the induced flux potential) of $F$ as
\begin{equation}
    \|F\|^2=\int_{\Sigma_g}F\wedge\star F=\begin{pmatrix}m&n\end{pmatrix}\mathbb{A}\begin{pmatrix}
        m\\n
    \end{pmatrix}\;,
\end{equation}
with $\mathbb{A}$ a $2g\times 2g$ real, positive definite symmetric matrix given by
\begin{equation}\label{e:mathbbA}
    \mathbb{A}=i\begin{pmatrix}
        2\Omega(\Omega-\bar\Omega)^{-1}\bar\Omega&-(\Omega+\bar\Omega)(\Omega-\bar\Omega)^{-1}\\
        -(\Omega+\bar\Omega)(\Omega-\bar\Omega)^{-1}&2(\Omega-\bar\Omega)^{-1}
    \end{pmatrix}\;,    
\end{equation}
where $\Omega$ is the period matrix.

\section{A quick overview of 1-cycle decay through tachyon condensation.}\label{app:tachyon}
     
    In this appendix  we give some basic overview of handle decay in Riemann surfaces, i.e., $\Sigma_g\to\Sigma_{g-1}$ via tachyon condensation. For more details, see \cite{Saltman:2004jh,Adams:2005rb}. We will work in the weak coupling and large volume limit ($g_s\ll1$ and $\mathcal{V}\gg \ell_s^2$), so that string interactions and corrections can be safely ignored. As we will discuss later this section, this is the regime towards which a compactification on $\Sigma_g$ with $g\geq 2$ naturally evolves. From \eqref{e.spin bord}, $\Omega^{\rm Spin}_1=\langle\mathbb{S}_p^1 \rangle\simeq \mathbb{Z}_2$, so 1-cycles need anti-periodic boundary conditions in order to shrink them to nothing (otherwise a spin defect would need to be included\footnote{From \cite{McNamara:2019rup}, in type the IIA $\mathbb{S}^1_p$ class in $\Omega_1^{\rm Spin}$ is killed through a $\mathbb{R}/\mathbb{Z}_2\times\mathbb{S}^1_p$ defect consisting in an O8-plane wrapped along the circle (in the bulk, not in the boundary to nothing). For type IIB, the T-dual of the defect has a $(\mathbb{R}\times\mathbb{S}^1_p)/\mathbb{Z}_2$ geometry, with two O7-planes located at each end of the interval. These introduced defects are stringy by nature but, being located in the bulk of the bordism, should still allow for a smooth bordism to nothing after them.}).

     As shown in Appendix \ref{app:RiemannianSurfaces}, Ricci flow results in $\Sigma_g$ evolving towards having constant-curvature. For large volume, Gauss-Bonnet theorem implies $\alpha'\mathcal{R}_n=2\pi(1-g)\frac{\alpha'}{\mathcal{V}}\ll1$, so that we can consider the topology change locus neighborhood to be approximately flat, locally looking like a SUSY-breaking Scherk-Schwarz circle \cite{Scherk:1978ta,Scherk:1979zr,Kounnas:1989dk} of radius $\mathsf{R}$ with anti-periodic boundary conditions times a small interval, parameterized by $(\theta,r)\in\mathbb{S}^1\times(r_0-\epsilon,r_0+\epsilon)$. The world-sheet energy of the $n$-th twisted sector is given by
     \begin{equation}
         \frac{m_n^2}{m_{\rm s}^2}=-\frac{1}{2}+n^2\frac{\mathsf{R}^2}{\alpha'}\;,
     \end{equation}
     so that for $\mathsf{R}<\sqrt{2\alpha'}$ the spectrum will have tachyonic states, resulting in a instability in the internal geometry. As shown in footnote \ref{fn:stab}, a flux potential automatically stabilizes the complex structure moduli at finite values, so $\mathsf{R}<\sqrt{2\alpha'}$ is energetically obstructed (though as seen in section \ref{ss:IIB Riemann}, topology changes are energetically favored, at least in the studied settings). As we will argue now, this tachyon condensation results in a barrier that prevents propagation trough the $\mathbb{S}^1\times(r_0-\epsilon,r_0+\epsilon)$ tube, effectively changing the topology of our internal geometry.

     To do so, a (zero ghost picture) tachyon vertex is added to the $(1,1)$ local world-sheet action,
     \begin{equation}\label{e:tachyon}
         S\supseteq \int \dd^2\sigma \dd\vartheta^+\dd\vartheta^- T(X)\;,\quad \text{where}\quad T(X)=e^{k_0X^0}\hat{T}(R)\cos(\omega\tilde{\Theta})\,,
     \end{equation}
     with $X^0$, $R$ and $\tilde{\Theta}$ are the superspace coordinates associated to time $t$, radial direction $r$ and the T-dual $\bar\theta$ to $\theta$ \cite{Adams:2005rb}, with $\hat{T}(R)$ complex as we have both winding and anti-winding modes in $T(X)$. It is further assumed that $\hat{T}(R)$ is localized at $r=r_0$, where the tube radius is the smallest, with exponentially decaying profile in the $r$ direction away from there. This results in a (classical) bosonic potential in the world-sheet action with the form
         \begin{align}
             U(X)&=(\partial T)^2\left\{\left[-k_0^2\cos^2(\omega \tilde{\theta})+\omega^2\sin^2(\omega\tilde{\theta})\right]\hat{T}(R)^2+\cos^2(\omega\tilde{\theta})(\partial_r\hat{T})^2\right\}e^{2k_0X^0}            
         \end{align}

     Now, for $\omega^2\gg k_r^2>k_0^2$, this results in a \emph{classical} barrier for all values of $\tilde\theta$ and $r$ over which $\hat{T}$ has support, precisely when $\partial_r\hat{T}$ becomes significant. This prevents the different world-sheet modes from propagating along the tube, effectively disconnecting the two sides of the handle and changing the topology of the internal manifold. This qualitative behavior remains after quantum effects are considered, see sections 2 and 3 from \cite{Adams:2005rb}.

     As a final note, consider the following. Due to $\hat{T}$ having a maximum at $r_0$, $U(X)$ will have a minimum there (in general more, depending on the explicit shape of the profile), resulting in isolated vacua in which the $\Theta$ and $R$ directions have been removed, with the resulting vacua being subcritical (this will result in additional bulk tachyons). This is evident from the invariance of the Witten index ${\rm Tr}[(-1)^{\mathsf{F}}e^{-\beta H}]$, which for supersymmetric sigma models such as this one is given by the Euler characteristic of the manifold our fields take values one. Now, as after the handle loss $\chi (\Sigma_{g-1})=\chi(\Sigma_g)+2$, then the new vacua will precisely have $n_F^{(0)}-n_B^{(0)}=2$, as expected from the pair of bosonic d.o.f.'s lost.

\vspace*{.5cm}
\bibliographystyle{JHEP2015}
\bibliography{ref}

\end{document}